\documentclass[aps,prb,notitlepage,twocolumn,longbibliography]{revtex4-2}
\usepackage[latin9]{inputenc}
\usepackage{array}
\usepackage{amsmath}
\usepackage{amssymb}
\usepackage{graphicx}
\usepackage{esint}
\providecommand{\tabularnewline}{\\}
\usepackage{natbib}
\usepackage{amsfonts}
\usepackage{hyperref}

\begin{document}
\title{Origin of anomalous magnetotransport in kagome superconductors \emph{A}V$_{3}$Sb$_{5}$ (\emph{A}=K,Rb,Cs) }
\author{A. E. Koshelev}
\affiliation{Materials Science Division, Argonne National Laboratory, Lemont,
Illinois 60439}
\affiliation{Dept. of Physics and Astronomy, University of Notre Dame, Notre Dame, Indiana, 46656}
\author{R. Chapai}
\affiliation{Materials Science Division, Argonne National Laboratory, Lemont,
	Illinois 60439}
\author{D. Y. Chung}
\affiliation{Materials Science Division, Argonne National Laboratory, Lemont,
	Illinois 60439}
\author{J. F. Mitchell}
\affiliation{Materials Science Division, Argonne National Laboratory, Lemont,
	Illinois 60439}
\author{U. Welp}
\affiliation{Materials Science Division, Argonne National Laboratory, Lemont,
	Illinois 60439}
\date{\today}
\begin{abstract}
Multiple anomalous features in electronic spectra of metals with kagome lattice
structure -- van Hove singularities, Dirac points, and flat bands -- 
imply that materials containing this structural motif may lie at a nexus of topological and correlated electron physics. 
Due to the prospects of such exceptional electronic behavior, the recent  discovery of superconductivity coexisting with charge-density wave (CDW) order in the layered kagome metals \emph{A}V$_{3}$Sb$_{5}$ (\emph{A}=K,Rb,Cs) has attracted considerable attention.
Notably, these archetypal kagome metals express unconventional magnetotransport behavior, including an unexpected linear-in-$H$ diagonal resistivity at low fields, and an even more peculiar, nonmonotonic sign-changing behavior of the Hall resistivity, which has been speculated to arise from a chiral CDW. 
We argue here that this unusual magnetotransport 
derives not from such unconventional phenomena, but rather from the unique fermiology of the \emph{A}V$_{3}$Sb$_{5}$ materials. Specifically, it is caused by 
a large, concave hexagonal Fermi surface sheet formed in the close proximity to the van Hove singularities, which is backfolded into a small hexagonal sheet and two large triangular sheets in the CDW state. 
We introduce and analyze a model of the electronic structure of these Fermi surface sheets that allows for a full analytical treatment within Boltzmann kinetic theory 
and that enables semi-quantitative fits of our transport data. 
Specifically, we find that the anomalous magnetotransport behavior is caused by the confluence of strong reduction of the Fermi velocity near the van Hove singularities located near the vertices of the hexagonal sheet and sharp corners in Fermi surface generated by the CDW reconstruction. 
Our analytical approach not only explains the anomalous magnetotransport in the kagome superconductors but also can be extended to a variety of metallic systems hosting singular features in their Fermi surfaces.
\end{abstract}
\maketitle

\section{Introduction \label{sec:intro}}

Kagome lattices composed of networks of corner-sharing triangles have been
recognized as a fertile ground to realize novel electronic and
magnetic phenomena. Metallic materials with layered kagome-lattice
crystal structure display multiple anomalous features in their electronic
band structure including Dirac points, van Hove singularities, and
flat bands \citep{GuoPhysRevB.80.113102,WangPhysRevB.87.115135,MazinNatComm14}.
These features inspired several theoretical proposals for exotic electronic
states \citep{KoPhysRevB.79.214502,YuPhysRevB.85.144402,WangPhysRevB.87.115135,KieselPhysRevLett.110.126405,MazinNatComm14,FengSciBull21,DennerPhysRevLett.127.217601,LinPhysRevB.104.045122,ScammellNatComm23}
such as bond order, loop current state, chiral charge-density wave
or unconventional superconductivity. Against this background,
the recent discovery of kagome-lattice superconductors \emph{A}V$_{3}$Sb$_{5}$
with \emph{A}=K,Rb,Cs \citep{OrtizPhysRevMaterials.3.094407} attracted
considerable attention and triggered extensive research activity, see recent reviews \citep{ChenChinhysB2022,NeupertNatPhys2022,JiangNatSciRev22,HuNPJQM2023}.
The dominant structural motif in these layered materials are kagome
nets of vanadium atoms formed around antimony honeycomb lattices. These
materials display two electronic instabilities\citep{ChenChinhysB2022,NeupertNatPhys2022,JiangNatSciRev22,HuNPJQM2023}: 
a charge-density wave transition
at $T_{\mathrm{CDW}}\!=80\!-\!100$K followed by superconducting transition
at $T_{\mathrm{SC}}\!=1\!-\!2.7$K (in particular, for CsV$_{3}$Sb$_{5}$,
$T_{\mathrm{CDW}}\!=\!94$K and $T_{\mathrm{SC}}\!=\!2.7$K). The
CDW instability develops simultaneously at three wave vectors (3Q CDW).
The possible realizations are 2$\times$2 reconstructions in the form of star of David, inverse star of David
(or trihexagonal) patterns, or alternating combination of these states\citep{OrtizPhysRevX.11.041030}.
In addition, experimental indications of chiral charge order and time-reversal symmetry breaking  have
been reported \citep{JiangNatMat2021,WangPhysRevB.104.075148,ShumiyaPhysRevB.104.035131,yu2021ArXiv,XuNatPhys2022,KhasanovPhysRevResearch.4.023244,MielkeNat2022,LeNature2024}. 
The exact nature of the CDW order and its relation to superconductivity remain open issues.

The band structure of the kagome superconductors experimentally determined
by angle-resolved photoemission spectroscopy (ARPES) \citep{ChoPhysRevLett.127.236401,NakayamaPhysRevB.104.L161112,KangNatPhys2022,HuNatComm2022,KatoCommMat2022,LuoPhysRevLett.128.036402,LuoPhysRevB.105.L241111,LuoNatComm2022,HuPhysRevB.106.L241106}
and quantum oscillations \citep{OrtizPhysRevX.11.041030,FuPhysRevLett.127.207002,ShresthaPhysRevB.105.024508,ZhangPhysRevB.106.195103,HuangPhysRevB.106.064510,BroylesPhysRevLett.129.157001,ChapaiPhysRevLett.130.126401}
is qualitatively consistent with electronic band-structure calculations
based on density-functional theory (DFT)\citep{FuPhysRevLett.127.207002,TanPhysRevLett.127.046401,OrtizPhysRevX.11.041030,LaBollitaPhysRevB.104.205129,JeongPhysRevB.105.235145,KangNatPhys2022,TanNPJQM2023}.
The Fermi surface (FS) is rather complicated and contains several
sheets. In the unfolded hexagonal Brillouin zone (BZ) without CDW
order the main pockets are: (i) a warped cylindrical pocket around the $\Gamma$
point derived from $Sb$ 5p-orbitals, 
(ii) a large concave hole hexagon derived from 3d-orbitals
of V, and (iii) convex triangular pockets around $K$ points. While the Dirac
points and flat bands are separated from the Fermi energy, the van
Hove points are very close \citep{ChoPhysRevLett.127.236401,LaBollitaPhysRevB.104.205129,JeongPhysRevB.105.235145,KangNatPhys2022,HuNatComm2022}
and affect many physical properties of these materials. In particular, the van Hove singularities are expected to have a pronounced influence on both CDW \citep{WangPhysRevB.87.115135,KieselPhysRevLett.110.126405,DennerPhysRevLett.127.217601,LinPhysRevB.104.045122,FerrariPhysRevB.106.L081107,ScammellNatComm23} and superconducting \citep{HirschPhysRevLett.56.2732,DzyaloshinskiiZhETF87,MarkiewiczJPCS97,IrkhinPhysRevLett.89.076401,WangPhysRevB.87.115135,KieselPhysRevLett.110.126405,LuoNatComm2023} instabilities.
The 3Q CDW order reconstructs the Fermi-surface structure \citep{TanPhysRevLett.127.046401,LuoPhysRevLett.128.036402,TanNPJQM2023,ChapaiPhysRevLett.130.126401}, 
with the large hexagon pocket being the most significantly transformed.

The kagome superconductors are characterized by unusual magnetotransport behavior
below the CDW transition\citep{YangSciAdv20,YuPhysRevB.104.L041103,ZhouPhysRevB.105.205104,HuangPhysRevB.106.064510,ZhengNatComm2023}.
The magnetoresistance exhibits a linear field dependence at low fields up to $\sim\!1$ tesla and at higher fields a downward curvature develops followed by a second region of quasilinear growth with smaller slope.  
The Hall resistivity is negative at small magnetic fields but shows a pronounced nonmonotonic behavior and changes sign in fields of several kG. At fields higher than 5 tesla the behavior becomes linear again with the slope strongly dependent on temperature. 
The unusual small-field behavior of the Hall resistivity has been attributed to the so-called anomalous Hall effect\citep{YangSciAdv20,YuPhysRevB.104.L041103,ZhengNatComm2023}, 
which may be caused by chiral or time-reversal symmetry breaking states \cite{NagaosaRevModPhys.82.1539} .

Here we propose that these unusual features in magnetotransport mostly originate from unique features of the Fermi surface of kagome metals. In particular, the unusual transport behavior is induced by the proximity of the Fermi level to van Hove singularities. In this paper, we present a combined theoretical and experimental study of the magnetotransport of CsV$_3$Sb$_5$ crystals. 
We demonstrate that the anomalous behavior mainly originates from the large hexagonal Fermi pocket in the pristine Brillouin zone due to (i) proximity of its corners to the van Hove points and (ii) its overall concave shape.  
This observation implies that the proximity to the van Hove singularity has a profound influence on transport behavior in these materials, as it 
has been recognized before for other systems, see, e.g., \citep{MaharajPhysRevB.96.045132,HermanPhysRevB.99.184107,MaoPhysRevB.104.024501}.

We introduce a simplified minimum model for this Fermi surface allowing for a full analytical consideration. We evaluate the magnetoconductivity both in the pristine phase and the low-temperature 3Q CDW state, where the large hexagon is transformed into a small hexagon and two large triangular sheets. The reconstructed Fermi surface is characterized by sharp corners which enhances the small-field magnetoresistivity. Our model is motivated by our recent observation of a series of break-down orbits in high-field quantum oscillation measurements \citep{ChapaiPhysRevLett.130.126401}, by the observation that the magnetoresistance increases strongly at the CDW transition (see e.g., Ref.\ \citep{HuangPhysRevB.106.064510} and discussion below), and by recent DFT calculations \cite{TanNPJQM2023}.
The model provides a qualitative description of the behavior for both the diagonal and Hall conductivity. It also  accounts for the strong increase of the magnetoresistance in the CDW state. 

Although the large hexagon is nominally a hole pocket, the small-field linear Hall
conductivity is \emph{negative}. The reason for this anomalous feature
is the velocity reduction near the van Hove points leading to suppression
of the hole contribution to the Hall conductivity near the corners. As
a result, the concave hexagon sides provide a dominant negative contribution.
Strong velocity variations along the Fermi surfaces also lead to strong
field dependences of both conductivity components. An important feature
of these field dependences is the existence of two magnetic field
scales originating from van Hove singularities: the lower scale is set by the two effective masses determining
the saddle-point electronic spectrum at the van Hove point while the
upper scale logarithmically diverges when the Fermi energy approaches
the van Hove energy. The reconstruction of the large hexagon pocket
does not change qualitatively the low-field nonmonotonic behavior
of the Hall conductivity. 
On the other hand, this reconstruction generates sharp corners in the new Fermi
pockets leading to enhancement of the low-field diagonal magnetoconductivity with linear magnetic field dependence. This linear magnetoconductivity naturally appears in materials with charge and
spin density waves having a nonideal nesting. In such materials the
reconstruction of the Fermi surface in the folded Brillouin zone leads
to the formation of Fermi pockets with sharp corners. The interruption of the
smooth orbital motion of quasiparticles along the Fermi surface at
these corners leads to linear magnetoconductivity\citep{PippardBook,FentonSchofieldPRL05,Koshelev2013}
and also to strongly field-dependent Hall conductivity\citep{LinMillisPhysRevB05}.
The rounding of the corners due to finite CDW gap leads to the crossover from linear to quadratic field dependence of the diagonal conductivity when magnetic fields drops below the typical value proportional to the gap.
Besides the low-field linear behavior, the field dependence of the
diagonal conductivity is characterized by an intermediate $1/H$ field
dependence between the two field scales set by the van Hove singularity
and crosses over to standard $1/H^{2}$ at higher fields. 
We demonstrate that the Hall conductivity from the reconstructed pockets qualitatively reproduce shapes of the experimental field dependences.

This paper is organized as follows. In section \ref{sec:AnomMC}, we illustrate the anomalous behavior of the magnetoconductivity that we aim to explain.
In section \ref{sec:LargeHex}, we introduce the model for the large hexagon pocket on which our calculations of anomalous magnetotransport are based. Subsequently, in section \ref{sec:FoldCDW}, we consider the reconstructed Fermi surface arising due to the onset of the 3Q CDW order emphasizing a small hexagonal Fermi surface sheet near the zone centered surrounded by large triangular sheets.
Subsequently, in section \ref{sec:OrbInt}, we introduce the orbital-integral approach for the calculation of the magnetoconductivity,  which is based on the relaxation-time approximation of the Boltzmann transport equation. 
The theoretical analysis of the magnetoconductivity arising from the large hexagonal hole Fermi surface sheet in the unfolded Brillouin zone is presented in section \ref{sec:LHsigmaH}, while in section \ref{sec:MagCondReconstr}, we compute the magnetoconductivity components for the reconstructed Fermi surface consisting of small hexagon triangle and two large triangular pockets and in section \ref{sec:ModifMCReconstr}, we analyze the modification of the magnetoconductivity due to the CDW order. In section \ref{sec:ModelExpData}, we present our experimental magnetotransport data and their qualitative modeling.

\section{Anomalous behavior of magnetoconductivity in C\MakeLowercase{s}V$_{3}$S\MakeLowercase{b}$_{5}$ in CDW state}
\label{sec:AnomMC}

\begin{figure}
	\includegraphics[width=3.4in]{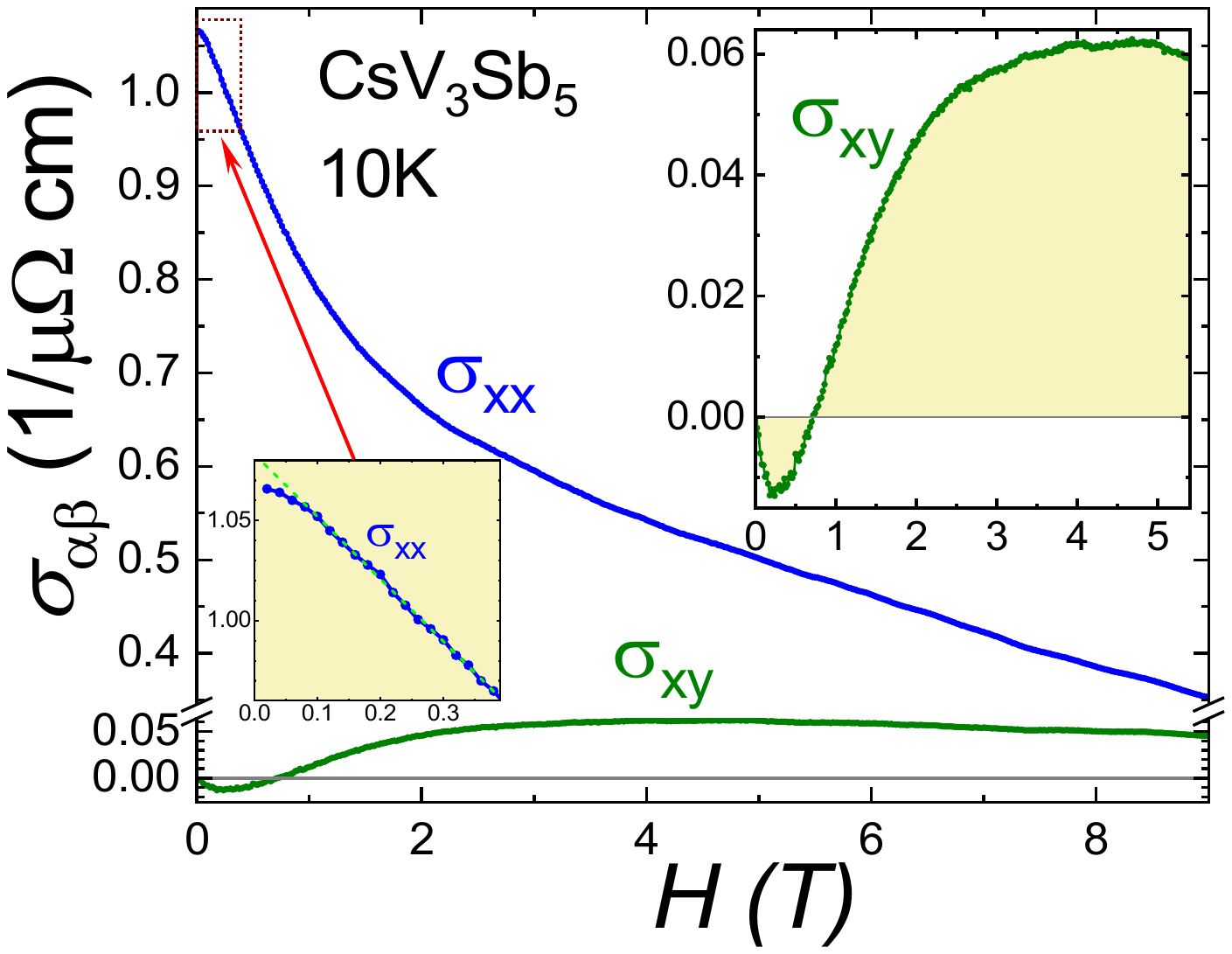}
	\caption{The diagonal and Hall conductivities of CsV$_{3}$Sb$_{5}$ at 10 K.
		The upper-right inset shows the anomalous behavior of $\sigma_{xy}(H)$ in the expanded vertical scale and the lower-left inset zooms into the low-field behavior of $\sigma_{xx}(H)$ illustrating the crossover between the quadratic and linear regimes near 0.1 T. The dashed green line shows linear fit within the range 0.15 -- 0.5 T. }
	\label{Fig:sab10K}
\end{figure}
As an illustration of the anomalous behavior, we show in Fig.~\ref{Fig:sab10K} the magnetic field dependences of the diagonal and Hall conductivities for CsV$_{3}$Sb$_{5}$ at 10 K.
These conductivities are extracted from the experimentally measured diagonal and Hall resistivities, as discussed in more detail below, in section \ref{sec:ModelExpData}. 
The standard quadratic field dependence of $\sigma_{xx}$
for the magnetic field $H$ applied along the $c$ axis is limited
to very small fields, below 0.1 T, as illustrated in the lower-left inset of Fig.\ \ref{Fig:sab10K}. At higher fields up to $\sim$1 T, the magnetoconductivity
has linear field dependence followed by upward
curvature and a crossover in the range 1.5 - 3 T to a second region of quasilinear variation. The behavior
of the Hall conductivity $\sigma_{xy}$ is more complicated.
In low fields $\sigma_{xy}$ is negative. However, at fields
\textless{} 0.3 tesla it displays a nonmonotonic dependence and changes
sign. At even higher field $\sigma_{xy}$ passes through a smooth
maximum. 

The unusual small-field behavior of $\sigma_{xy}$ has been
attributed to the so-called anomalous Hall effect \citep{YangSciAdv20,YuPhysRevB.104.L041103,ZhengNatComm2023,WangPhysRevB.108.035117}, i.e., by the manifestation
of the spontaneous Hall effect,
which may arise without magnetic field in parity-breaking states, such as chiral CDW, or time-reversal symmetry breaking such as loop-current order \cite{NagaosaRevModPhys.82.1539}. 
In this interpretation, the region near the inflection
point around 1 tesla is treated as a linear background. After subtraction
of this background, the curve $\sigma_{xy}(H)$ acquires a plateaulike
region in the range 0.5 -- 1.5 tesla, which is interpreted as the anomalous
Hall contribution. This interpretation, however, is not straightforward.
Strictly speaking, in the case of a single chiral domain, a Hall
conductivity should be finite even in zero magnetic field. In the scenario
where the zero-field Hall conductivity averages to zero due to multiple
domains with opposite chiralities, and the plateau appears because
the magnetic field selects a definite chirality, one would expect
a hysteretic behavior, which was not observed. 
Despite of these rather obvious issues, the ``anomalous Hall effect'' interpretation is widely cited in the literature as one of experimental indications for the realization of a chiral CDW state in the kagome family \emph{A}V$_{3}$Sb$_{5}$.

We demonstrate that the overall behavior of the conductivity components, especially the Hall conductivity, can be more naturally and straightforwardly explained as arising from the fermionic properties of the large hexagon pocket. 
In the next section, we present a simple model for this pocket which we employ in our analytic calculations of magnetoconductivity.

\section{Model for the large concave hexagon pocket}
\label{sec:LargeHex}

\begin{figure}
	\includegraphics[width=3.4in]{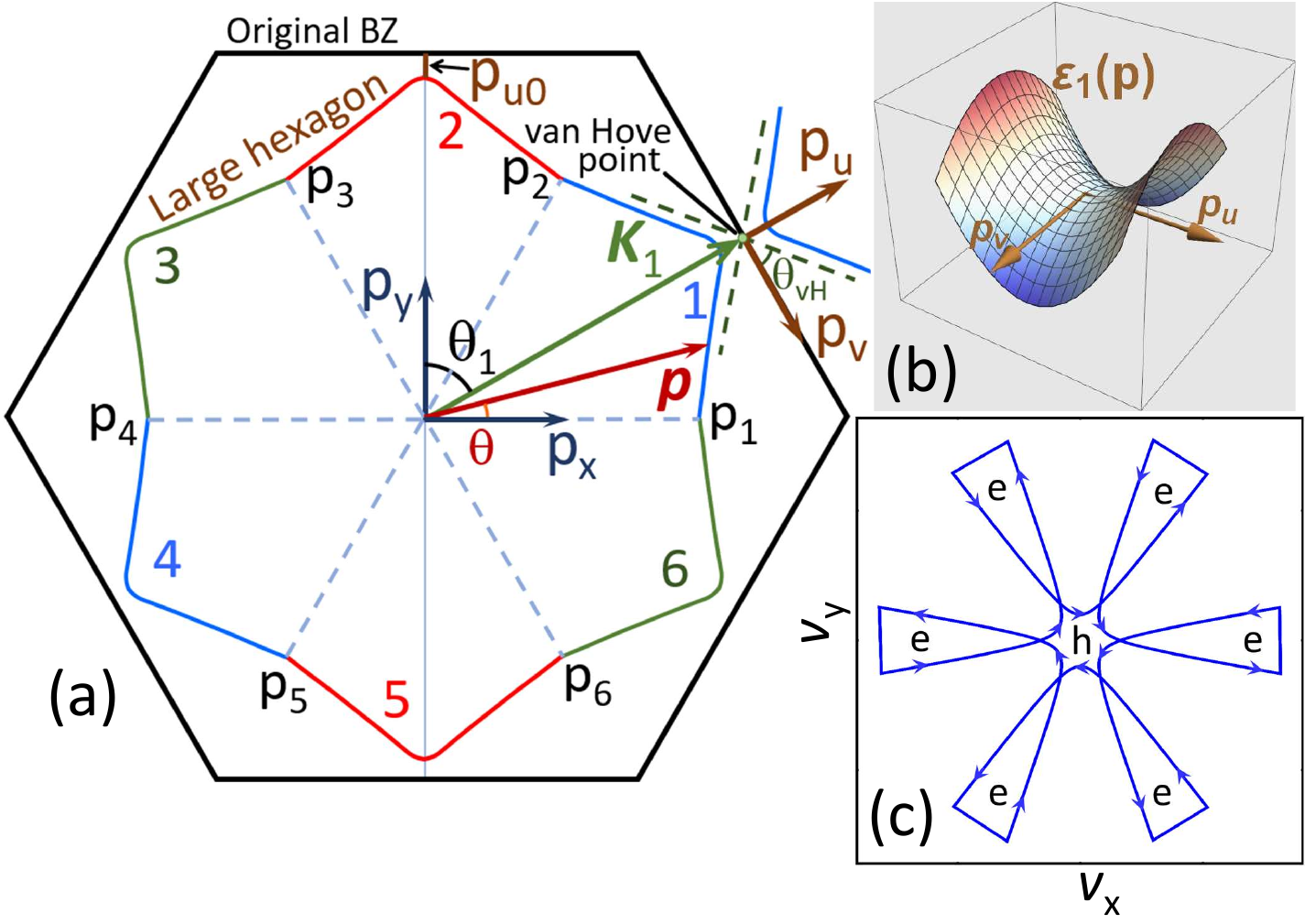}
	\caption{(a)The large concave hexagon Fermi pocket approximated by the hyperbolic
		segments.(b)Saddle-point spectrum near the van Hove point. (c)The velocity contour corresponding to this Fermi surface,
		which determines the linear Hall term\citep{OngPhysRevB.43.193},
		Eq.~\eqref{eq:SxyLinArea}. The central small hexagon corresponds
		to locations near the van Hove points and gives the hole contribution
		to the Hall conductivity, while the six outside ``petals'' give
		electronic contribution. }
	\label{Fig:LargeHex}
\end{figure}
In this section, we introduce a model for the large concave hexagon pocket in the unreconstructed original Brillouin zone, which we will use to evaluate the components of conductivity in magnetic field. 
As illustrated in Fig.~\ref{Fig:LargeHex}(a), this pocket is mostly determined by
the van Hove points which are located near the six wave vectors $\boldsymbol{K}_{j}$
with 
\begin{subequations}
	\begin{align}
		K_{j,x}\! & =K\cos\left(-\frac{\pi}{6}+\frac{\pi}{3}j\right),\label{eq:Kx}\\
		K_{j,y} & \!=K\sin\left(-\frac{\pi}{6}+\frac{\pi}{3}j\right),\label{eq:Ky}
	\end{align}
\end{subequations}
where $j\!=\!1,2,\ldots,6$, $K\!=\!2\pi/(\sqrt{3}a)$
is half of the basic reciprocal-lattice vector for the unfolded
BZ, and $a$ is the lattice parameter (5.45\AA \ for CsV$_{3}$Sb$_{5}$\citep{FuPhysRevLett.127.207002}). 
Note that, due to BZ periodicity, the vectors $\boldsymbol{K}_{j}$ and $\boldsymbol{K}_{j+3}\!=\!-\boldsymbol{K}_{j}$ point to the same van Hove point meaning that there are only three nonequivalent van Hove points. 
In the vicinity of these wave vectors, $\left|\boldsymbol{p}-\boldsymbol{K}_{j}\right|\ll K$,
we can expand momentum $\boldsymbol{p}$ using the local basis, $\boldsymbol{p}=\boldsymbol{K}_{j}\!+p_{u}\boldsymbol{e}_{l}\!+p_{v}\boldsymbol{e}_{t}$,
where $\boldsymbol{e}_{l}$ and $\boldsymbol{e}_{t}$ are the unit
vectors along and perpendicular to $\boldsymbol{K}_{j}$. 
The local momenta $(p_{v,}p_{u})$ are connected with the global momenta $(p_{x},p_{y})$ as
\begin{subequations}
	\begin{align}
p_{v}&\!=\!\left(p_{x}\!-K_{j,x}\right)\cos\theta_{j}\!+\!\left(p_{y}\!-\!K_{j,y}\right)\sin\theta_{j},\label{eq:pv-rotLH}\\
p_{u}&\!=\!-\!\left(p_{x}\!-\!K_{j,x}\right)\sin\theta_{j}\!+\!\left(p_{y}\!-\!K_{j,y}\right)\cos\theta_{j},\label{eq:pu-rotLH}
	\end{align}
\end{subequations} 
where $\theta_{j}\!=\!\left(\pi/3\right)(j\!-\!2)$ is the rotation angle of the local frame, see Fig. \ref{Fig:LargeHex}(a). 
In the local coordinates, the electronic spectrum has a saddle-point shape illustrated in Fig.~\ref{Fig:LargeHex}(b),
\begin{equation}
	\varepsilon_{j}(\boldsymbol{p}) =\frac{p_{u}^{2}}{2m_{u}}-\frac{p_{v}^{2}}{2m_{v}},\label{eq:SddlPntSpectrLargeHex}
\end{equation}
where $m_{u}$ and $m_{v}$ are the effective masses. The shape of
the Fermi surface near $\boldsymbol{K}_{j}$ is determined by the equation
$\varepsilon_{j}(\boldsymbol{p})\!=\!\varepsilon_{\mathrm{vH}}$, where $\varepsilon_{\mathrm{vH}}$
is the Fermi energy measured with respect to the van Hove point. In
a 3D layered metal it depends on the c-axis momentum $p_{z}$. At
$\varepsilon_{\mathrm{vH}}\!=\!0$, the Fermi surfaces are given by straight
van Hove lines, $p_{u}\!=\!\pm p_{v}/\sqrt{r_{m}}\!=\!\pm\tan\theta_{\mathrm{vH}}\,p_{v}$,
where the mass ratio $r_{m}\!=\!m_{v}/m_{u}$ determines the van Hove
angle as $\tan\theta_{\mathrm{vH}}\!=\!1/\sqrt{r_{m}}$. For reference,
the mass ratio $r_{m}$ and other key parameters used in the paper
are listed in Table \ref{Tbl:param}. At a finite separation between the
van Hove point and Fermi level $\varepsilon_{\mathrm{vH}}$, the Fermi surfaces
are given by hyperbola branches located near every vector $\boldsymbol{K}_{j}$,
\begin{equation}
	p_{F,u}(p_{v})\!=\!-\sqrt{p_{u0}^{2}\!+\!p_{v}^{2}/r_{m}}\!=\!-\sqrt{p_{v0}^{2}\!+\!p_{v}^{2}}/\sqrt{r_{m}}\label{eq:HypFS}
\end{equation}
with $p_{u0}\!=\!\sqrt{2m_{u}\varepsilon_{\mathrm{vH}}}$ being the distance
between the hexagon corner and van Hove point, see Fig.\ \ref{Fig:LargeHex}(a), and $p_{v0}\!=\!\sqrt{2m_{v}\varepsilon_{\mathrm{vH}}}\!=\!\sqrt{r_{m}}p_{u0}$.
The minus sign in Eq.~\eqref{eq:HypFS} corresponds to the hexagon
inside the first BZ. Due to the BZ periodicity, the branches located
near opposite wave vectors $\boldsymbol{K}_{j}$ and $-\boldsymbol{K}_{j}$
are, in fact, two branches of the same hyperbola. The hexagon is concave
if $\theta_{\mathrm{vH}}>30^{\circ}$ corresponding to $r_{m}<3$.
From the ARPES and DFT data for CsV$_{3}$Sb$_{5}$ \citep{OrtizPhysRevX.11.041030,KangNatPhys2022},
we estimate $\theta_{\mathrm{vH}}\approx39^{\circ}$ giving $\sqrt{r_{m}}\approx1.23$
and $r_{m}\approx1.51$. We will use $r_{m}=1.5$ in the illustrative
plots. The Fermi velocity components can be represented as
\begin{subequations}
	\begin{align}
		v_{u} & =\frac{p_{u}}{m_{u}}=-\frac{\sqrt{p_{v0}^{2}+p_{v}^{2}}}{\sqrt{m_{u}m_{v}}},\label{eq:vu}\\
		v_{v} & =-\frac{p_{v}}{m_{v}},\label{eq:vv}
	\end{align}
\end{subequations}
giving the total in-plane velocity at the Fermi surface
\begin{equation}
	v=\!\sqrt{\frac{p_{u}^{2}}{m_{u}^{2}}+\frac{p_{v}^{2}}{m_{v}^{2}}}=\!\sqrt{v_{u0}^{2}\!+\frac{p_{v}^{2}}{m_{v}^{2}}\left(1\!+\!r_{m}\right)}.\label{eq:vtot}
\end{equation}
The key feature that strongly influences the magnetotransport behavior is
that the velocity drops linearly on approaching the van Hove point,
$v\! \simeq \! p_{v}\sqrt{1\!+\!r_{m}}/m_{v}$. The minimum value $v_{u0}\!=\! p_{u0}/m_{u}$
is realized at the hexagon corner. Exactly at the van Hove point for
$\varepsilon_{\mathrm{vH}}\!=\!0$, the velocity vanishes.

To reveal the  qualitative behavior, we will use a simple approximate
model for the electronic spectrum in the form of a piecewise function composed
of hyperbolic segments
\begin{align}
	\varepsilon(p_{x},p_{y}) & =\varepsilon_{l}(\boldsymbol{p}),\  \mathrm{for}\,\frac{\pi}{3}(l\!-\!1)\!<\!\theta\!<\!\frac{\pi}{3}l,\label{eq:PieceSpectrumLargeHex}
\end{align}
where $\theta$ is the polar angle of the momentum $p_{x}\!=\!p\cos\theta$,
$p_{y}\!=\!p\sin\theta$. This approximation means that the Fermi
surface is composed of hyperbola branches located between the momenta
$p_{l}$ and $p_{l+1}$, see Fig.~\ref{Fig:LargeHex}(a). We only
need this approximation in the vicinity of the Fermi surface. The
piecewise hyperbolic approximation allows for a full analytical evaluation
of the conductivity components in arbitrary magnetic fields. The price paid
for the model simplicity is the jumps of the Fermi velocities at the
boundaries between the regions. Nevertheless, in spite of its simplicity,
the model captures the essential anomalous features of magnetoconductivity.  
As suggested by the shape of Fermi surface
in Eq.~\eqref{eq:HypFS}, calculations can be conveniently carried out using
the hyperbolic parametrization,
\begin{equation}
	p_{v}\!=\!-p_{v0}\sinh t,\ p_{u}\!=\!-p_{u0}\cosh t,\label{eq:HypParam}
\end{equation}
which automatically places the momentum at the Fermi surface
within every segment. Here the hyperbolic parameter $t$ is located
within the interval $-t_{b}<t<t_{b}$. At the boundaries between the
segments, we have the geometric relation $\sqrt{3}|p_{v}|+|p_{u}|=K$
yielding the equation for the limiting hyperbolic parameter $t_{b}$,
\begin{equation}
	\sqrt{3r_{m}}\sinh t_{b}\!+\cosh t_{b}=\kappa,\label{eq:kappa}
\end{equation}
with $\kappa\!=\!K/p_{u0}\gg1$, which has the following exact solution
\begin{align}
	\sinh t_{b}= & \frac{\sqrt{3r_{m}}\kappa\!-\!\sqrt{\kappa^{2}\!+\!3r_{m}\!-\!1}}{3r_{m}-1}.\label{eq:tb}
\end{align}
In the limit of very large $\kappa$ this gives a simpler approximate
relation $\exp (t_{b})\approx2\kappa/(\sqrt{3r_{m}}\!+\!1)$, i.~e.,
the parameter $t_{b}$ diverges logarithmically on approaching the
van Hove point, $t_{b}\propto\ln\left(K/p_{u0}\right)$ for $p_{u0}\rightarrow0$.
\begin{table*}
	\begin{tabular}{|>{\raggedright}p{2.3in}|c|}
		\hline 
		Quantity & Notation and definition\tabularnewline
		\hline 
		\hline 
		Mass ratio for saddle-point spectrum & $r_{m}\!=\!m_{v}/m_{u}$\tabularnewline
		\hline 
		Reduced magnetic field & $\omega_{h}\!=\!|e|H\tau/(c\sqrt{m_{u}m_{v}})$\tabularnewline
		\hline 
		Lower magnetic field scale & $H_{0}\!=\!c\sqrt{m_{u}m_{v}}/(|e|\tau)$\tabularnewline
		\hline 
		Upper magnetic field scale & $H_{b}\!=\!2t_{b}H_{0}$\tabularnewline
		\hline 
		Conductivity slice scale & $S_{0}\!=\!3\tau p_{u0}^{2}/\sqrt{m_{u}m_{v}}$\tabularnewline
		\hline 
		Ratio of reciprocal-lattice vector and separation from van Hove point & $\kappa\!=\!K/p_{u0}$\tabularnewline
		\hline 
		Boundary hyperbolic parameter & $\sinh t_{b}=\frac{\sqrt{3r_{m}}\kappa\!-\!\sqrt{\kappa^{2}\!+\!3r_{m}\!-\!1}}{3r_{m}-1}$\tabularnewline
		\hline 
		Crossing-point hyperbolic parameter & $\tanh t_{c}=1/\sqrt{3r_{m}}$\tabularnewline
		\hline 
	\end{tabular}\caption{Definitions of the key parameters used throughout the paper.}
	\label{Tbl:param}
\end{table*} 

\section{CDW reconstruction of the large hexagon in the folded Brillouin zone\label{sec:FoldCDW}}

\begin{figure*}
	\includegraphics[width=2.5in]{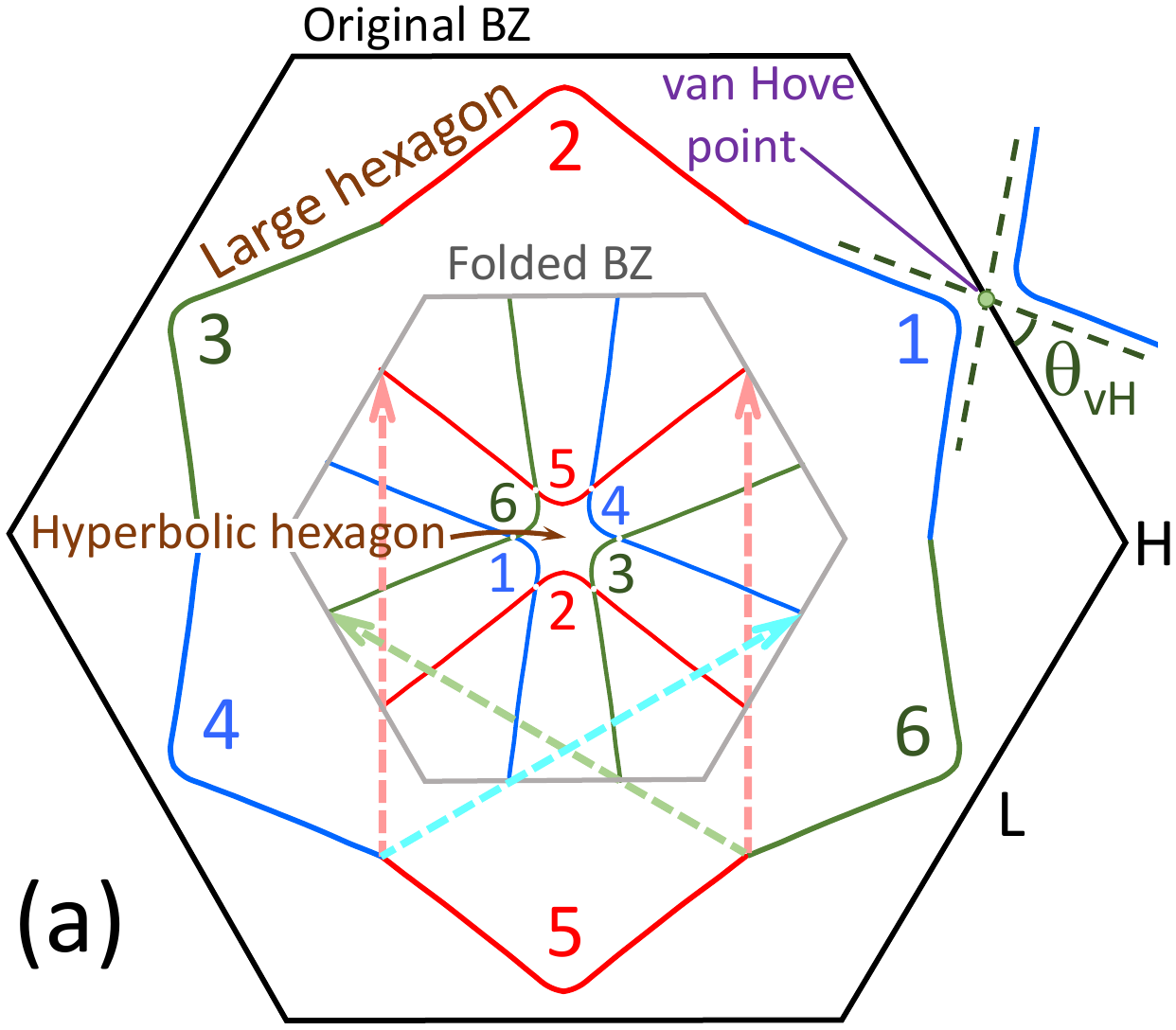}\includegraphics[width=1.8in]{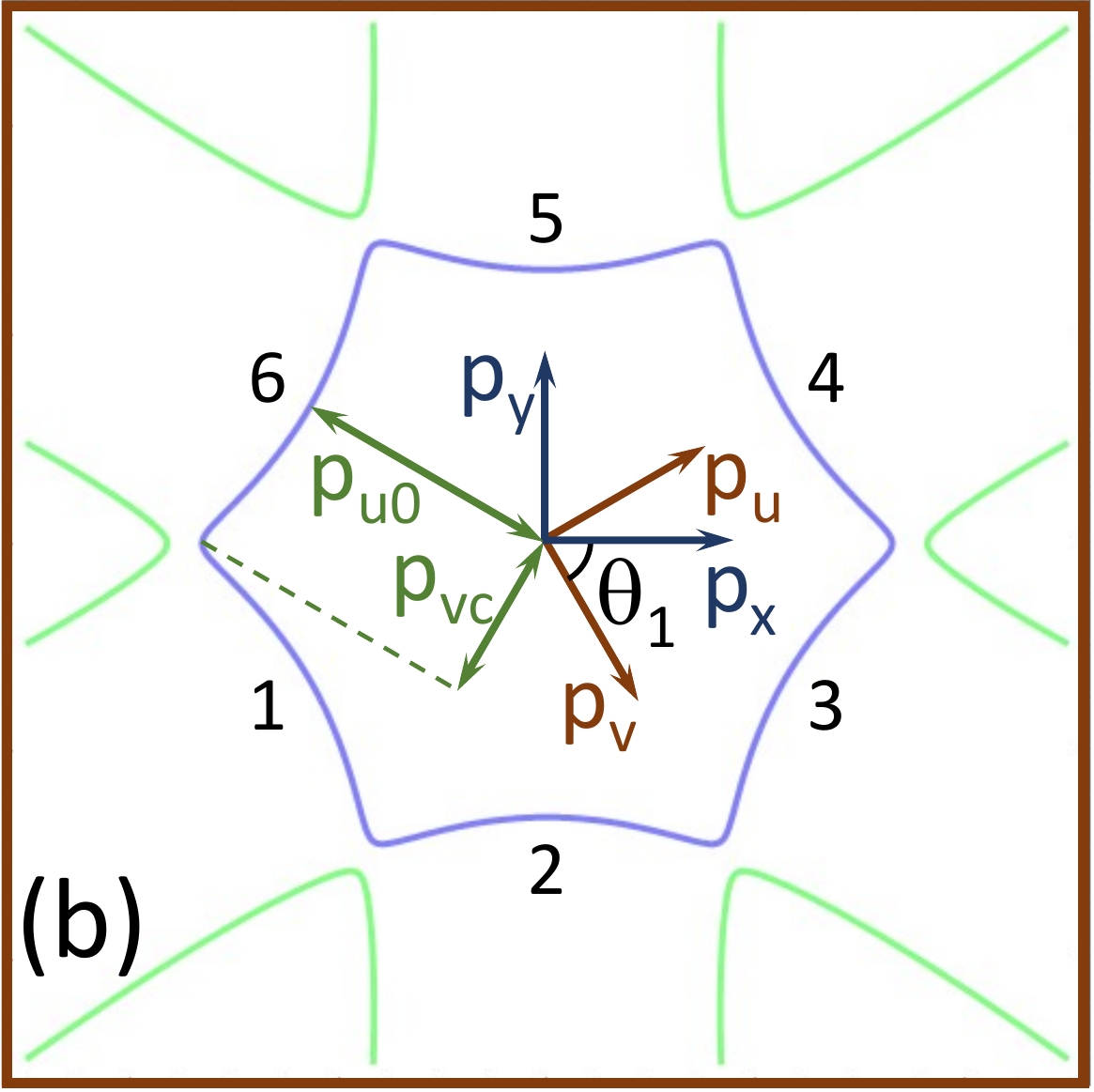}\includegraphics[width=2.5in]{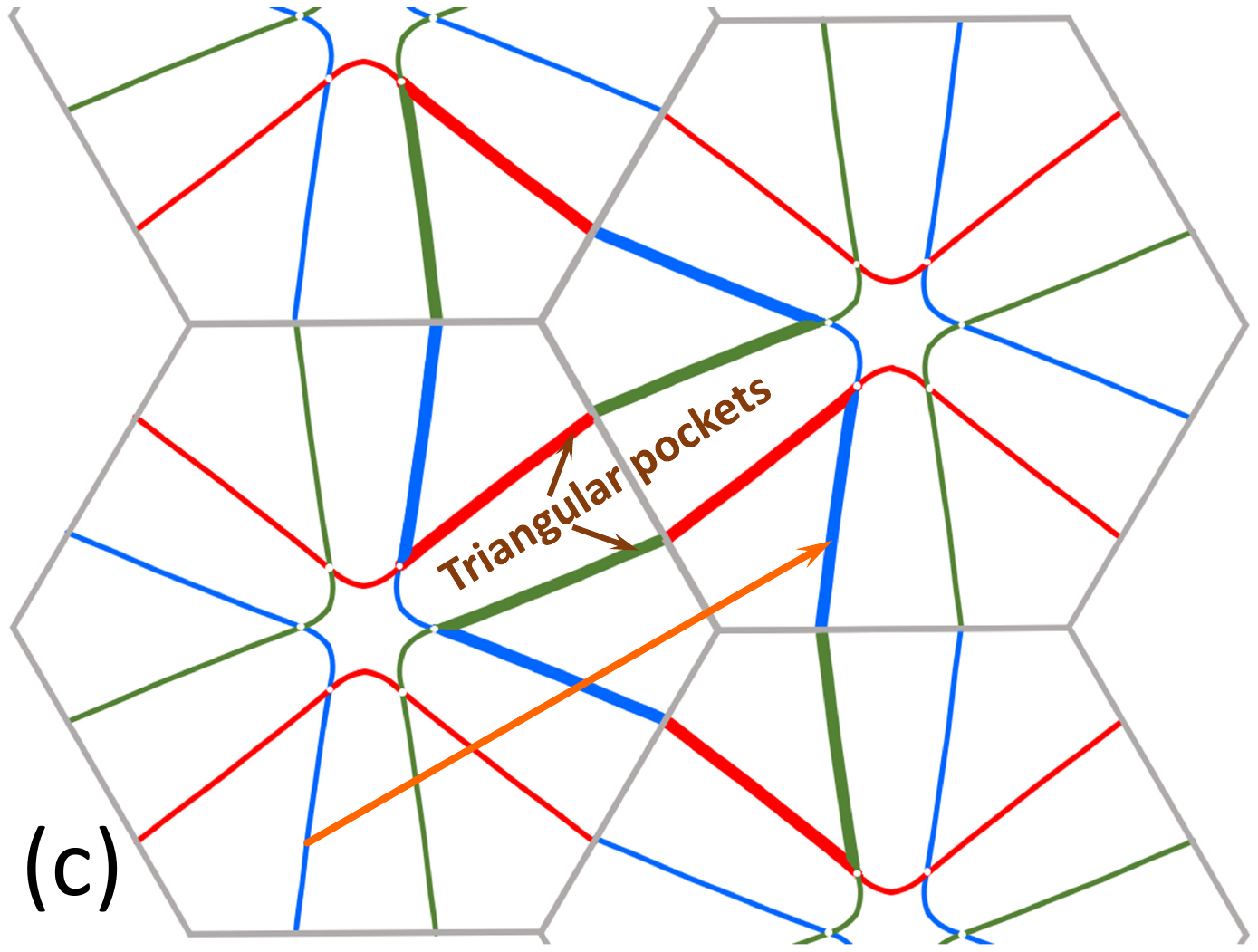}
	\caption{(a) The large concave hexagon in the original Brillouin zone has vertices
		close to the three van Hove points. The segments corresponding to the hyperbola
		branches are color coded. The 3Q CDW reconstruction leads to the formation of
		a small ``hyperbolic hexagon'' Fermi pocket in the folded Brillouin
		zone and outside branches which form two large triangles in the extended
		Brillouin zone, as illustrated in (c). The dashed arrows in (a) show the
		CDW wave vectors transferring the colored segments of the original large
		hexagon into the folded Fermi surface. The van Hove points are transferred to the center.
		The sharpness of the ``hyperbolic
		hexagon'' and of the triangles corners is due to the small value of CDW
		gap. (b) Zoom into the ``hyperbolic hexagon'' Fermi pocket which
		appears due to CDW reconstruction at the crossing points.  
		The definitions of the local coordinates $\left(p_{v},p_{u}\right)$ for the first segment  
		as well as the two key geometrical parameters
		$p_{u0}$ and $p_{vc}$ are also illustrated. (c)Triangular pockets
		formed in the extended Brillouin zone. The orange arrow shows the reciprocal
		lattice vector connecting identical branches. Coloring emphasizes
		that Fermi surfaces of these pockets are composed from the pieces belonging
		to different hyperbolic branches. }
	\label{Fig:HypHexagon}
\end{figure*}
The 3Q CDW order leads to the folding of the Brillouin zone. The folded BZ
is given by a hexagon with four times smaller area, as shown
in Fig.~\ref{Fig:HypHexagon}(a). 
During this folding, all three van Hove points are displaced to the BZ center  
and the segments of the large hexagon Fermi surface shown in Fig.\ \ref{Fig:LargeHex}(a) are shifted to the center, $\boldsymbol{p}\rightarrow \boldsymbol{p}-\boldsymbol{K}_j$, and reconstructed at
the crossings \citep{ChapaiPhysRevLett.130.126401}. 
Three crossing hyperbolas form the electronic hyperbolic-hexagon (HH) pocket at the
center, as illustrated in Figs.\ \ref{Fig:HypHexagon}(a,b). The outside branches
form two holelike triangular pockets in the extended BZ, see Fig.~\ref{Fig:HypHexagon}(c).
This reconstruction affects magnetotransport mostly by modifying the pattern of quasiparticle orbital motion in the magnetic field. On the one hand, the formation of sharp corners and their rounding due to the opening of CDW gaps affects the magnetoconductivity at small magnetic fields. On the other hand, modification of pocket areas affects magnetotransport at high magnetic fields. 

Let us consider first the structure of the small hyperbolic-hexagon
pocket. The HH pocket is composed of three crossing hyperbolas rotated
60$^{\circ}$ with respect to each other. This corresponds to six crossing segments. Near the crossing points,
the spectra are reconstructed due to the CDW order. 
We will maintain the same segment numbering for the original and displaced branches, see Fig.\ \ref{Fig:HypHexagon}(a).
For every segment one can again introduce a local rotated coordinate system $\left(p_{v},p_{u}\right)$.
Since hyperbolas are now located near the center, in the relation between the local and global momenta, Eqs.\ \eqref{eq:pv-rotLH} and \eqref{eq:pu-rotLH}, we have to remove the shifts by the vectors $\boldsymbol{K}_j$.  
In these local coordinates, the saddle-point hyperbolic spectrum is
again given by Eq.~\eqref{eq:SddlPntSpectrLargeHex}. At a finite
positive shift of the Fermi level from the van Hove point, $\varepsilon_{\mathrm{vH}}\!>\!0$,
the hyperbolic Fermi-surface branches are given by the Eq.\ \eqref{eq:HypFS}.
The parameter $p_{u0}\!=\!\sqrt{2m_{u}\varepsilon_{\mathrm{vH}}}$
becomes the minimal distance between the hyperbola and origin, see
Fig.~\ref{Fig:HypHexagon}(b). Contrary to the large hexagon, the
Fermi surface of the small hexagon is very accurately described by the
hyperbolic branches due to its immediate proximity to the van Hove points.
That is why we use the name ``hyperbolic hexagon'' for this pocket.
For our calculations, the Fermi surface shape in Eq.~\eqref{eq:HypFS}
can be again conveniently parametrized in terms of hyperbolic functions, Eq.\ \eqref{eq:HypParam}.

Neglecting the reconstruction due to opening of CDW gap, a hyperbolic branch crosses with the
neighboring branch at the angle of 30$^{\circ}$ with respect to the
$p_{u}$ axis. This corresponds to the condition $|p_{u}|=\sqrt{3}|p_{v}|$
and gives the relation between the $p_{v}$ coordinate of the crossing
point and $p_{u0}$,
\begin{align}
	p_{vc} & =\frac{p_{u0}}{\sqrt{3\!-\!1/r_{m}}}.\label{eq:pvc}
\end{align}
In the hyperbolic parametrization defined by Eq.~\eqref{eq:HypParam},
the crossing takes place at $t=t_{c}$ with $\tanh t_{c}=1/\sqrt{3r_{m}}$,
yielding
\begin{align}
	t_{c} & =\frac{1}{2}\ln\frac{\sqrt{3r_{m}}+1}{\sqrt{3r_{m}}-1}.\label{eq:tc-rm-def}
\end{align}
We will present the final results for the conductivity components of the
reconstructed Fermi surface using this parameter together with the
parameter $t_{b}$ in Eq.~\eqref{eq:tb}. For a finite CDW gap, sharp
corners at the crossings become rounded, see Fig. \ref{Fig:HypHexagon}(b). 

In Appendix \ref{App:HHarea}, we evaluate the HH area $A_{\mathrm{HH}}$
and effective mass $m_{\mathrm{HH}}$, which can be probed by magnetic oscillations
\citep{FuPhysRevLett.127.207002,ShresthaPhysRevB.105.024508,ZhangPhysRevB.106.195103,HuangPhysRevB.106.064510,BroylesPhysRevLett.129.157001,ChapaiPhysRevLett.130.126401}
\begin{subequations}
	\begin{align}
		A_{\mathrm{HH}} & =6\sqrt{r_{m}}t_{c}p_{u0}^{2},\label{eq:HexArea}\\
		m_{\mathrm{HH}} & =\frac{1}{2\pi}\frac{dA_{\mathrm{HH}}}{d\varepsilon_{F}}=\frac{6}{\pi}t_{c}\sqrt{m_{u}m_{v}}.
		\label{eq:HexEffMass}
	\end{align}
\end{subequations}
Clearly, the HH area is always somewhat larger
than the area of an ideal hexagon with the same $p_{u0}$, $A_{\mathrm{hex}}=2\sqrt{3}p_{u0}^{2}\approx3.464p_{u0}^{2}$,
corresponding to the inequality $\sqrt{3r_{m}}\ln\frac{\sqrt{3r_{m}}\!+\!1}{\sqrt{3r_{m}}\!-\!1}>2$.
The relations in Eqs.~\eqref{eq:HexArea} and \eqref{eq:HexEffMass}
allow for the estimation of the key parameters $p_{u0}$ and $m_{u}$ from
the experimental data. If we take the oscillation frequency $F_{\gamma}=226$
T attributed to hexagon pocket cross section at the top of the Brillouin
zone \citep{ChapaiPhysRevLett.130.126401}, we estimate for the extremal
cross section $A_{\mathrm{HH}}\!=\!0.0216$ \AA$^{-2}$ and $p_{u0}\!=0.076$
\AA$^{-1}$, which is $\sim6$\% of the size of the original
unfolded Brillouin zone, $k_{BZ}\!=\frac{4\pi}{\sqrt{3}a}=\!1.33$ \AA$^{-1}$.
With experimental value $m_{\mathrm{HH}}\!=\!0.71$$m_{e}$ and $r_m\!=\!1.51$, we estimate from Eq.\ \eqref{eq:HexEffMass} $m_{u}\!=\!0.59m_{e}$ and $m_{v}\!=\!0.9m_{e}$.

\begin{figure}
	\includegraphics[width=3.2in]{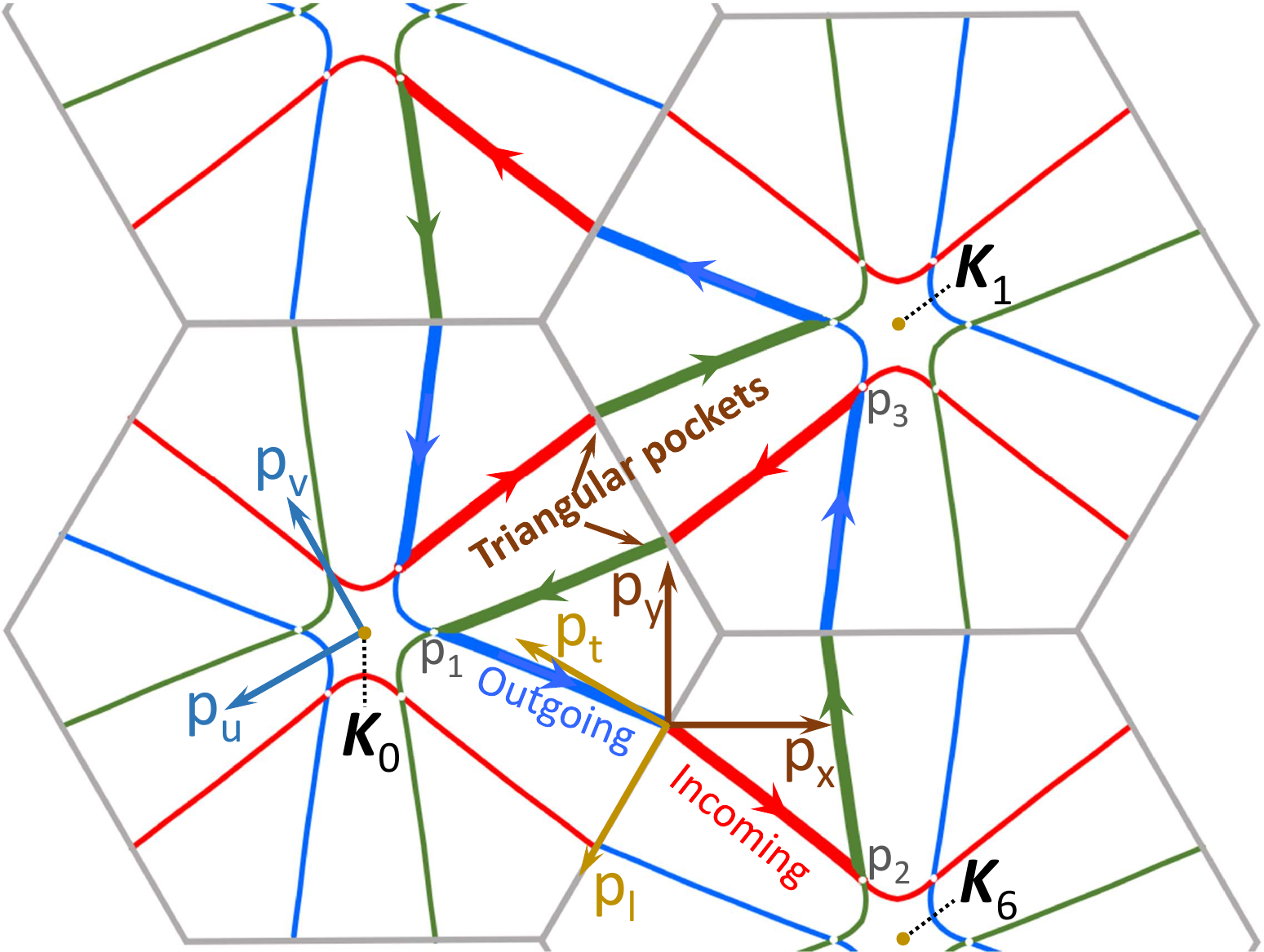}
	\caption{Geometry of the triangular pockets for the reconstructed Fermi surface.
		Every side of the triangle is composed of the incoming and outgoing segments
		belonging to different hyperbolas 
		which can be indexed by the location of its center $\boldsymbol{K}_m$ and rotation angle $\theta_j$. 
		Three axes sets used in the paper
		are shown: laboratory frame $\left(p_{x},p_{y}\right)$, hyperbolic
		basis $\left(p_{u},p_{v}\right)$, and transverse and longitudinal
		directions $\left(p_{t},p_{l}\right)$ for selected side segment.}
	\label{Fig:TriangPocketAxes}
\end{figure}

Let us consider now the configuration of the outside triangular pockets.
In our model, each triangular pocket is composed of six hyperbola
branches with different centers and rotations, see Fig.~\ref{Fig:HypHexagon}(c)
and \ref{Fig:TriangPocketAxes}. Each side of the triangle contains two segments belonging to different hyperbolas. We will call these
branches as ``outgoing'' and ``incoming'', referring to the direction
of orbital motion with respect to the hyperbolic hexagon pocket, see Fig.~\ref{Fig:TriangPocketAxes}.  The
momentum located at one of the branches of the triangular Fermi surface
can be presented as
\begin{subequations}
	\begin{align}
		p_{x} & =K_{m,x}+p_{v}\,\cos\theta_{j}-p_{u}\,\sin\theta_{j},\label{eq:TriParam-px}\\
		p_{y} & =K_{m,y}+p_{v}\,\sin\theta_{j}+p_{u}\,\cos\theta_{j},\label{eq:TriParam-py}
	\end{align}
\end{subequations}
where $\theta_{j}\!=\!\left(\pi/3\right)(j\!-\!2)$,
the components of the vectors $\boldsymbol{K}_{m}$ are defined in
Eqs.~\eqref{eq:Kx} and \eqref{eq:Ky}, and we also added zero vector
$K_{0,x}\!=\!K_{0,y}\!=\!0$. For example, the lower triangle is located
between the vectors $\boldsymbol{K}_{0}$, $\boldsymbol{K}_{1}$,
and $\boldsymbol{K}_{6}$, see Fig.\ \ref{Fig:TriangPocketAxes} and its corresponding indices $m,j$ are
summarized in Table \ref{Table:TrianIndices}. 
We will use the same
parametrization for hyperbolic coordinates $p_{v}$ and $p_{u}$ as
for the large hexagon in Eq.~\eqref{eq:HypParam}. The difference
is that for the outgoing branches the hyperbolic parameter $t$ varies
in the limits $t_{c}\!<\!t\!<\!t_{b}$ and for the incoming branches it
varies in the limit $-t_{b}\!<t<\!-t_{c}$. 

The area of the triangular pocket is evaluated in appendix \ref{sec:TArea} as
\begin{align}
	A_{\mathrm{T}}\!= & p_{u0}^{2}\left[\frac{\sqrt{3}}{2}\left(3r_{m}\sinh^{2}t_{b}\!-\!\cosh^{2}t_{b}\right)\!-\!3\sqrt{r_{m}}\left(t_{b}\!-\!t_{c}\right)\right]\nonumber\\
	\simeq & \frac{\sqrt{3}}{2}K^{2}\frac{\sqrt{3r_{m}}\!-\!1}{\sqrt{3r_{m}}\!+\!1}.\label{eq:TriArea} 
\end{align}
This area mostly determines the high-field asymptotics of the Hall conductivity, as will be discussed below.  Note that the total area of two triangular pockets is more than two times smaller than the area of the large hexagon.
We approximately evaluate the effective mass of the triangular pocket as
\begin{equation}
m_{T}\!\approx\!\frac{3}{\pi}\left(t_{b}\!-\!t_{c}\right)\sqrt{m_{u}m_{v}}.
\label{eq:TriMass} 
\end{equation}
It is larger than the effective mass of the hyperbolic hexagon in Eq.\ \eqref{eq:HexEffMass} due to the presence of the large factor $t_b$. 
\begin{table}
	\begin{tabular}{|c|c|c|c|}
		\hline 
		$m,j$ & 1 & 2 & 3\tabularnewline
		\hline 
		\hline 
		out & 0,4 & 6,6 & 1,2\tabularnewline
		\hline 
		in & 6,5 & 1,1 & 0,3\tabularnewline
		\hline 
	\end{tabular}\caption{The indices $m,j$ in Eqs.~\eqref{eq:TriParam-px} and \eqref{eq:TriParam-py}
		defining central vectors $\boldsymbol{K}_{m}$ and rotation angles
		$\theta_{j}$ for for six branches in the lower triangular pocket,
		see Fig.~\ref{Fig:TriangPocketAxes}.}
	\label{Table:TrianIndices}
\end{table}

\section{Magnetoconductivity via orbit integrals \label{sec:OrbInt}}

Transverse magnetoconductivity of most metallic materials originates from classical orbital mechanism, which arises due to the bending of quasiparticle trajectories by the magnetic field \cite{PippardBook,AbrikosovBook}.
We assume that the system can be described by the quasiclassical Boltzmann equation within the relaxation-time approximation. In this case, the conductivity tensor for the magnetic field along the $z$
axis can be represented as \citep{ShockleyPhysRev50,Koshelev2013}
\begin{equation}
\sigma_{\alpha\beta}=2e^{2}\sum_{\mathrm{pockets}}\int\frac{dp_{z}}{(2\pi)^{3}}S_{\alpha\beta}(p_{z}),\label{CondGen}
\end{equation}
where $S_{\alpha\beta}(p_{z})$ describes the contribution from a
single orbit at a single $p_{z}$ slice, which for closed orbits can
be written as
\begin{align}
S_{\alpha\beta}\! & =\!\varsigma_{\alpha\beta}\frac{c}{|e|H}\left[1\!-\!\exp\left(\!-\frac{H_{O}}{H}\right)\right]^{-1}\nonumber \\
\times & \!\varointctrclockwise\frac{dp}{v}v_{\beta}\!\varointctrclockwise_{p}\frac{dp^{\prime}}{v^{\prime}}v_{\alpha}^{\prime}\exp\left(\!-\!\int_{p}^{p^{\prime}}\frac{dp^{\prime\prime}}{v^{\prime\prime}}\frac{c}{|e|H\tau}\right),\label{pzSliceFullPer}
\end{align}
where $v$ and $\tau$ are the Fermi velocity and the isotropic scattering
time, respectively, $\varsigma_{\alpha\alpha}=1$, $\varsigma_{xy}=-\varsigma_{yx}=1\left(-1\right)$
for an electron (hole) pocket, 
\[
H_{O}=\frac{c}{|e|}\varointctrclockwise\frac{dp}{v}\frac{1}{\tau}
\]
is the field at which the time required to complete the full orbit
is equal to the scattering time. For brevity, we will call the quantity
$S_{\alpha\beta}$ a conductivity slice. All integrals are performed
along fixed-$p_{z}$ orbits along the Fermi surface and $\varointctrclockwise$
notates the counterclockwise integration over the whole orbit. 
\footnote{Note that the counterclockwise integration direction in Eq.\ \eqref{pzSliceFullPer} coincides with direction of orbital motion for hole pockets while for electron pockets the orbital motion is clockwise.}
This ``tube-integral'' presentation provides a convenient basis for the analysis
of the conductivity when the Fermi velocity $v$ has sharp features
leading to complicated magnetic field dependences. A similar orbital
approach has been used for modeling of anisotropic magnetotransport
of cuprate superconductors \citep{AbdelJawadNPhys06,AyresNat2021,FangNatPhys2022}.

In the case of a Fermi surface with $m$-fold symmetry, which can be
split into $m$ equivalent segments each having mirror symmetry, we derive
in appendix \ref{sec:App-mfold} a convenient presentation for the
conductivity slices containing only integrals for a single segment,
\begin{subequations} 
\begin{align}
S_{xx}^{(m)}\! & =\frac{m}{2}\frac{c}{|e|H}\left\{ \mathcal{G}_{tt}\!+\!\mathcal{G}_{ll}\!-\!\mathrm{Re}\!\left[\frac{\left(\mathcal{R}_{t}\!+\!\imath\mathcal{R}_{l}\right)^{2}}{\exp\left(-\imath\frac{2\pi}{m}\right)\!-\!\eta}\right]\!\right\} ,\label{eq:mfoldSxx}\\
S_{xy}^{(m)}\! & =\!\varsigma_{xy}\!\frac{m}{2}\frac{c}{|e|H}\left\{ \mathcal{G}_{lt}\!-\!\mathcal{G}_{tl}\!+\!\mathrm{Im}\!\left[\frac{\left(\mathcal{R}_{t}\!+\!\imath\mathcal{R}_{l}\right)^{2}}{\exp\left(-\imath\frac{2\pi}{m}\right)\!-\!\eta}\right]\!\right\} ,\label{eq:mfoldSxy}
\end{align}
\end{subequations}
where the indices $t$ and $l$ refer to transverse
and longitudinal projections of Fermi velocities inside the segments
and the quantities $\eta$, $\mathcal{R}_{s}$ and $\mathcal{G}_{sr}$
are defined in terms of segment integrals as 
\begin{subequations}
\begin{align}
	\eta& =\exp\left(\!-\int\limits _{p_{\mathrm{i}}}^{p_{\mathrm{f}}}\frac{dp^{\prime}}{v^{\prime}}\frac{c}{|e|H\tau}\right),\label{eq:mfoldQdef-1}\\
\mathcal{R}_{s} & \!=\!\int\limits _{p_{\mathrm{i}}}^{p_{\mathrm{f}}}\frac{dp^{\prime}}{v^{\prime}}v_{s}^{\prime}\exp\left(\!-\int\limits _{p_{\mathrm{i}}}^{p^{\prime}}\frac{dp^{\prime\prime}}{v^{\prime\prime}}\frac{c}{|e|H\tau}\right),\label{eq:mfoldRsDef-1}\\
\mathcal{G}_{sr} & \!=\!\int\limits _{p_{\mathrm{i}}}^{p_{\mathrm{f}}}\!\frac{dp}{v}v_{s}\int\limits _{p}^{p_{\mathrm{f}}}\frac{dp^{\prime}}{v^{\prime}}v_{r}^{\prime}\exp\left(\!-\!\int\limits _{p_{\mathrm{i}}}^{p^{\prime}}\frac{dp^{\prime\prime}}{v^{\prime\prime}}\frac{c}{|e|H\tau}\right)
\label{eq:mfoldGsrDef}
\end{align}
\end{subequations}
with $s,r=l,t$. Here the momenta $p_{\mathrm{i}}$
and $p_{\mathrm{f}}$ define segment limits. The terms with the functions
$\mathcal{G}_{sr}$ and $\mathcal{R}_{s}$ represent the contributions
from the integration over $p$ and $p^{\prime}$ in Eq.~\eqref{pzSliceFullPer}
belonging to the same and different segments of the Fermi surface,
respectively. 

Generally, the calculation of the conductivity using the above orbital-integral method requires knowledge of the electronic spectrum and local scattering time. 
Although electronic spectra can be obtained through ab-initio DFT calculations, detailed microscopic information about scattering times is typically unavailable requiring some simplifying assumptions. Moreover, calculations with a fully realistic electronic spectrum can only be performed numerically.  
In our case, however, using the simple minimum model of the large hexagon pocket outlined in Section \ref{sec:LargeHex} and its subsequent reconstruction discussed in Section \ref{sec:FoldCDW}, we can analytically evaluate all orbital integrals and derive closed analytical results for the conductivity slices in Eqs.\ \eqref{eq:mfoldSxx} and \eqref{eq:mfoldSxy} for both original and reconstructed Fermi sheets. This will allow us to perform a qualitative analysis of the conductivity behavior in a magnetic field. 

\section{Magnetoconductivity from large concave hexagon in unfolded Brillouin
zone\label{sec:LHsigmaH}}

The anomalous behavior of magnetotransport in kagome superconductors
is observed at low temperatures, when the Fermi surface is reconstructed
by the CDW order \cite{YangSciAdv20,YuPhysRevB.104.L041103,ZhengNatComm2023,WangPhysRevB.108.035117}. Nevertheless, we consider first the magnetoconductivity
for the large concave hexagon in the unfolded Brillouin zone, because
it provides a convenient point of reference and already has a nonstandard behavior
due to the proximity to the van Hove singularities. In particular,
we will see that the Hall conductivity actually has a shape similar
to the experimental one in Fig. \ref{Fig:sab10K} pointing to its van
Hove origin. Also, the behavior of the Hall conductivity in this case
is of general interest, since it represents an almost textbook example
of a hole-type Fermi surface with \emph{negative} sign of $\sigma_{xy}$
at small fields. 

We use the general results derived in Appendix \ref{sec:App-mfold}
for conductivity slices $S_{x\beta}$ of an arbitrary Fermi pocket
with $m$-fold symmetry, as summarized in Eqs.~\eqref{eq:mfoldSxx} and
\eqref{eq:mfoldSxy}. For the hexagon hole Fermi pocket ($m=6$),
these results become 
\vspace{-0.2in}
\begin{widetext}
\vspace{-0.2in}
\begin{subequations}
\begin{align}
S_{xx}^{\mathrm{LH}}= & \frac{3c}{|e|H}\left\{ \mathcal{G}_{vv}^{\mathrm{LH}}\!+\!\mathcal{G}_{uu}^{\mathrm{LH}}-\frac{\left(\frac{1}{2}-\!\eta_{\mathrm{LH}}\right)\left[(\mathcal{R}_{v}^{\mathrm{LH}})^{2}\!-\!(\mathcal{R}_{u}^{\mathrm{LH}})^{2}\right]\!-\!\sqrt{3}\mathcal{R}_{v}^{\mathrm{LH}}\mathcal{R}_{u}^{\mathrm{LH}}}{1-\eta_{\mathrm{LH}}+\eta_{\mathrm{LH}}^{2}}\right\} ,\label{eq:SxxResultLH}\\
S_{xy}^{\mathrm{LH}}= & \!-\!\frac{3c}{|e|H}\left\{ -2\mathcal{G}_{vu}^{\mathrm{LH}}\!+\frac{\frac{\sqrt{3}}{2}\left[\left(\mathcal{R}_{v}^{\mathrm{LH}}\right)^{2}\!-\!\left(\mathcal{R}_{u}^{\mathrm{LH}}\right)^{2}\right]+\left(\frac{1}{2}-\!\eta_{\mathrm{LH}}\right)2\mathcal{R}_{v}^{\mathrm{LH}}\mathcal{R}_{u}^{\mathrm{LH}}}{1-\eta_{\mathrm{LH}}+\eta_{\mathrm{LH}}^{2}}\right\} .\label{eq:SxyResultLH}
\end{align}
\end{subequations}
Here, the parameter $\eta_{\mathrm{LH}}$ and segment
integrals $\mathcal{R}_{k}^{\mathrm{LH}}$ and $\mathcal{G}_{km}^{\mathrm{LH}}$
are defined by Eqs.~\eqref{eq:mfoldQdef-1}, \eqref{eq:mfoldRsDef-1},
and \eqref{eq:mfoldGsrDef}. The indices $k\!=\!v,u$ and $m\!=\!v,u$
correspond to the velocity projections to the directions of the local
rotated hyperbolic coordinates, which in this case correspond to the
transverse and longitudinal directions of the segments. 
Therefore, the $p$ and $p^{\prime}$ integration over the hyperbolic
segments in Eqs.~\eqref{eq:mfoldQdef-1}, \eqref{eq:mfoldRsDef-1},
and \eqref{eq:mfoldGsrDef} can be analytically
carried out expanding the velocity over the local rotated basis and using the hyperbolic parametrization defined in Eq.\ \eqref{eq:HypParam}, as
described in Appendix \ref{subsec:Integration-hyperbolicLH}. For the parameter $\eta_{\mathrm{LH}}$
in Eq.~\eqref{eq:mfoldQdef-1}, we obtain 
\begin{equation}
\eta_{\mathrm{LH}}=\exp\left(-2t_{b}/\omega_{h}\right),\label{eq:nuLH}
\end{equation}
where $t_{b}$ is defined in Eq.~\eqref{eq:tb} and 
\begin{equation}
\omega_{h}\!\equiv\!\omega_{c}\tau\!=\!H/H_{0}\!=\!|e|H\tau/(c\sqrt{m_{u}m_{v}})\label{eq:wh}
\end{equation}
is the reduced magnetic field with the field scale $H_{0}\!=\!c\sqrt{m_{u}m_{v}}/(|e|\tau)$.
The ratio in the exponent $2t_{b}/\omega_{h}$ is the ratio of the
time to pass one hexagon segment $t_{\mathrm{segm}}\!=\!2ct_{b}\sqrt{m_{u}m_{v}}/(|e|H)$
to the scattering time $\tau$. This ratio defines the second field
scale, $H_{b}\!=\!2t_{b}H_{0}\!\simeq2\ln\frac{2K/p_{u0}}{\sqrt{3r_{m}}\!+\!1}H_{0}$,
which is much larger than $H_{0}$ and diverges on approaching the
van Hove singularity for $p_{u0}\!\rightarrow\!0$. The integrals
$\mathcal{R}_{k}^{\mathrm{LH}}$ in Eq.~\eqref{eq:mfoldRsDef-1}
are evaluated as $\mathcal{R}_{v}^{\mathrm{LH}}=p_{u0}G_{bv}$, $\mathcal{R}_{u}^{\mathrm{LH}}=-p_{v0}G_{bu}$,
with the reduced functions \begin{subequations}
\begin{align}
G_{bv} & =\frac{1}{1-\omega_{h}^{-2}}\left[-\!\cosh t_{b}\!+\frac{1}{\omega_{h}}\sinh t_{b}\!+\!\left(\cosh t_{b}\!+\frac{1}{\omega_{h}}\sinh t_{b}\right)\eta_{\mathrm{LH}}\right],\label{eq:Gbv}\\
G_{bu} & =\frac{1}{1-\omega_{h}^{-2}}\left[\sinh t_{b}\!-\frac{1}{\omega_{h}}\cosh t_{b}\!+\!\left(\sinh t_{b}\!+\frac{1}{\omega_{h}}\cosh t_{b}\right)\eta_{\mathrm{LH}}\right].\label{eq:Gbu}
\end{align}
\end{subequations}In turn, the same-segment integrals $\mathcal{G}_{km}^{\mathrm{LH}}$
defined in Eq.~\eqref{eq:mfoldGsrDef} are evaluated as $\mathcal{G}_{vv}^{\mathrm{LH}}=p_{u0}^{2}K_{bvv}$,
$\mathcal{G}_{uu}^{\mathrm{LH}}=p_{v0}^{2}K_{buu}$, and $\mathcal{G}_{vu}^{\mathrm{LH}}=-p_{u0}p_{v0}\mathcal{K}_{bvu}$
with\begin{subequations}
\begin{align}
\mathcal{K}_{bvv}= & \frac{1}{1\!-\!\omega_{h}^{-2}}\left[ -\left(\cosh t_{b}\!+\!\frac{1}{\omega_{h}}\sinh t_{b}\right)G_{bv}-\!\frac{\sinh\left(2t_{b}\right)-2t_{b}}{2\omega_{h}}\right] ,\label{eq:Kbvv}\\
\mathcal{K}_{buu}= & \frac{1}{1\!-\!\omega_{h}^{-2}}\left[ \left(\sinh t_{b}\!+\!\frac{1}{\omega_{h}}\cosh t_{b}\right)G_{bu}\!-\!\frac{\sinh\left(2t_{b}\right)\!+2t_{b}}{2\omega_{h}}\right] ,\label{eq:Kbuu}\\
\mathcal{K}_{bvu}= & \frac{1}{1\!-\!\omega_{h}^{-2}}\left[ \frac{\frac{1}{\omega_{h}}\!-\!\eta_{\mathrm{LH}}
	\left(\sinh t_{b}\!+\frac{1}{\omega_{h}}\cosh t_{b}\right)
	\left(\cosh t_{b}\!+\frac{1}{\omega_{h}}\sinh t_{b}\right)}{1\!-\omega_{h}^{-2}}+\!t_{b}\right] .\label{eq:Kbvu}
\end{align}
\end{subequations}
Note that, in spite of the denominators
$1\!-\!\omega_{h}^{-2}$ in the above formulas, all functions have
regular behavior for $\omega_{h}\rightarrow1$. 

For numerical evaluations, we rewrite the results in Eqs.~\eqref{eq:SxxResultLH}
and \eqref{eq:SxyResultLH} in the dimensionless form as 
\begin{subequations}
\begin{align}
S_{\alpha\beta}^{\mathrm{LH}} & \!=\!S_{0}F_{\alpha\beta}^{\mathrm{LH}},\label{eq:SLHabReduced}\\
F_{xx}^{\mathrm{LH}} & =\frac{1}{\omega_{h}}\left[\mathcal{K}_{bvv}\!+r_{m}\mathcal{K}_{buu}\!-\frac{\left(\frac{1}{2}-\!\eta_{\mathrm{LH}}\right)\left(G_{bv}^{2}\!-\!r_{m}G_{bu}^{2}\right)\!+\!\sqrt{3r_{m}}G_{bv}G_{bu}}{1-\eta_{\mathrm{LH}}+\eta_{\mathrm{LH}}^{2}}\right],\label{eq:FLHxx}\\
F_{xy}^{\mathrm{LH}} & =-\frac{1}{\omega_{h}}\left[-2\sqrt{r_{m}}\mathcal{K}_{buv}\!+\frac{\frac{\sqrt{3}}{2}\left(G_{bv}^{2}\!-\!r_{m}G_{bu}^{2}\right)\!-\left(\frac{1}{2}-\!\eta_{\mathrm{LH}}\right)2\sqrt{r_{m}}G_{bv}G_{bu}}{1-\eta_{\mathrm{LH}}+\eta_{\mathrm{LH}}^{2}}\right].\label{eq:FLHxy}
\end{align}
\end{subequations}
\end{widetext}
Here 
\begin{equation}
S_{0}\!=\!3\tau p_{u0}^{2}/\sqrt{m_{u}m_{v}}\label{eq:S0Def}
\end{equation}
the overall scale for $S_{\alpha\beta}$, which we will use throughout
the paper. The dimensionless functions $F_{\alpha\beta}^{\mathrm{LH}}$
of the reduced magnetic field $\omega_{h}$ also depend on two parameters,
the mass ratio $r_{m}$ and the limiting hyperbolic parameter $t_{b}$
which in turn is determined by $r_{m}$ and the ratio $\kappa\!=\!K/p_{u0}$
in Eq.~\eqref{eq:tb}. The latter parameter diverges on approaching
the van Hove point for $\varepsilon_{\mathrm{vH}}\!\rightarrow\!0$. 
At large $t_b$ the functions $F_{\alpha\beta}^{\mathrm{LH}}$ scale as $\exp(2t_b)\! \propto \! \kappa^2$ meaning that the main scale of $S_{\alpha\beta}^{\mathrm{LH}}$ is 
$S_{0}\kappa^{2}\!=\!3\tau K^{2}/\sqrt{m_{u}m_{v}}$ which does not depend on $p_{u0}$.
As a consequence, the slices $S_{\alpha\beta}^{\mathrm{LH}}$ only weakly depend on the distance to the van Hove point $\varepsilon_{\mathrm{vH}}$.
We can also note that the product $S_{0}H_{0}=3p_{u0}^{2}c/|e|$ does not depend
on the scattering rate and effective masses and is only determined
by the distance between the LH corner and the van Hove point $p_{u0}$.
As follows from Eq.~\eqref{CondGen}, the total contribution of the
large hexagon to the conductivity components are obtained by the integration
of the slices $S_{\alpha\beta}^{\mathrm{LH}}$ over the c-axis momentum
$p_{z}$. Most importantly, the distance to the van Hove point $p_{u0}$
depends on $p_{z}$ and this distance determines both $S_{0}$ and
$t_{b}$. 

The key feature determining the behavior of the magnetotransport is the slowing
down of the orbital motion near the van Hove points, see Eq.~\eqref{eq:vtot}.
To understand the behavior of the large-hexagon magnetoconductivity
at the qualitative level, we will study the dependences $F_{\alpha\beta}^{\mathrm{LH}}(\omega_{h})$
at the different values of $p_{u0}$ (or, equivalently, the parameter
$\kappa=K/p_{u0}$). A quick inspection of the above results suggests
that for $\kappa\gg$1 these dependences are characterized by two
scales of the magnetic field corresponding to $\omega_{h}\sim1$ and
$\omega_{h}\sim2t_{b}\gg1$, where the second scale logarithmically
diverges when the Fermi energy approaches the van Hove point. 

Consider first the behavior of the diagonal function $F_{xx}^{\mathrm{LH}}$.
Using the low-field expansions of the reduced functions $G_{bk}$
and $\mathcal{K}_{bkk}$ listed in Appendix \ref{subsec:Integration-hyperbolicLH},
in Eqs.~\eqref{eq:GbvAs}, \eqref{eq:GbuAs}, \eqref{eq:KbvvAs},
and \eqref{eq:KbuuAs}, we obtain the small-field asymptotics of $F_{xx}^{\mathrm{LH}}$,
\begin{align}
F_{xx}^{\mathrm{LH}}\! & \simeq\!\frac{r_{m}\!+\!1}{2}\sinh\left(2t_{b}\right)\!+\left(r_{m}\!-\!1\right)t_{b}\nonumber \\
 & -\frac{\omega_{h}}{2}\left(\sqrt{3}\sinh t_{b}\!-\!\sqrt{r_{m}}\cosh t_{b}\right)^{2}.\label{eq:FLHxxSmallH}
\end{align}
The finite linear term here is an artifact of the approximate model,
it appears due to velocity jumps at the boundaries between the segments.
Numerically, this term is rather small. At higher magnetic fields
$\omega_{h}\gg1$, using asymptotics in Eqs.~\eqref{eq:GbvsInt}
and \eqref{eq:KbssInt} also listed in Appendix \ref{subsec:Integration-hyperbolicLH},
we obtain a simple approximate result
\begin{align}
F_{xx}^{\mathrm{LH}}\simeq & \frac{\kappa^{2}}{2\omega_{h}}\frac{1-\eta_{\mathrm{LH}}^{2}}{1-\eta_{\mathrm{LH}}+\eta_{\mathrm{LH}}^{2}}.\label{eq:SxxLargeHexHighH}
\end{align}
As follows from the definition of $\eta_{\mathrm{LH}}$ in Eq.~\eqref{eq:nuLH},
nominally, this function describes a crossover between $1/\omega_{h}$
behavior for $1\ll\omega_{h}\ll2t_{b}$ and $1/\omega_{h}^{2}$ for
$\omega_{h}\gg2t_{b}$. However, since $t_{b}\sim\ln\kappa$ and the
logarithm is a slowly growing function, the first asymptotics is only
pronounced for extremely large $\kappa$. The highest-field asymptotics
of $F_{xx}^{\mathrm{LH}}$ is $F_{xx}^{\mathrm{LH}}\simeq2\kappa^{2}t_{b}/\omega_{h}^{2}$. 

\begin{figure}
\includegraphics[width=3.2in]{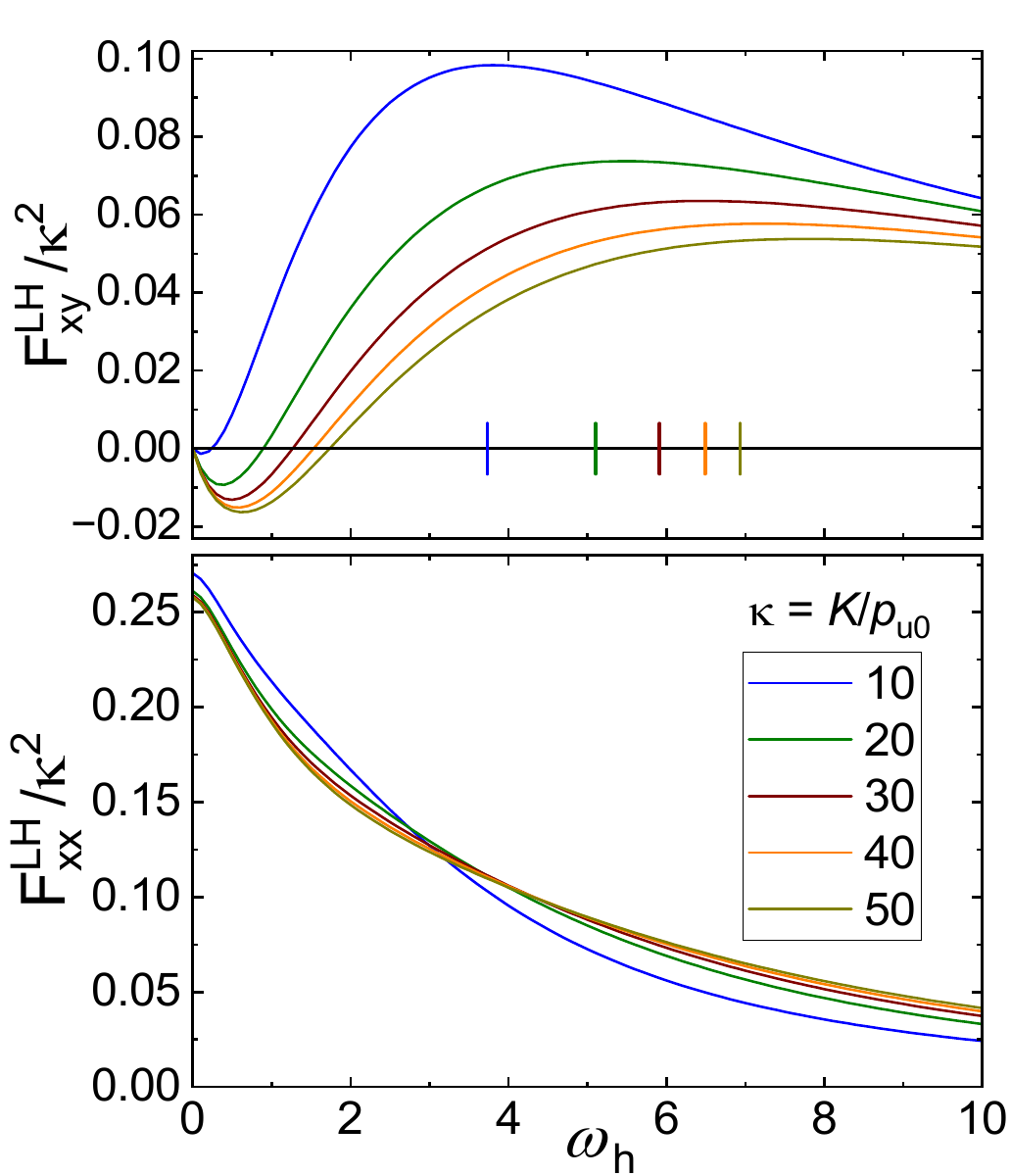}
\caption{The dependences of the functions $F_{\alpha\beta}^{\mathrm{LH}}$
on the reduced magnetic field $\omega_{h}$. These functions determine
the slice contributions to the diagonal and Hall conductivities and
are defined in Eqs.~\eqref{eq:FLHxx} and \eqref{eq:FLHxy}. The plots
are made for for $r_{m}=1.5$ and different values of the parameter
$\kappa=K/p_{u0}$, characterizing proximity to the van Hove point.
The vertical bars in the upper plot mark the crossover fields $\omega_{h}=2t_{b}\propto\ln\kappa$.}
\label{Fig:FLHab}
\end{figure}

The full field dependence at the van Hove point, $\kappa\rightarrow\infty$,
is
\begin{align}
F_{xx}^{\mathrm{LH}}/\kappa^{2}\! & \rightarrow\!\frac{1}{\left(\sqrt{3r_{m}}\!+\!1\right)^{2}}\left\{ \left[1\!-\frac{\omega_{h}^{2}}{\left(1\!+\!\omega_{h}\right)^{2}}\right]\left(1\!+\!r_{m}\right)\!\right.\nonumber \\
 & \left.\!-\frac{\omega_{h}/2}{\left(1\!+\!\omega_{h}\right)^{2}}\left(\sqrt{3}\!-\!\sqrt{r_{m}}\right)^{2}\right\} .\label{eq:FLHxx-vH}
\end{align}
In this case, the high-field decay remains $\propto1/\omega_{h}$
in the whole field range $\omega_{h}\gg1$ without crossover to $1/\omega_{h}^{2}$
limit. Therefore, the intermediate $1/\omega_{h}$ asymptotics is
the distinct qualitative feature induced by van Hove singularities.
Plots of $F_{xx}^{\mathrm{LH}}(\omega_{h})$ for $r_{m}=1.5$ and different
values of $\kappa$ are shown in the lower panel of Fig.~\ref{Fig:FLHab}. 

The behavior of the Hall component $F_{xy}^{\mathrm{LH}}$ in Eq.~\eqref{eq:FLHxy}
is more peculiar. The linear Hall term in small magnetic fields,
following from the low-field limits of the functions $G_{bs}$ and
$\mathcal{K}_{buv}$ (see Eqs.~\eqref{eq:GbvAs}, \eqref{eq:GbuAs},
and \eqref{eq:KbuvAs} in Appendix \ref{subsec:Integration-hyperbolicLH}),
can be evaluated as
\begin{align}
	F_{xy}^{\mathrm{LH}}\simeq & \!-\!\omega_{h}
	\Bigg[\frac{1}{2}\left(\sqrt{3}\sinh t_{b}\!-\!\sqrt{r_{m}}\cosh t_{b}\right)\nonumber\\
	\times&\left(\sinh t_{b}\!+\!\sqrt{3r_{m}}\cosh t_{b}\right)\!-\!2\sqrt{r_{m}}t_{b}\Bigg]\label{eq:SxyLHLinH}\\
	\approx & -\frac{\kappa^{2}\omega_{h}}{2}\frac{\sqrt{3}-\sqrt{r_{m}}}{1+\sqrt{3r_{m}}}.\nonumber 
\end{align}
The linear Hall component is \emph{negative}, in spite of the hole nature
of the large hexagon pocket (its Fermi surface surrounds empty states).
It is straightforward to understand the origin of this unusual feature.
Due to its concave shape, the Fermi surface is composed of pieces with both curvatures.
Since the regions with positive curvature are located near the van Hove
points where the Fermi velocity is small, their contribution is
suppressed resulting in the dominating electron-like (negative) contribution of the sections with negative curvature. A simple general geometric interpretation
of the linear Hall magnetoconductivity has been suggested by Ong \citep{OngPhysRevB.43.193},
who demonstrated that in the two-dimensional case $\sigma_{xy}$ is proportional
by the area $A_{l}$ swept by the vector mean-free path $\boldsymbol{l}=\tau\boldsymbol{\upsilon}$
during a full orbital cycle, $A_{l}=\frac{1}{2}\oint\left(l_{y}dl_{x}\!-\!l_{x}dl_{y}\right)$.
For constant scattering time $A_{l}$ is proportional to the the area
$A_{v}$ swept by the velocity, $A_{l}=\tau^{2}A_{v}$ with 
\begin{equation}
A_{v}=\frac{1}{2}\oint\left(v_{y}dv_{x}\!-\!v_{x}dv_{y}\right)\label{eq:Av}
\end{equation}
and in our case we have
\begin{align}
S_{xy} & =\!\frac{|e|H\tau^{2}}{c}A_{v}\label{eq:SxyLinArea}
\end{align}
in the small-field limit.
Note that areas swept clockwise and counterclockwise contribute to
$A_{v}$ with opposite signs. Figure \ref{Fig:LargeHex}(c) illustrates
the velocity contour corresponding to the large concave hexagon Fermi
surface shown in Figure \ref{Fig:LargeHex}(a). It is composed of
the central small hexagon corresponding to the regions near the van Hove
points and six outside ``petals''. The area inside the small hexagon
is swept clockwise corresponding to holelike contribution, while total
area inside the petals which are swept counterclockwise gives electron
contribution to $A_{v}$ and $S_{xy}$. As the Fermi energy approaches
the van Hove point, the central hexagon shrinks and contribution from
the ``petals'' dominates yielding the total negative sign of the
Hall term.

The field range for the linear behavior, however, is rather narrow.
The regions near the van Hove points also induce strong field dependence
of $F_{xy}^{\mathrm{LH}}$. At higher magnetic fields $\omega_{h}\gg1$,
using approximate limits for the functions $G_{bs}$ and $\mathcal{K}_{buv}$
listed in Eqs.~\eqref{eq:GbvsInt} and \eqref{eq:KbuvInt}, we
obtain a simple approximate result,
\begin{align}
F_{xy}^{\mathrm{LH}}\simeq & -\frac{\kappa^{2}}{2\omega_{h}}\left[\frac{\sqrt{3}\!-\!\sqrt{r_{m}}}{1\!+\!\sqrt{3r_{m}}}\!-\frac{\sqrt{3}\eta_{\mathrm{LH}}}{1\!-\!\eta_{\mathrm{LH}}\!+\!\eta_{\mathrm{LH}}^{2}}\right].
\label{eq:eq:SxyLH-HighH}
\end{align}
This function describes the crossover between the two $1/\omega_{h}$
dependences with opposite signs,
\begin{widetext}
\begin{align}
F_{xy}^{\mathrm{LH}} & \!\simeq\frac{\kappa^{2}}{2\left(1\!+\!\sqrt{3r_{m}}\right)\omega_{h}}\!\times \!
\begin{cases}
-\left(\sqrt{3}\!-\!\sqrt{r_{m}}\right),\!\!\!\! &\!\mathrm{for}\,1\!\ll\omega_{h}\ll\!2t_{b}\\
4\sqrt{r_{m}},\! &\!\mathrm{for}\,\omega_{h}\gg2t_{b}
\end{cases}.\label{eq:FLHxyAsymp}
\end{align}
\end{widetext}
This means that $F_{xy}^{\mathrm{LH}}$ changes sign.
It is known that the highest-field asymptotics of the Hall conductivity
does not depend on the shape of the Fermi surface and is determined only
by the type and density of the carriers \citep{AbrikosovBook}. In
our case this corresponds to the relation
\begin{align}
S_{xy}^{\mathrm{LH}} & \simeq\frac{A_{\mathrm{LH}}}{|e|H/c},\label{eq:SxyLargeH}
\end{align}
where $A_{\mathrm{LH}}$ is the area of the large-hexagon pocket,
which can be evaluated as $A_{\mathrm{LH}}\approx\frac{6\sqrt{r_{m}}K^{2}}{1+\sqrt{3r_{m}}}$
for the Fermi energy at the van Hove point. The high-field asymptotics
in Eq.~\eqref{eq:FLHxyAsymp} reproduces this result. On the other
hand, the intermediate-filed asymptotics in Eq.~\eqref{eq:FLHxyAsymp}
corresponds to 
\begin{align}
S_{xy}^{\mathrm{LH}} & \simeq\frac{A_{\mathrm{LH}}-A_{\mathrm{hex}}}{|e|H/c},\label{eq:SxyIntH}
\end{align}
where $A_{\mathrm{hex}}\approx\frac{3\sqrt{3}}{2}K^{2}$ is the area
inside the ideal hexagon composed of straight lines connecting the
van Hove points. In Fig.~\ref{Fig:FLHab} we also present plots of
$F_{xy}^{\mathrm{LH}}(\omega_{h})$ for different values of $\kappa$.
We see that the field dependences are nonmonotonic and $F_{xy}^{\mathrm{LH}}$
changes sign at the magnetic field progressively increasing with $\kappa$.
At higher fields $F_{xy}^{\mathrm{LH}}$ has a smooth maximum. We
immediately notice that these shapes resemble the shape of experimental
curve $\sigma_{xy}(H)$ for CsV$_{3}$Sb$_{5}$ shown in Fig.~\ref{Fig:sab10K}
strongly suggesting that the unusual behavior of the Hall conductivity
in this material has van Hove origin. The absence of anomalous behavior above the CDW transition is most likely related by high scattering rates leading to large magnetic-field scales. 
In addition, it is clear that the
magnetotransport in this material at low temperatures is also affected
by the Fermi surface reconstruction by CDW, which we analyze in the
next section.

\section{Magnetoconductivity from reconstructed pockets\label{sec:MagCondReconstr}}

\subsection{Hyperbolic hexagon pocket\label{subsec:MagCondHH}}

As discussed in Section \ref{sec:FoldCDW}, reconstruction of the large hexagon pocket by the 3Q CDW order leads to formation of small hyperbolic hexagon sheet near the BZ center and two large triangular sheets, see Fig.\ \ref{Fig:HypHexagon}.
In this section, we consider the contribution of the hyperbolic hexagon to the diagonal conductivity
neglecting the CDW gaps near the branch crossings. In this
case, the interruption of the smooth motion along the orbit at the sharp corners
leads to linear magnetoconductivity \citep{PippardBook,FentonSchofieldPRL05,Koshelev2013}.
The general result for the diagonal and Hall conductivity slices is
given by Eqs.~\eqref{eq:SxxResultLH} and \eqref{eq:SxyResultLH},
where the large-hexagon label $\mathrm{LH}$ has to be replaced by
the hyperbolic-hexagon label $\mathrm{HH}$. The only differences
are in the values of parameter $\eta_{\mathrm{HH}}$ and the segment
integrals $\mathcal{R}_{s}^{\mathrm{HH}}$ and $\mathcal{G}_{sr}^{\mathrm{HH}}$.
In calculations, we will employ the same hyperbolic parametrization as for the large hexagon, Eq.\ \eqref{eq:HypParam}.
Two differences with respect to the large hexagon are 
(i)the crossing of
hyperbolas is determined by the parameter $t_{c}$ in Eq.~\eqref{eq:tc-rm-def},
which is much smaller than the parameter $t_{b}$ in Eq.~\eqref{eq:tb} and (ii) the counterclockwise direction of motion along HH Fermi surface corresponds to change of the hyperbolic parameter from $t_c$ to $-t_c$.

The $p$ and $p^{\prime}$ integration over hyperbolic segments in
Eq.~\eqref{pzSliceFullPer} for $S_{xx}$ and $S_{xy}$ can be analytically
carried out for every segment $[p_{j},p_{j+1}]$ expanding the velocity
over the local rotated basis and using hyperbolic parametrization.
The calculation of segment integrals is virtually identical to the
large-hexagon case, as described in Appendix \ref{subsec:Integration-hyperbolicLH}.
The only differences are the range of the hyperbolic parameter, $-t_{c}\!<\!t\!<\!t_{c}$
and the opposite direction of integration with respect to parameter $t$ 
leading to the opposite signs of the integrals $\mathcal{R}_{v}^{\mathrm{HH}}$
and $\mathcal{G}_{vu}^{\mathrm{HH}}$ in comparison with  $\mathcal{R}_{u}^{\mathrm{LH}}$
and $\mathcal{G}_{vu}^{\mathrm{LH}}$. Therefore, the parameter $\eta_{\mathrm{HH}}=\exp\left(\!-\!\int_{p_{1}}^{p_{2}}\frac{dp^{\prime\prime}}{v^{\prime\prime}}\frac{c}{|e|H\tau}\right)$
is evaluated as
\begin{equation}
\eta_{\mathrm{HH}}\!=\!\exp\left(-2t_{c}/\omega_{h}\right)\label{eq:nuHH}
\end{equation}
with $t_{c}$ being defined in Eq.~\eqref{eq:tc-rm-def}. The parameter
in the exponent $2t_{c}/\omega_{h}$ is the ratio of the time to pass
one hexagon segment $t_{\mathrm{segm}}=2ct_{c}\sqrt{m_{u}m_{v}}/|e|H$
to the scattering time $\tau$. Contrary to the case of the large
hexagon considered in the Sec.~\ref{sec:LHsigmaH}, this ratio does
not define an additional field scale, since $t_{c}\! \sim \! 1$. The integrals
$\mathcal{R}_{k}^{\mathrm{HH}}$ are evaluated as $\mathcal{R}_{v}^{\mathrm{HH}}\!=\!-p_{u0}G_{cv}$,
$\mathcal{R}_{u}^{\mathrm{HH}}\!=\!-p_{v0}G_{cu}$, where the functions
$G_{cs}$ are obtained from the functions $G_{bs}$ in Eqs.~\eqref{eq:Gbv}
and \eqref{eq:Gbu} by replacements $t_{b}\rightarrow t_{c}$ and
$\eta_{\mathrm{LH}}\rightarrow\eta_{\mathrm{HH}}$. Correspondingly,
the same-segment integrals $\mathcal{G}_{sr}^{\mathrm{HH}}$ are evaluated
as $\mathcal{G}_{vv}^{\mathrm{HH}}\!=\!p_{u0}^{2}K_{cvv}$, $\mathcal{G}_{uu}^{\mathrm{HH}}\!=\!p_{v0}^{2}K_{cuu}$,
and $\mathcal{G}_{vu}^{\mathrm{HH}}\!=\!p_{u0}p_{v0}\mathcal{K}_{cvu}$
and the functions $K_{csr}$ are obtained from the functions $K_{bsr}$
in Eqs.~\eqref{eq:Kbvv}, \eqref{eq:Kbuu}, and \eqref{eq:Kbvu}
using the same substitutions $t_{b}\rightarrow t_{c}$ and $\eta_{\mathrm{LH}}\rightarrow\eta_{\mathrm{HH}}$.
These results yield the following reduced presentation for the HH
conductivity slices
\begin{subequations}
\begin{equation}
	S_{\alpha\beta}^{\mathrm{HH}} =S_{0}F_{\alpha\beta}^{\mathrm{HH}},\label{eq:SabHHReduced}
\end{equation}	
\begin{widetext}
\begin{align}
F_{xx}^{\mathrm{HH}} & =\frac{1}{\omega_{h}}\left[\mathcal{K}_{cvv}\!+\!r_{m}\mathcal{K}_{cuu}-\frac{\left(\frac{1}{2}-\!\eta_{\mathrm{HH}}\right)\left(G_{cv}^{2}\!-\!r_{m}G_{cu}^{2}\right)\!-\!\sqrt{3r_{m}}G_{cv}G_{cu}}{1-\eta_{\mathrm{HH}}+\eta_{\mathrm{HH}}^{2}}\right],\label{eq:FHHxx}\\
F_{xy}^{\mathrm{HH}} & =\frac{1}{\omega_{h}}\left[-2\sqrt{r_{m}}\mathcal{K}_{cvu}\!+\frac{\frac{\sqrt{3}}{2}\left(G_{cv}^{2}\!-\!G_{cu}^{2}\right)+\!2\left[\frac{1}{2}-\!\eta_{\mathrm{HH}}\right]G_{cv}G_{cu}}{1-\eta_{\mathrm{HH}}+\eta_{\mathrm{HH}}^{2}}\right].\label{eq:FHHxy}
\end{align}
\end{widetext}
\end{subequations}
Here, the reduced magnetic field
$\omega_{h}$ and the scale $S_{0}$ are defined in Eq.~\eqref{eq:wh}
and Eq.~\eqref{eq:S0Def}, respectively, and, also, listed in Table
\ref{Tbl:param}. 

We observe that the shape of the HH magnetoconductivity is entirely
determined by the mass ratio $r_{m}$, since neither the magnetic-field
scale nor the reduced function $F_{\alpha\beta}^{\mathrm{HH}}$ in
Eqs.~\eqref{eq:FHHxx} and \eqref{eq:FHHxy} depend on the shift
of the Fermi energy with respect the van Hove point $\varepsilon_{\mathrm{vH}}$.
This means that only the scale of the HH conductivity $S_{0}$ depends
on the Fermi energy. As a consequence, the $p_{z}$ integration in
Eq.~\eqref{CondGen} does not change the shape of the HH contribution
to the total conductivity. This means that the HH contribution to
the conductivity in Eq. \eqref{CondGen} has the form similar to Eq.~\eqref{eq:SabHHReduced},
\begin{equation}
\sigma_{xx}^{\mathrm{HH}}=\sigma_{0}^{\mathrm{HH}}F_{xx}^{\mathrm{HH}}(\omega_{h},r_{m})\label{eq:sigmaHH_xx}
\end{equation}
with 
\begin{equation}
\sigma_{0}^{\mathrm{HH}}\!=\!2e^{2}\!\int\!\frac{dp_{z}}{(2\pi)^{3}}S_{0}(p_{z})\!=\frac{3e^{2}\tau}{4\pi^{3}\hbar^{2}\sqrt{m_{u}m_{v}}}\!\int\! dk_{z}p_{u0}^{2}.\label{eq:sigma0}
\end{equation}
The last formula is written in real units and, in order to make the
units more obvious, we assumed that $p_{u0}$ has the dimension of
momentum and $k_{z}\!=\!p_{z}/\hbar$ has the dimension of wave vector.
The $k_{z}$ integration in the last formula is performed over the
$c$-axis extend of the HH pocket. The maximum possible range is determined
by the size of the Brillouin zone $2\pi/\mathsf{c}$ set by the $c$-axis
lattice parameter $\mathsf{c}$. For the product $\sigma_{0}^{\mathrm{HH}}H_{0}$,
we obtain
\begin{equation}
\sigma_{0}^{\mathrm{HH}}H_{0}=\frac{3e^{2}}{4\pi^{3}\hbar^{2}}\frac{c}{|e|}\int\!dk_{z}p_{u0}^{2}.
\label{eq:ScaleProd}
\end{equation}
Since the pocket area $A_{\mathrm{HH}}$ is proportional to $p_{u0}^{2}$,
\eqref{eq:HexArea}, and the pocket volume $V_{\mathrm{HH}}\!=\!\int dk_{z}A_{\mathrm{HH}}$
determines the total carrier density in the pocket $n_{\mathrm{HH}}\!=\!2V_{\mathrm{HH}}/(2\pi)^{3}$,
the product $\sigma_{0}^{\mathrm{HH}}H_{0}$ is proportional to $n_{\mathrm{HH}}$,
\[
\sigma_{0}^{\mathrm{HH}}H_{0}=\frac{1}{2\pi\sqrt{r_{m}}t_{c}}\frac{e^{2}}{\hbar}\Phi_{0}n_{\mathrm{HH}}.
\]
We can also note that the ratio of the product $\sigma_{0}^{\mathrm{HH}}H_{0}$
to the MO frequency $F_{\mathrm{HH}}=\frac{c\hbar}{2\pi|e|}A_{\mathrm{HH}}$,
\begin{equation}
\frac{\sigma_{0}^{\mathrm{HH}}H_{0}}{F_{\mathrm{HH}}}=\frac{3e^{2}}{2\pi^{2}\hbar}
\frac{\int\!dk_{z}k_{u0}^{2}}{A_{\mathrm{HH}}}
\label{eq:ScaleRelatHH}
\end{equation}
gives the possibility to evaluate the c-axis extend of the HH pocket from
experimental data and provides a consistency check.

Let us consider the asymptotic behavior of $F_{xx}^{\mathrm{HH}}(\omega_{h})$.
The asymptotic expansions of the reduced functions $G_{cs}$,$K_{css}$
are the same as for $G_{bs},K_{bss}$ in Eqs.~\eqref{eq:GbvAs},
\eqref{eq:GbuAs}, \eqref{eq:KbvvAs}, and \eqref{eq:KbuuAs} with
the substitution $t_{b}\rightarrow t_{c}$. Using these expansions,
in the low-field limit $\omega_{h}\ll1$, we derive
\begin{align}
F_{xx}^{\mathrm{HH}}(\omega_{h}) & \simeq\left(r_{m}\!-\!1\right)t_{c}+\frac{r_{m}\!+\!1}{2}\sinh2t_{c}\!\nonumber \\
 & -\frac{\omega_{h}}{2}\left(\sqrt{3}\sinh t_{c}\!+\!\sqrt{r_{m}}\cosh t_{c}\right)^{2}.\label{eq:SxxLowH}
\end{align}
In particular, the zero-field value can be directly related with the
mass ratio as
\begin{align*}
F_{xx}^{\mathrm{HH}}(0) & =\frac{\sqrt{3r_{m}}\left(1\!+\!r_{m}\right)}{3r_{m}-1}+\!\left(r_{m}\!-\!1\right)\tanh^{-1}\!\left(\frac{1}{\sqrt{3r_{m}}}\right)
\end{align*}
giving the zero-field HH partial conductivity
\[
\sigma_{xx}^{\mathrm{HH}}(0)\!=\!\frac{3e^{2}\tau F_{xx}^{\mathrm{HH}}(0)}{4\pi^{3}\hbar^{2}\sqrt{m_{u}m_{v}}}\int\!dk_{z}p_{u0}^{2}.
\]
As expected, at low fields the conductivity has linear magnetic-field
dependence caused by sharp corners. When CDW gaps are taken
into account, the behavior at very small magnetic fields becomes quadratic
and then crosses over to a linear dependence at the field scale set
by the CDW gap \citep{Koshelev2013}. Deviation from linearity starts
at the field $\omega_{h}\sim2t_{c}$ or $H\sim2c\sqrt{m_{u}m_{v}}t_{c}/e\tau$,
at which the time to pass one HH segment becomes comparable with the
scattering time.

In the high-field limit $\omega_{h}\gg1$, we obtain
\begin{equation}
F_{xx}^{\mathrm{HH}}(\omega_{h})\simeq\frac{1}{\omega_{h}^{2}}\left[-\left(r_{m}\!-\!1\right)t_{c}
+\!\frac{r_{m}\!+\!1}{2}\sinh2t_{c}\right],
\label{eq:SxxHighH}
\end{equation}
meaning that the HH contribution decays $\propto H^{-2}$ at high fields. 
\begin{figure}
\includegraphics[width=3.4in]{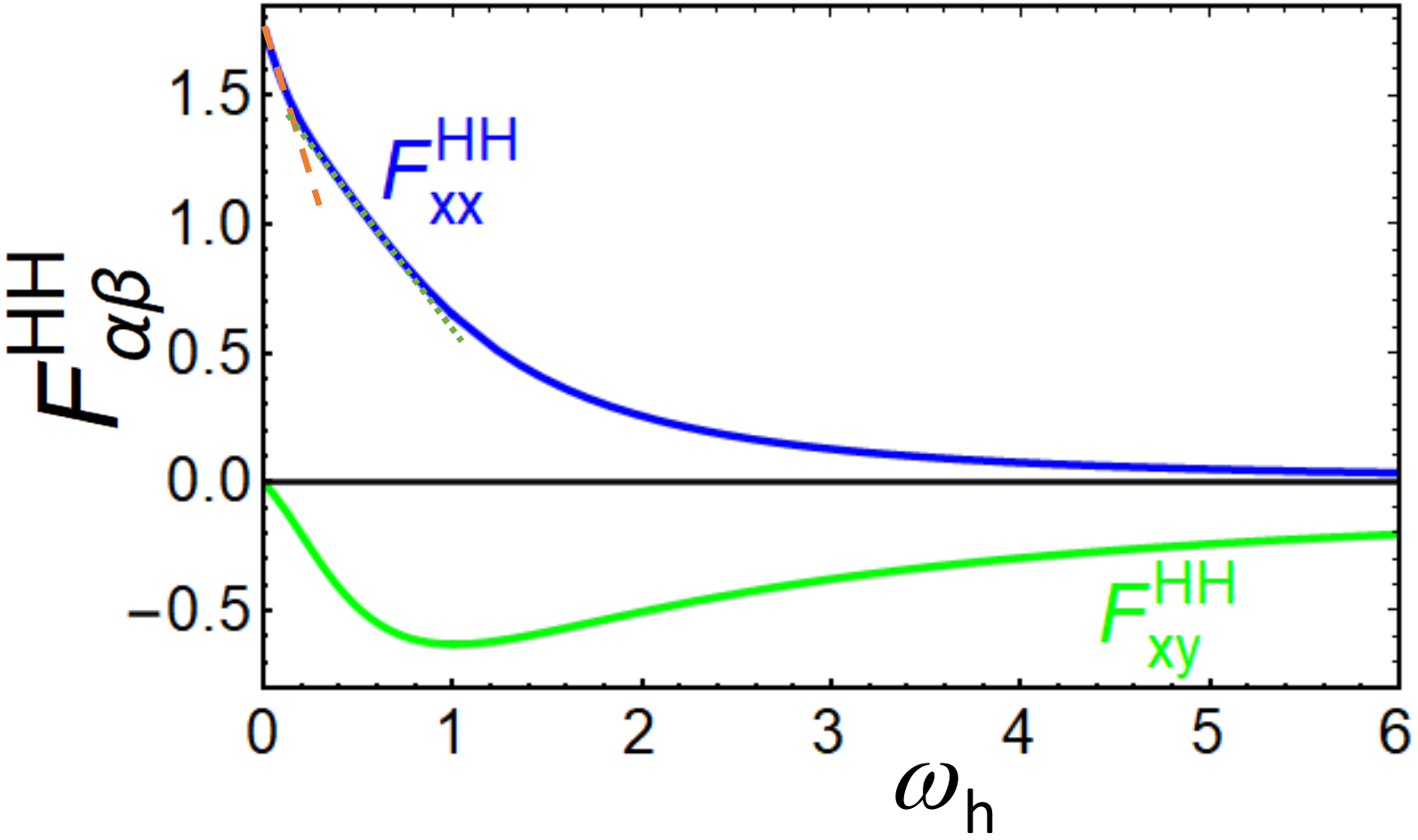}
\caption{The field dependences of reduced functions, $F_{xx}^{\mathrm{HH}}$
and $F_{xy}^{\mathrm{HH}}$, following from Eqs.~\eqref{eq:FHHxx}
and \eqref{eq:FHHxy} computed using $r_{m}\!=\!1.5$. These functions
determine shapes of the HH contributions to the diagonal and Hall
conductivities. The orange dashed and green dotted lines mark initial linear drop and further quasi-linear decrease of $F_{xx}^{\mathrm{HH}}$. }
\label{Fig:FHHab-H}
\end{figure}
The full shape of the field dependence of $F_{xx}^{\mathrm{HH}}$
is illustrated in Fig.~\ref{Fig:FHHab-H} for $r_{m}=1.5$. We
see that the initial linear dependence breaks down at field $\omega_{h}\sim0.3$
and is followed by the second region of close-to-linear dependence
with smaller slope in the range $0.4\lesssim\omega_{h}\lesssim1$.
At higher fields, the function $F_{xx}^{\mathrm{HH}}(\omega_{h})$
approaches the asymptotics in Eq.~\eqref{eq:SxxHighH}.

\begin{figure}[ht]
	\includegraphics[width=3.4in]{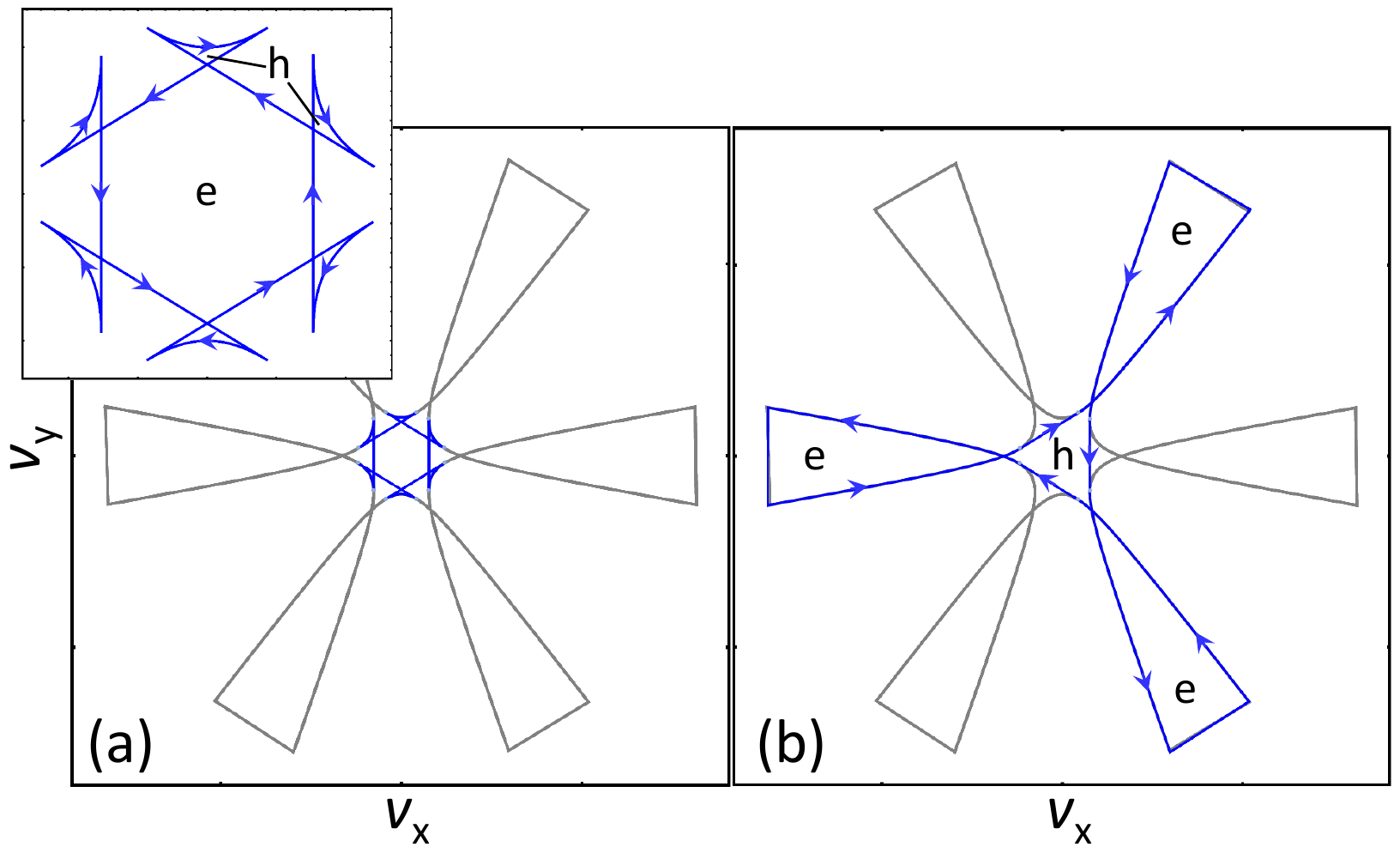}
	\caption{The velocity contours for the reconstructed Fermi pockets for $r_{m}=1.5$
		and $\kappa=20$: (a) hyperbolic hexagon and (b)one of the triangular
		pockets. These contours define velocity areas $A_{v}$ which determines
		the linear Hall term at small magnetic fields according to Eq.~\eqref{eq:SxyLinArea}.
		The main plot in panel (a) shows the HH velocity contour with respect
		to the large hexagon while the inset shows this contour in larger
		scale. }
	\label{Fig:VelContHH-T}
\end{figure}

The HH contribution to the Hall conductivity is given by $\sigma_{xy}^{\mathrm{HH}}=\sigma_{0}^{\mathrm{HH}}F_{xy}^{\mathrm{HH}}(\omega_{h})$,
where the scale $\sigma_{0}^{\mathrm{HH}}$ is defined in Eq.~\eqref{eq:sigma0}.
In the low-field limit, using the asymptotics $\mathcal{K}_{cvu}\simeq-\omega_{h}^{2}\left(t_{c}-\omega_{h}\right),\,\mathrm{for}\,\omega_{h}\rightarrow0$,
we obtain
\begin{align}
F_{xy}^{\mathrm{HH}} & \simeq\omega_{h}\left[2\sqrt{r_{m}}t_{c}\!-\frac{\sqrt{3}}{2}\left(r_{m}+1\right)\right].\label{eq:SyxSmallH}
\end{align}
 The linear Hall coefficient is negative corresponding to electron
pocket. The slice contribution to the linear Hall conductivity is
given by 
\begin{equation}
S_{xy}^{\mathrm{HH}}  =6\frac{|e|H}{c}\tau^{2}v_{u0}v_{v0}\left[t_{c}\!-\frac{\sqrt{3}}{4}\frac{r_{m}\!+\!1}{\sqrt{r_{m}}}\right].
\end{equation}
 The velocity contour determining the sign and magnitude of the linear
Hall term in Eq.~\eqref{eq:SxyLinArea} is shown in the inset of
Fig.~\ref{Fig:VelContHH-T}(a). We can see that the HH velocity contour
is composed of small and smooth hyperbolic branches connected by relatively
long straight lines corresponding to velocity jumps at the corners.
An interesting observation is that the presence of velocity discontinuities
on the Fermi surface does not invalidate the construction and Eq.~\eqref{eq:SxyLinArea}
remains valid. The dominating contribution to the velocity area $A_{v}$
is coming from the regular electronic hexagon in the center.

In the high-field limit, using the asymptotics $\mathcal{K}_{cvu}\simeq-(1/2)\left[\sinh(2t_{c})-2t_{c}\right],\,\mathrm{for}\,\omega_{h}\gg1$,
we find
\begin{equation}
F_{xy}^{\mathrm{HH}} \simeq-\frac{2\sqrt{r_{m}}}{\omega_{h}}t_{c}\label{eq:SyxHighH}
\end{equation}
This corresponds to universal relation between the conductivity slice
and cross section area $A_{\mathrm{HH}}$ in Eq.~\eqref{eq:HexArea},
$S_{xy}^{\mathrm{HH}}\!=-\!A_{\mathrm{HH}}/(|e|H/c)$, which is similar
to Eq.~\eqref{eq:SxyLargeH}. Correspondingly, $\sigma_{xy}^{\mathrm{HH}}$
is connected with quasiparticle density inside the pocket by a simple
relation\citep{AbrikosovBook}
\begin{equation}
\sigma_{xy}^{\mathrm{HH}}=-\frac{e^{2}}{\pi\hbar}\frac{\Phi_{0}n_{\mathrm{HH}}}{H}.\label{eq:sigmaHHxyHighH}
\end{equation}
The full field dependence of $F_{xy}^{\mathrm{HH}}$ is illustrated
in Fig.~\ref{Fig:FHHab-H} for $r_{m}\!=\!1.5$. This function remains
negative for the whole field range, corresponding to an electron pocket.
Its absolute value reaches maximum $|F_{xy}^{\mathrm{HH}}|\approx0.64$
at $\omega_{h}\approx1$ and it exceeds $F_{xx}^{\mathrm{HH}}$ roughly
at the same field. 

\subsection{Triangular pockets}

In this section, we evaluate conductivity slices for the triangular
pockets described in Sec.~\ref{sec:FoldCDW} and illustrated in Figs.~\ref{Fig:HypHexagon}(c)
and \ref{Fig:TriangPocketAxes}. Using the general results for $m$-fold
symmetric 2D slices in Eqs.~\eqref{eq:mfoldSxx} and \eqref{eq:mfoldSxy},
we can write the conductivity slices for a triangular pocket, $m=3$, as
\begin{subequations}
\begin{flalign}
&S_{xx}^{\mathrm{T}} \!=\frac{3}{2}\frac{c}{|e|H}\!
\left\{\! \mathcal{G}_{tt}^{\mathrm{T}}\!+\!\mathcal{G}_{ll}^{\mathrm{T}}
\!-\!\mathrm{Re}\!\left[\frac{\left(\mathcal{R}_{t}^{\mathrm{T}}\!+\!\imath\mathcal{R}_{l}^{\mathrm{T}}\right)^{2}}{\exp\!\left(-\imath\frac{2\pi}{3}\right)\!-\!\eta_{T}}\right]\!
\right\},
\label{eq:SxxTri}\\
&S_{xy}^{\mathrm{T}} \!=\!-\!\frac{3}{2}\frac{c}{|e|H}\!
\left\{\! \mathcal{G}_{lt}^{\mathrm{T}}\!-\!\mathcal{G}_{tl}^{\mathrm{T}}
\!+\!\mathrm{Im}\!\left[\!\frac{\left(\mathcal{R}_{t}^{\mathrm{T}}\!+\!\imath\mathcal{R}_{l}^{T}\right)^{2}}
{\exp\!\left(-\imath\frac{2\pi}{3}\right)\!-\!\eta_{T}}\right]\!
\right\},
\label{eq:SxyTri}
\end{flalign}
\end{subequations}
where the parameter $\eta_{T}\!=\!\exp\left(\!-\mathcal{Q}_{\mathrm{T}}\right)$
and segment integrals $\mathcal{R}_{s}^{\mathrm{T}}$ and $\mathcal{G}_{sr}^{\mathrm{T}}$
are defined similar to Eqs.~\eqref{eq:mfoldQdef-1}, \eqref{eq:mfoldRsDef-1},
and \eqref{eq:mfoldGsrDef}. Contrary to the hexagon pockets, the indices
$s,r\!=\!l,t$ corresponding to the longitudinal and transverse components
of velocity do not coincide with the hyperbolic indices $u$ and $v$,
see Fig.~\ref{Fig:TriangPocketAxes}. The calculations described
in Appendix \ref{app:TrianPockets} lead to the following presentation
for the conductivity slices for one triangular pocket,
\begin{subequations}	
\begin{align}
S_{\alpha\beta}^{\mathrm{T}} & =\frac{S_{0}}{2}F_{\alpha\beta}^{\mathrm{T}},\label{eq:SabTri}\\
F_{xx}^{\mathrm{T}} & =\frac{1}{\omega_{h}}\Bigg\{2\mathcal{K}_{vv}(t_{b},t_{c})\!+\!2r_{m}\mathcal{K}_{uu}(t_{b},t_{c})\nonumber \\
- & \!\left.\mathrm{Re}\!\left[\!\frac{\exp\left(-\imath\frac{\pi}{3}\right)\left(\mathcal{G}_{bc}^{2}\!+\!\mathcal{G}_{cb}^{2}\right)\!+\!2\sqrt{\eta_{T}}\mathcal{G}_{bc}\mathcal{G}_{cb}}{\exp\left(-\imath\frac{2\pi}{3}\right)-\eta_{T}}\right]\right\} ,\label{eq:SxxTriResult}\\
F_{xy}^{\mathrm{T}} & =-\frac{1}{\omega_{h}}\Bigg\{2\sqrt{r_{m}}\left[\mathcal{K}_{vu}(t_{b},t_{c})\!+\!\mathcal{K}_{vu}(-t_{c},-t_{b})\right]\nonumber \\
+ & \!\left.\mathrm{Im}\!\left[\frac{\exp\left(-\imath\frac{\pi}{3}\right)\left(\mathcal{G}_{bc}^{2}\!+\!\mathcal{G}_{cb}^{2}\right)\!+\!2\sqrt{\eta_{T}}\mathcal{G}_{bc}\mathcal{G}_{cb}}{\exp\left(-\imath\frac{2\pi}{3}\right)\!-\!\eta_{T}}\right]\right\} .\label{eq:SxyTriResult}
\end{align}
\end{subequations}
Here the hyperbolic limits $t_{b}$ and $t_{c}$
are defined in Eqs.~\eqref{eq:tb} and \eqref{eq:tc-rm-def}, respectively,
see also Table \ref{Tbl:param},
\begin{equation}
\eta_{T}=\exp\left[-\frac{2(t_{b}\!-\!t_{c})}{\omega_{h}}\right],\label{eq:nuT}
\end{equation}
$\mathcal{G}_{bc}$ and $\mathcal{G}_{cb}$ are the complex functions
defined as 
\begin{equation}
\mathcal{G}_{bc}\!=G_{bcv}\!-\!\imath\sqrt{r_{m}}G_{bcu},\ \mathcal{G}_{cb}\!=G_{cbv}\!-\!\imath\sqrt{r_{m}}G_{cbu},
\label{eq:ComplexGbcGcb}
\end{equation}
where the functions $G_{bck}$ and $G_{cbk}$ with $k=u,v$ are given by
\begin{equation}
	G_{bck}\!=\!G_{k}(t_{b},t_{c}),\ G_{cbk}\!=\!G_{k}(-t_{c},-t_{b}),\label{eq:GbckGcbkDef}
\end{equation}
with
\begin{widetext}
\begin{subequations}
\begin{align}
G_{v}(t_{2},t_{1})= & \frac{\left(\cosh t_{2}\!+\frac{1}{\omega_{h}}\sinh t_{2}\right)\exp\left(\!-\frac{t_{2}-t_{1}}{\omega_{h}}\right)-\!\left(\cosh t_{1}\!+\frac{1}{\omega_{h}}\sinh t_{1}\right)}{1-\omega_{h}^{-2}},\label{eq:Gvt2t1Result}\\
G_{u}(t_{2},t_{1})= & \frac{\left(\sinh t_{2}\!+\frac{1}{\omega_{h}}\cosh t_{2}\right)\exp\left(\!-\frac{t_{2}-t_{1}}{\omega_{h}}\right)-\!\left(\sinh t_{1}\!+\frac{1}{\omega_{h}}\cosh t_{1}\right)}{1-\omega_{h}^{-2}},\label{eq:Gut1t2Result}
\end{align}
\end{subequations}
see Appendix \ref{subsec:Integration-hyperbolicLH} for details.
The functions $\mathcal{K}_{ss}(t_{b},t_{c})$ in the first line of
Eq.~\eqref{eq:SxxTriResult} determining the same-segment contributions
to $S_{xx}^{\mathrm{T}}$ are evaluated in Appendix \ref{app:TrianPockets}
as
\begin{subequations}
\begin{align}
\mathcal{K}_{vv}(t_{b},t_{c})= & \frac{1}{1\!-\!\omega_{h}^{-2}}\left[-\left(\cosh t_{b}\!+\!\frac{1}{\omega_{h}}\sinh t_{b}\right)G_{cbv}\!-\!\frac{\cosh\left(2t_{b}\right)\!-\!\cosh\left(2t_{c}\right)}{4}\right.\nonumber \\
 & \left.-\frac{\sinh\left(2t_{b}\right)\!-\!\sinh\left(2t_{c}\right)\!-\!2\left(t_{b}\!-\!t_{c}\right)}{4\omega_{h}}\right],\label{eq:Kvvtbtc}\\
\mathcal{K}_{uu}(t_{b},t_{c})= & \frac{1}{1\!-\!\omega_{h}^{-2}}\left[\left(\sinh t_{b}\!+\!\frac{1}{\omega_{h}}\cosh t_{b}\right)G_{cbu}\!-\!\frac{\cosh\left(2t_{b}\right)\!-\!\cosh\left(2t_{c}\right)}{4}\right.\nonumber \\
 & \!\left.-\frac{\sinh\left(2t_{b}\right)\!-\!\sinh\left(2t_{c}\right)\!+\!2\left(t_{b}\!-\!t_{c}\right)}{4\omega_{h}}\right],\label{eq:Kuutbtc}
\end{align}
\end{subequations} and the off-diagonal function $\mathcal{K}_{vu}(t_{b},t_{c})$
in the first line of the Hall term in Eq.~\eqref{eq:SxyTriResult}
is given by
\begin{align}
\mathcal{K}_{vu}(t_{b},t_{c}) & =\frac{1}{1\!-\!\omega_{h}^{-2}}\left[\frac{\frac{1}{\omega_{h}}\!-\!\exp\left(\!-\frac{t_{b}-t_{c}}{\omega_{h}}\right)\left(\sinh t_{b}\!+\frac{1}{\omega_{h}}\cosh t_{b}\right)\left(\cosh t_{c}\!-\frac{1}{\omega_{h}}\sinh t_{c}\right)}{1\!-\!\omega_{h}^{-2}}\right.\nonumber \\
+ & \!\left.\frac{\sinh\left(2t_{b}\right)\!+\!\sinh\left(2t_{c}\right)}{4}\!+\frac{t_{b}\!-\!t_{c}}{2}\!-\frac{\cosh\left(2t_{b}\right)\!-\!\cosh\left(2t_{c}\right)}{4\omega_{h}}\right].\label{eq:Kvutbtc}
\end{align}
\end{widetext}
Two features strongly influence the shape of magnetoconductivity components.
The first feature are sharp corners related to FS reconstruction.
The second feature is related to proximity of these corners to van
Hove singularities. As a consequence, the Fermi velocity reduces on
approaching the corners. In fact, the velocity at the corner is only
slightly larger than the minimum velocity of the HH pocket $v_{u0}$.
The consequences of this velocity reduction are similar to those for
the large hexagon, discussed in Sec.~\ref{sec:LHsigmaH}. For
example, the Hall conductivity for the triangular pocket is negative
at small magnetic fields in spite of its hole nature but it has strong
field dependence and changes sign. 

Let us discuss now the asymptotic behavior of the conductivity components
for the triangular pockets. We start with the diagonal component.
For small fields, $\omega_{h}\ll1$, using asymptotics of the functions
$G_{bck}$, $G_{cbk}$, $\mathcal{K}_{kk}(t_{b},t_{c})$ listed in
Appendix \ref{app:TrianPockets}, Eqs.~\eqref{eq:GbckSmallH}, \eqref{eq:GcbkSmallH},\eqref{eq:KvvbcSmallH},
and \eqref{eq:KuubcSmallH}, we obtain the zero-field result
\begin{align}
F_{xx}^{\mathrm{T}}(0) & =\!\frac{1}{2}\big\{ \left(r_{m}\!+\!1\right)\left[\sinh\left(2t_{b}\right)\!-\!\sinh\left(2t_{c}\right)\right]\nonumber\\
&\!+2\left(r_{m}\!-\!1\right)\left(t_{b}\!-\!t_{c}\right)\big\} 
\label{eq:FxxTriZeroH}
\end{align}
and the linear term
\begin{align}
F_{xx}^{\mathrm{T}}(\omega_{h})\!-\!F_{xx}^{\mathrm{T}}(0) & \simeq-\frac{\omega_{h}}{2}\Bigg[\left(\sqrt{3}\sinh t_{c}\!+\!\sqrt{r_{m}}\cosh t_{c}\right)^{2}\nonumber\\
+&\left(\sqrt{3}\sinh t_{b}\!-\!\sqrt{r_{m}}\cosh t_{b}\right)^{2}\Bigg].\label{eq:FxxTriLinH}
\end{align}
Here the first term is caused by the sharp corners of triangles. 
It is identical with the linear term for the hyperbolic hexagon in Eq.\ \eqref{eq:SxxLowH}. 
The second term appears due to velocity jumps at the matching point between
incoming and outgoing branches and is artifact of the model approximations.

As in the case of the large hexagon, the field dependence of the triangular
pockets is characterized by the two field scales corresponding to
$\omega_{h}\sim1$ and $\omega_{h}\sim t_{b}-t_{c}$. The upper scale
is somewhat smaller than the one for the large hexagon. In the limit
$\omega_{h}\gg1$, using high-field limits in Eqs.~\eqref{eq:GbckHighH},\eqref{eq:GcbkHighH},
\eqref{eq:KvvbcHighH}, and \eqref{eq:KuubcHighH}, we obtain a remarkably
simple approximate result for the diagonal term, similar to Eq.~\eqref{eq:SxxLargeHexHighH},
\begin{align}
F_{xx}^{\mathrm{T}} & \simeq\frac{\kappa^{2}}{2\omega_{h}}\frac{1-\eta_{T}^{2}}{1+\eta_{T}+\eta_{T}^{2}},\label{eq:FxxTIntAsymp}
\end{align}
which also describes crossover between $1/H$ and $1/H^{2}$ decays
of diagonal conductivity at $\omega_{h}\sim2(t_{b}\!-\!t_{c})$. At
highest fields, for $\omega_{h}\gg2(t_{b}\!-\!t_{c})$, we have $F_{xx}^{\mathrm{T}}\simeq2\kappa^{2}(t_{b}\!-\!t_{c})/(3\omega_{h}^{2})$.
This is roughly three times smaller than the corresponding asymptotic limit
for the large hexagon. 

Let us consider the asymptotic limits of the Hall component $F_{xy}^{\mathrm{T}}$
in Eq.~\eqref{eq:SxyTriResult}. Using asymptotic limits of the functions
$G_{bck}$, $G_{cbk}$, $\mathcal{K}_{uv}(t_{b},t_{c})$ for $\omega_{h}\ll1$
presented in Eqs.~\eqref{eq:GbckSmallH}, \eqref{eq:GcbkSmallH}, and \eqref{eq:KvubcSmallH}, we obtain the linear term at small
fields
\begin{align}
	&F_{xy}^{\mathrm{T}}=  -\omega_{h}\bigg[\frac{1}{2}\left(\sqrt{3}\sinh t_{b}\!-\!\sqrt{r_{m}}\cosh t_{b}\right) \label{eq:FxyTriSmall}\\
	&\times\!\left(\sinh t_{b}\!+\!\sqrt{3r_{m}}\cosh t_{b}\right)
	\!-\!2\sqrt{r_{m}}\left(t_{b}\!-\!t_{c}\right)\!-\frac{\sqrt{3}}{2}\left(r_{m}\!+\!1\right)\!\bigg].\nonumber
\end{align}
As in the case of the large hexagon pocket, Eq.~\eqref{eq:SxyLHLinH}, the sign
of this term is \emph{negative} even though the triangles are nominally
hole-type pockets. The reason for this behavior is also very similar
to the case of large hexagon. The triangular pockets have regions
with both positive and negative curvature. In this context, the sharp
corners should be considered as the regions with extreme positive
curvature. However, these corners are located very close to van Hove
points where velocities are small and, as a consequence, they give
small contribution to the Hall term. The dominating negative contribution
is coming from far away regions with negative curvature. This is illustrated
by the velocity contour shown in Fig.~\ref{Fig:VelContHH-T}(b).
Velocity jumps at the corners from the small triangle at the center. The
area of this triangle contributes to the total velocity area $A_{v}$
in Eq.~\eqref{eq:Av} with positive sign. The larger areas of three
outside ``blades'' contribute to $A_{v}$ with negative sign yielding
the net negative velocity contour area.

At higher field, $\omega_{h}\gg1$, using Eqs.~\eqref{eq:GbckHighH},\eqref{eq:GcbkHighH},
and \eqref{eq:KvubcHighH}, we derive the following approximate result
\begin{widetext}
\begin{align}
F_{xy}^{\mathrm{T}} & \simeq\!-\frac{1}{\omega_{h}}\left[\frac{\kappa}{2}\left(\sqrt{3}\cosh t_{b}\!-\!\sqrt{r_{m}}\sinh t_{b}\!-\frac{\sqrt{3}\eta_{T}\kappa}{1\!+\!\eta_{T}\!+\!\eta_{T}^{2}}\right)\!+\!2\sqrt{r_{m}}\left(t_{b}\!-\!t_{c}\right)\right]\label{eq:FxyTriInt}\\
 & \simeq\!-\frac{\kappa^{2}}{2\omega_{h}}\left(\frac{\sqrt{3}\!-\!\sqrt{r_{m}}}{1+\sqrt{3r_{m}}}\!-\frac{\sqrt{3}\eta_{T}}{1\!+\!\eta_{T}\!+\!\eta_{T}^{2}}\right).\nonumber 
\end{align}
It describes the crossover between the two $1/\omega_{h}$ dependences
\begin{align}
F_{xy}^{\mathrm{T}} & \!\simeq\!\frac{\kappa^{2}}{2(\sqrt{3r_{m}}\!+\!1)\omega_{h}}\!\times\!
\begin{cases}
-\left(\sqrt{3}\!-\!\sqrt{r_{m}}\right),\,\mathrm{for}\,1\!\ll\!\omega_{h}\!\ll\!2(t_{b}\!-\!t_{c})\\
\frac{2\sqrt{3}}{3}\left(\sqrt{3r_{m}}\!-\!1\right),\,\mathrm{for}\,\omega_{h}\!\gg\!2(t_{b}\!-\!t_{c})
\end{cases},
\label{eq:FxyTriAsymp}
\end{align}
\end{widetext}
which have opposite signs, similar to the large-hexagon
pocket in Eq.~\eqref{eq:FLHxyAsymp}. The large-field asymptotic
reproduces a universal relation between the Hall conductivity slice
and the triangle area $A_{\mathrm{T}}$ in Eq.~\eqref{eq:TriArea},
$S_{xy}^{\mathrm{T}}=\frac{A_{\mathrm{T}}}{|e|H/c}.$ Since the total
area of two triangles is more than two times smaller than the area
of the large hexagon, the Hall conductivity at high fields is also smaller
by the same factor. 

\section{Modification of magnetoconductivity by Fermi-surface reconstruction
due to CDW order \label{sec:ModifMCReconstr}}

As we have now the full analytical results for the conductivity slices for
all pockets in the reconstructed Fermi surface, we can analyze them and
compare with the behavior of the original large-hexagon pocket. The
total magnetoconductivity slice of the reconstructed Fermi surface
composed of hyperbolic hexagon and two triangular pockets is given
by
\begin{subequations}
\begin{align}
S_{\alpha\beta}^{\mathrm{R}}(\omega_{h}) & =S_{\alpha\beta}^{\mathrm{HH}}(\omega_{h})+2S_{\alpha\beta}^{\mathrm{T}}(\omega_{h})=S_{0}F_{\alpha\beta}^{\mathrm{R}}(\omega_{h}),\label{eq:SRab}\\
F_{\alpha\beta}^{\mathrm{R}}(\omega_{h}) & =F_{\alpha\beta}^{\mathrm{HH}}(\omega_{h})\!+\!F_{\alpha\beta}^{\mathrm{T}}(\omega_{h}).\label{eq:FRab}
\end{align}
\end{subequations}
Since we neglected corner rounding by the CDW order,
the total zero-field conductivity for the reconstructed FS, 
\begin{equation}
F_{xx}^{\mathrm{R}}(0)\!=\!F_{xx}^{\mathrm{HH}}(0)\!+\!F_{xx}^{\mathrm{T}}(0)\!=\!\frac{r_{m}\!+\!1}{2}\sinh\left(2t_{b}\right)\!+\!\left(r_{m}\!-\!1\right)t_{b}\label{eq:FRxx0}
\end{equation}
coincides with zero-field conductivity for the large hexagon pocket,
see Eq.~\eqref{eq:FLHxxSmallH}. In reality, there is a small correction caused by the Fermi-surface reconstruction by finite CDW gap. The total linear term in the conductivity slice is given by 
\begin{align}
F_{xx}^{\mathrm{R}}(\omega_{h})\!-\!F_{xx}^{\mathrm{R}}(0) & \simeq-\omega_{h}\Bigg[\left(\sqrt{3}\sinh t_{c}\!+\!\sqrt{r_{m}}\cosh t_{c}\right)^{2}\nonumber \\
 +&\frac{1}{2}\left(\sqrt{3}\sinh t_{b}\!-\!\sqrt{r_{m}}\cosh t_{b}\right)^{2}\Bigg].\label{eq:FxxRLinH}
\end{align}
Here the first term in the square brackets is caused by the sharp corners
of the reconstructed FS pockets. The HH pocket and the two triangles give
equal contributions to this term in spite of the overall small contribution
of the HH pocket. Note that this term only depends on the mass ratio $r_{m}$ and, using the definition of $t_{c}$ in Eq.\  \eqref{eq:tc-rm-def}, can be computed as $\left(\sqrt{3}\sinh t_{c}\!+\!\sqrt{r_{m}}\cosh t_{c}\right)^{2}=3\left(1\!+\!r_{m}\right)^{2}/\left(3r_{m}\!-\!1\right)$. 
The second term is identical to the large-hexagon pocket
in Eq.~\eqref{eq:FLHxxSmallH} and appears due to the small velocity
jumps at the matching points between the hyperbolic branches. 

\begin{figure}
\includegraphics[width=3.4in]{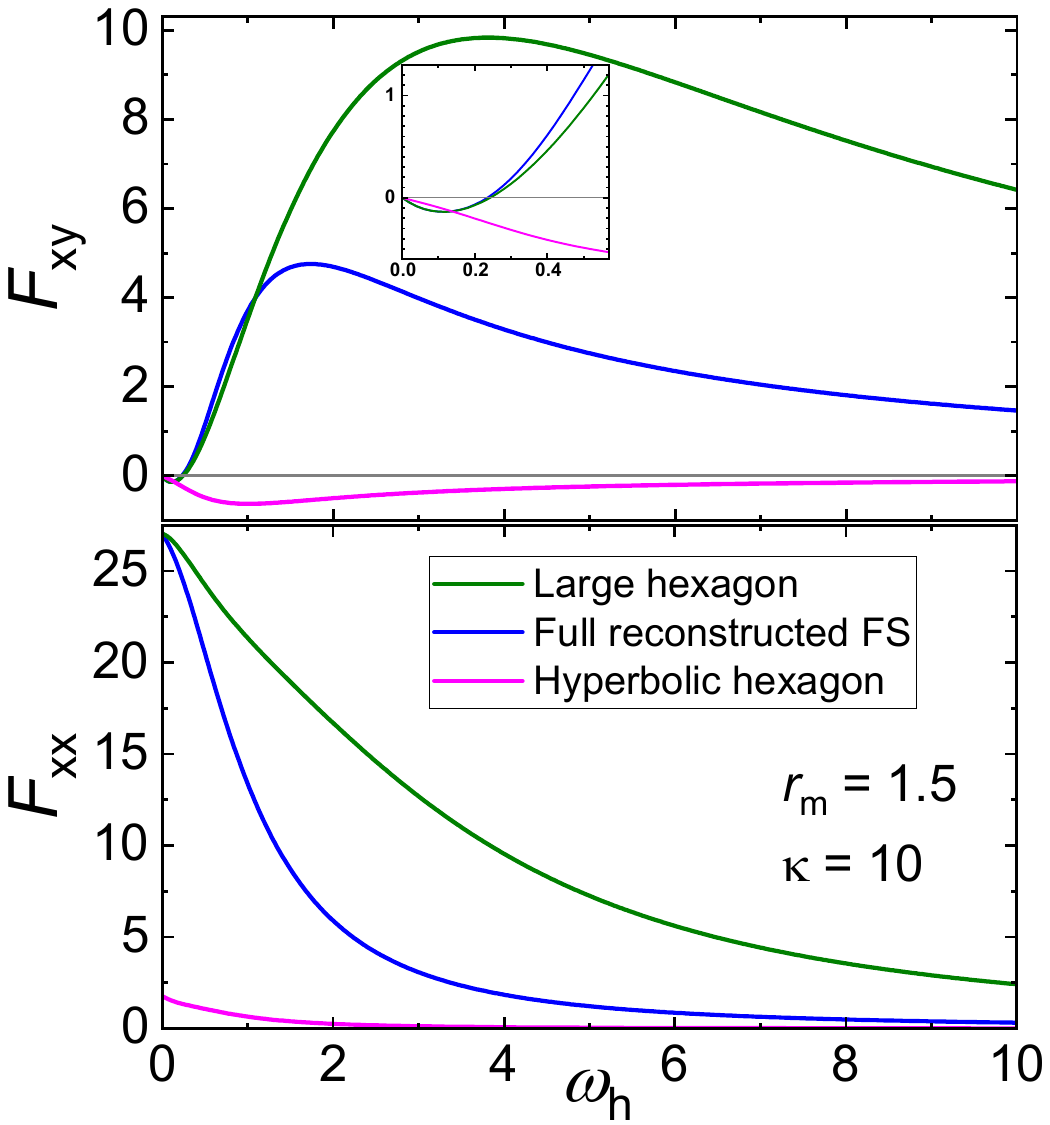}
\caption{The shapes of the field dependences of the reduced functions $F_{\alpha\beta}$
determining the conductivity components for the large hexagon and
reconstructed pockets. The functions are computed for the representative
parameters $r_{m}\! =\! 1.5$ and $\kappa\! =\! 10$. The inset in the upper panel
zooms into the nonmonotonic low-field behavior of $F_{xy}$.}
\label{Fig:FabReconstrComp-rm1_5k10}
\end{figure}

The full field dependence of the function $F_{xx}^{\mathrm{R}}(\omega_{h})$
is compared with the corresponding function $F_{xx}^{\mathrm{LH}}(\omega_{h})$
for representative parameters $r_{m}\!=\!1.5$ and $\kappa\!=\!10$
in the lower panel of Fig.~\ref{Fig:FabReconstrComp-rm1_5k10}. We
can see that $F_{xx}^{\mathrm{R}}(\omega_{h})$ has a sharper drop than
$F_{xx}^{\mathrm{LH}}(\omega_{h})$ with pronounced linear dependence
at small fields. This means that the reconstruction of the Fermi surface by
the 3Q CDW order significantly enhances small-field magnetoconductivity, in agreement
with experiment. 

Let us compare now the behavior of the Hall component for original
and reconstructed Fermi surfaces. The total linear term at small fields
\begin{align}
	F_{xy}^{\mathrm{R}}= & -\omega_{h}
	\Bigg[ \frac{1}{2}\left(\sqrt{3}\sinh t_{b}\!-\!\sqrt{r_{m}}\cosh t_{b}\right)\nonumber\\
	\times&\left(\sinh t_{b}\!+\!\sqrt{3r_{m}}\cosh t_{b}\right)\!-\!2\sqrt{r_{m}}t_{b}\Bigg] \label{eq:FxyRSmall}
\end{align}
is negative and coincides with the linear term for
the large hexagon pocket in Eq.~\eqref{eq:SxyLHLinH}. On the other
hand, at high magnetic fields, $\omega_{h}\!\gg\!2(t_{b}\!-\!t_{c})$,
the dominating contribution to total Hall conductivity is coming from
the triangular pockets, Eq.~\eqref{eq:FxyTriAsymp}. It is positive,
decays as $1/\omega_{h}$, and is more than two times smaller than the
Hall conductivity for large hexagon pocket in Eq.~\eqref{eq:FLHxyAsymp}. 

The full field dependences of the Hall components for original and
reconstructed Fermi surfaces are illustrated in the upper panel of
Fig.~\ref{Fig:FabReconstrComp-rm1_5k10}. We see that the shape of
the field dependence remains anomalous after the FS reconstruction.
Namely, the function $F_{xy}^{\mathrm{R}}(\omega_{h})$ is still nonmonotonic
and changes sign. In fact, the reconstruction has a very weak influence
on the low-field behavior. The most essential difference is that the
conductivity values at the maximum and above become significantly smaller. We also
note that even though the hyperbolic hexagon has an overall small
contribution, it strongly affects the linear slope at small fields,
see the inset in the upper panel of Fig.~\ref{Fig:FabReconstrComp-rm1_5k10}.

To avoid a possible confusion, we note that Fig.~\ref{Fig:FabReconstrComp-rm1_5k10} compares the behavior of the conductivity slices for original and reconstructed Fermi surfaces in the reduced form. In reality, these Fermi surfaces are realized in different temperature ranges, above and below the CDW transition temperature. In particular, strong difference in scattering rates leads to strong difference between the scales $S_0$ and $H_0$ in these temperature ranges.   

\section{Role of corner rounding by finite CDW gap in low-magnetic-field behavior \label{sec:CornerRoundCDW}}

So far, we considered the approximation of sharp corners at the branch crossing. A finite CDW gap rounds these corners. If the CDW gap is small in comparison with other energy scales, these roundings only modify the behavior of the magnetoconductivity at low magnetic fields. The corners emerge because band folding by the CDW wave vectors generates crossing quasiparticle branches. The CDW hybridization transforms the quasiparticle spectrum for two crossing branches $\xi_{1}(\boldsymbol{p})$
and $\xi_{2}(\boldsymbol{p})$ as 
\begin{equation}
	E_{\mathbf{p},\pm}=\frac{\xi_{+}}{2}\pm\sqrt{\frac{\xi_{-}^{2}}{4}+\Delta_{\mathrm{CDW}}^{2}},
	\label{Spectrum}
\end{equation}
where $\xi_{\pm}\!=\!\xi_{1}(\boldsymbol{p})\pm\xi_{2,}(\boldsymbol{p})$
and $\Delta_{\mathrm{CDW}}$ is the CDW gap parameter. Here we assume that the
branches cross at $\boldsymbol{p}=\boldsymbol{p}_{c}$, $\xi_{1}(\boldsymbol{p}_{c})\!=\!\xi_{2}(\boldsymbol{p}_{c})$. 
At this point, the gap in the quasiparticle spectrum opens equal to  $\Delta_{\mathrm{CDW}}$. The spectral gap, however, is only finite in a small region of the Fermi surface near $\boldsymbol{p}_{c}$.
The band splitting by CDW order has been directly observed by ARPES \citep{NakayamaPhysRevB.104.L161112,LuoNatComm2022,HuPhysRevB.106.L241106}.
The CDW hybridization rounds the corners at the crossing converting
them to high-curvature turning points. 

With the rounded corners, the asymptotics of magnetoconductivity for
$H\rightarrow0$ is quadratic $\sigma_{xx}(H)-\sigma_{xx}(0)\propto H^{2}$
and crosses over to a linear dependence at the magnetic field scale
proportional to $\Delta_{\mathrm{CDW}}$. The quantitative theoretical
description of this crossover has been elaborated in Ref.~\citep{Koshelev2013}.
The correction to the conductivity slice due to the reconstructed
crossing points can be represented as 
\begin{equation}
	S_{xx}^{(\mathrm{cr})}(H)\!-\!S_{xx}(0)\!=\frac{2n_{\mathrm{cr}}\left\langle v_{-,x}^{2}\right\rangle \tau\Delta_{\mathrm{CDW}}}{\left[\mathbf{v}_{1}\times\mathbf{v}_{2}\right]_{z}}G(H/H_{\Delta}),
	\label{DSxxHCross}
\end{equation}
where $n_{\mathrm{cr}}$ is the total number of crossing points ($6$
in our case), $\mathbf{v}_{1}$ and $\mathbf{v}_{2}$ are the velocities
of the two branches at the crossing, $v_{-,x}=v_{2,x}\!-\!v_{1,x}$
is the jump of the $x$ component of the velocity at the corner for
$\Delta_{\mathrm{CDW}}\!=\!0$ , $\left\langle \ldots\right\rangle $
means averaging over crossings, and
\begin{equation}
	H_{\Delta}=\frac{2c\Delta_{\mathrm{CDW}}}{e\tau\left[\mathbf{v}_{1}\times\mathbf{v}_{2}\right]_{z}},
	\label{eq:MagFieldScale}
\end{equation}
is the magnetic field scale set by the CDW gap $\Delta_{\mathrm{CDW}}$.
The reduced function $G(h)$ is defined by the integral
\begin{align}
	& G(h)=\int_{0}^{\infty}dx\int_{0}^{\infty}dy\exp\left(-y\right)\nonumber \\
	& \times\left(\frac{x^{2}-h^{2}y^{2}/4}{\sqrt{\left(x\!+\!h\frac{y}{2}\right)^{2}\!+\!1}\sqrt{\left(x\!-\!h\frac{y}{2}\right)^{2}\!+\!1}}-\frac{x^{2}}{x^{2}\!+\!1}\right)\label{eq:GhDef}
\end{align}
and has the following asymptotics 
\begin{equation}
	G(h)=\begin{cases}
		-\frac{3\pi}{16}h^{2} & \text{ for }h\ll1\\
		\frac{\pi}{2}-h & \text{ for }h\gg1
	\end{cases}.\label{eq:GhAsymp}
\end{equation}
We find an accurate approximation for this function in the form of
superposition of two hyperbolas
\begin{equation}
	G(h)\approx\frac{\pi}{2}-f\sqrt{a_{1}^{2}+h^{2}}-(1-f)\sqrt{a_{2}^{2}+h^{2}}\label{eq:GhAppr}
\end{equation}
with $f=0.8035$, $a_{1}=0.705$, $a_{2}=5.112$. This form is more
convenient for modeling of the data than the exact integral presentation
in Eq.~\eqref{eq:GhDef}. 

In the region $H\gg H_{\Delta}$, Eqs.~\eqref{DSxxHCross} and \eqref{eq:GhDef} give linear
magnetoconductivity
\begin{equation}
	S_{xx}^{(\mathrm{cr})}(H)\!-\!S_{xx}(0)\approx-n_{\mathrm{cr}}\left\langle v_{-,x}^{2}\right\rangle \frac{eH\tau^{2}}{c}.\label{eq:LinSxxCross}
\end{equation}
For the spectrum considered in this paper, we evaluate 
\begin{equation}
	\left\langle v_{-,x}^{2}\right\rangle =\frac{v_{v0}^{2}}{2}\left(\sqrt{3}\sinh t_{c}\!+\!\sqrt{r_{m}}\cosh t_{c}\right)^{2}.\label{eq:vxjump2Av}
\end{equation}
With this result, we can demonstrate that Eq.~\eqref{eq:LinSxxCross}
reproduces Eq.~\eqref{eq:SRab} when we substitute the first term
in Eq.~\eqref{eq:FxxRLinH} describing the contribution from the
sharp corners. Also, the $z$ component of the velocity cross product
in Eqs.~\eqref{DSxxHCross} and \eqref{eq:MagFieldScale} in our
case can be evaluated as
\begin{align}
	\left[\mathbf{v}_{1}\times\mathbf{v}_{2}\right]_{z}= & \frac{\sqrt{3}}{2}v_{v0}^{2}\frac{3r_{m}^{2}+1}{3r_{m}-1}.\label{eq:velCrossProd}
\end{align}
It is instructive to compare the CDW field scale $H_{\Delta}$ in
Eq.~\eqref{eq:MagFieldScale} with the field scale $H_{0}$ listed
in the Table \ref{Tbl:param}. Using Eq.~\eqref{eq:velCrossProd},
we obtain for the ratio of these two scales
\begin{align}
	\frac{H_{\Delta}}{H_{0}} & =\frac{2\Delta_{\mathrm{CDW}}}{\sqrt{m_{u}m_{v}}\left[\mathbf{v}_{1}\times\mathbf{v}_{2}\right]_{l}}\nonumber \\
	& =C_{m}\frac{\Delta_{\mathrm{CDW}}}{\varepsilon_{\mathrm{vH}}}
	\label{eq:HDCDWH0Ratio}
\end{align}
with $C_{m}\!=\!2\sqrt{r_{m}/3}\left(3r_{m}\!-\!1\right)/\left(3r_{m}^{2}\!+\!1\right)$.
We see that, up to a numerical factor, this ratio is given by the ratio
of the CDW gap and the distance between the van Hove energy and the
Fermi level $\varepsilon_{\mathrm{vH}}$. 

\section{Experimental data and qualitative modeling\label{sec:ModelExpData}}

\begin{figure}
	\includegraphics[width=3.4in]{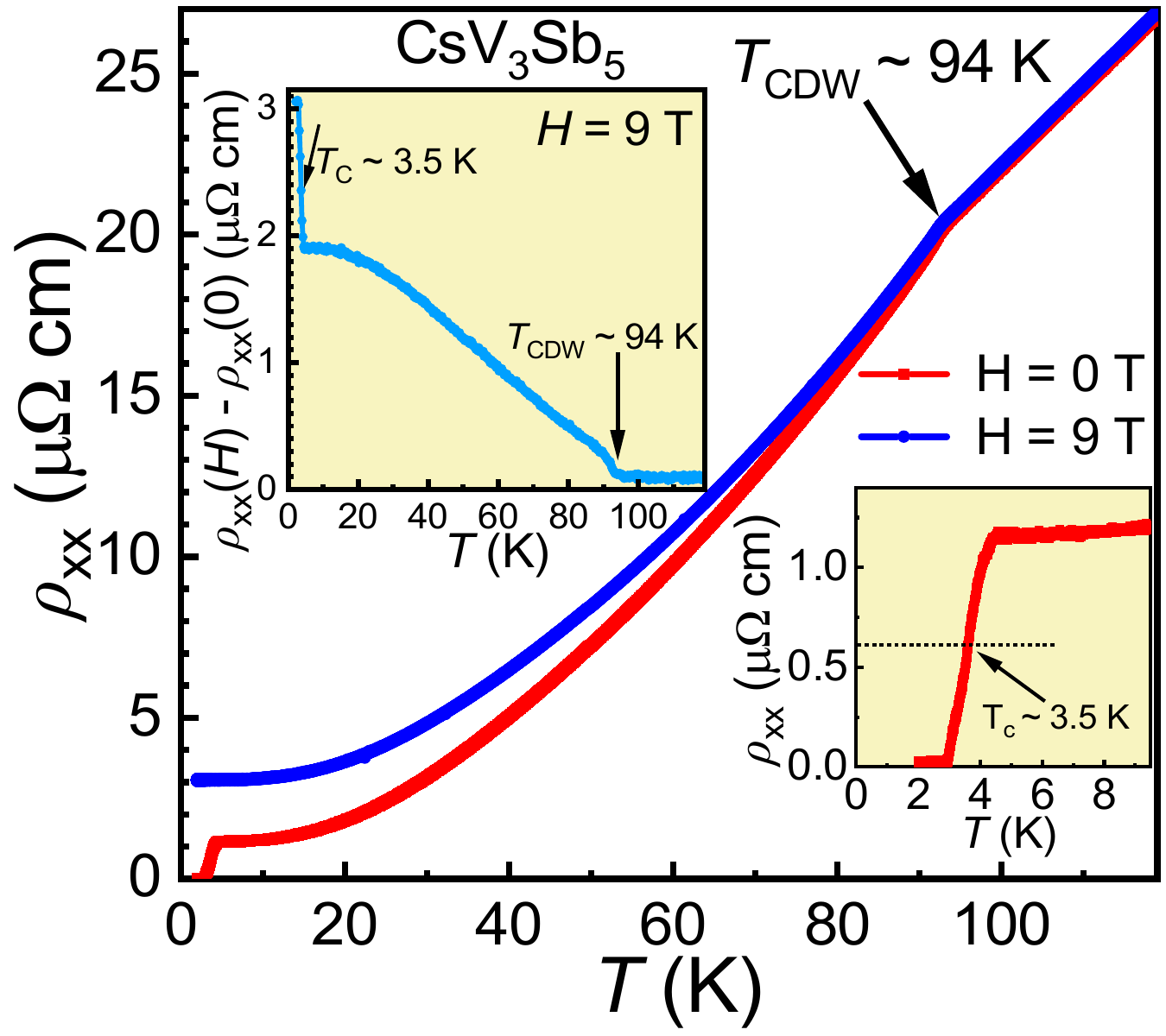}
	\caption{\emph{Main frame}: Temperature dependence of the longitudinal electrical resistivity
		$\rho_{xx}(T)$ of CsV$_{3}$Sb$_{5}$ displaying the CDW transition at $T_{\mathrm{CDW}}\! \approx \! 94$
		K and the superconducting transition at $T_{c}\!\approx \! 3.5$ K (defined as the midpoint of the transition, see the \emph{lower right inset}). Red curve is for $H\!=\!0$ T and blue curve is for $H\!=\!9$ T.
		\emph{Upper left inset}: The difference between zero-field data and data at 9 T presented
		in main frame.}
	\label{Fig:rhoxx-T-fixedHExp}
\end{figure}
\begin{figure*}
	\includegraphics[width=7in]{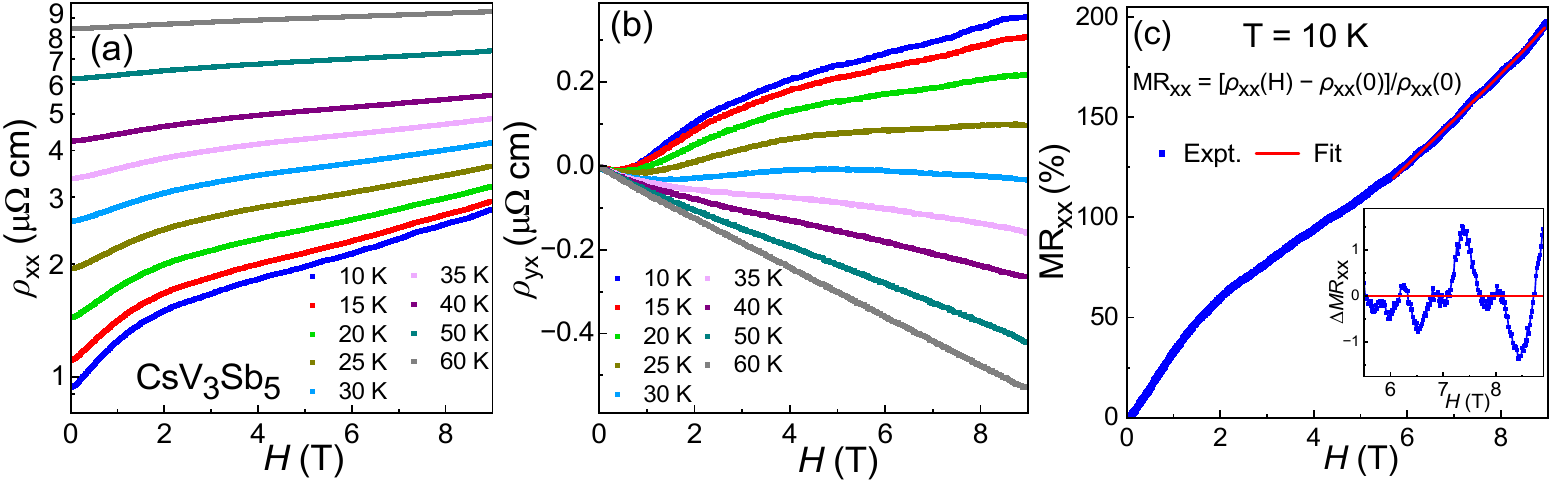}
	\caption{Magnetic-field dependences of the diagonal (a) and Hall resistivity (b) of CsV$_{3}$Sb$_{5}$ at indicated temperatures. These data are used to evaluate the diagonal and Hall
		conductivities presented in Fig.\ \ref{Fig:sab10K} and Fig.\ \ref{Fig:ExpMCondModel2sets}. The panel (c) shows the diagonal magnetoresistance MR$_{xx}$ at 10 K (blue dots). Red curve shows the third-order polynomial fit for the smooth background. Inset: Magnetoresistance above 5 T after background subtraction ($\Delta \mathrm{MR}_{xx}$). In addition to the anomalous magnetoresistance (described in the text), there are clear Shubnikov-de Haas quantum oscillations above 5 T in the background-subtracted data affirming the high quality of the sample.}
	\label{Fig:rhoabHExp}
\end{figure*}
Single crystals of CsV$_{3}$Sb$_{5}$ were grown using the
flux method as described in our earlier works \citep{ShresthaPhysRevB.105.024508,ChapaiPhysRevLett.130.126401}.
Our crystals have been extensively characterized by x-ray diffraction,
energy dispersive spectroscopy, bulk magnetization, magnetic torque,
and tunnel diode oscillator techniques \citep{ShresthaPhysRevB.105.024508,ChapaiPhysRevLett.130.126401}.
In this work, we focus on electrical transport. The electrical resistivity
and magnetotransport measurements were performed using a \emph{dc}-technique
in a Physical Properties Measurement System (DynaCool-PPMS, Quantum
Design) following a standard four-probe method with current applied
along the \emph{ab} plane and field along the \emph{c}-axis. For the
transverse magnetoresistance (\emph{MR}) and Hall resistivity, measurements
were carried out for both positive and negative field directions to
correct for misalignment of electrical contacts. A typical temperature
dependence of the in-plane electrical resistivity $\rho_{xx}(T)$
of CsV$_{3}$Sb$_{5}$ single crystal is shown in Fig.~\ref{Fig:rhoxx-T-fixedHExp}. 
The $\rho_{xx}(T)$ exhibits a metallic behavior and
displays two anomalies, one at $T_{\mathrm{CDW}}\sim 94$ K arising
due to charge density wave ordering and another transition at $T_{c}\!\sim\!3.5$K
corresponding to superconductivity, as presented in the lower right inset of Fig.\ \ref{Fig:rhoxx-T-fixedHExp}
(red curve). The CDW transition is seen as a kink in the $\rho_{xx}(T)$ curve and below the transition the resistivity is reduced. This most likely indicates the reduction of the scattering rate in the ordered state. 
These observations are consistent with previous reports
\citep{FuPhysRevLett.127.207002,HuangPhysRevB.106.064510,YuPhysRevB.104.L041103}. The
samples display a low residual resistivity $\sim\!  1.2\ \mu\Omega$
cm at $\sim\! 4$ K before entering the superconducting state,
reflecting their high quality.

Below $\sim\! 94$ K, there is a considerable \emph{MR}, as shown in Fig.~\ref{Fig:rhoxx-T-fixedHExp} (9 T,
blue curve). In the upper left inset of Fig.\ \ref{Fig:rhoxx-T-fixedHExp}, we plot the
difference between zero field data and the data at 9 T. Interestingly,
the \emph{MR} is 
sharply enhanced below the CDW transition at 94K.
This enhancement can be mostly attributed to the formation of the sharp corners in the reconstructed FS, as discussed in Section \ref{sec:ModifMCReconstr}. Also, \emph{MR} is enhanced due to the reduction of the scattering rate in the ordered state. 

Figure \ref{Fig:rhoabHExp}(a,b) shows the magnetoresistance and Hall resistivity
of CsV$_{3}$Sb$_{5}$ at various temperatures. The magnitude and shape
of magnetoresistance ($\rho_{xx}(H)$) and Hall resistivity ($\rho_{yx}(H)$)
observed in our single crystals is similar to that reported in previous studies \citep{OrtizPhysRevX.11.041030,HuangPhysRevB.106.064510,FuPhysRevLett.127.207002,YuPhysRevB.104.L041103}. In addition to large magnetoresistance $\mathrm{MR}_{xx}\!=\! [\rho_{xx}(H)\! -\! \rho_{xx}(0)]/\rho_{xx}(0)$, we observed clear Shubnikov-de Haas quantum oscillations (QOs) at low temperatures. 
Figure  \ref{Fig:rhoabHExp}(c) displays 
$\mathrm{MR}_{xx}$  
at 10 K, where the QOs are apparent above 5 T as seen more clearly in the background-subtracted data in the inset. 
The observation of QOs at low temperatures indicates a small scattering rate due to impurities in our samples which is also a prerequisite for the emergence of the anomalous features in the magnetotransport we are aiming to understand.

\begin{figure*}
\includegraphics[width=6.4in]{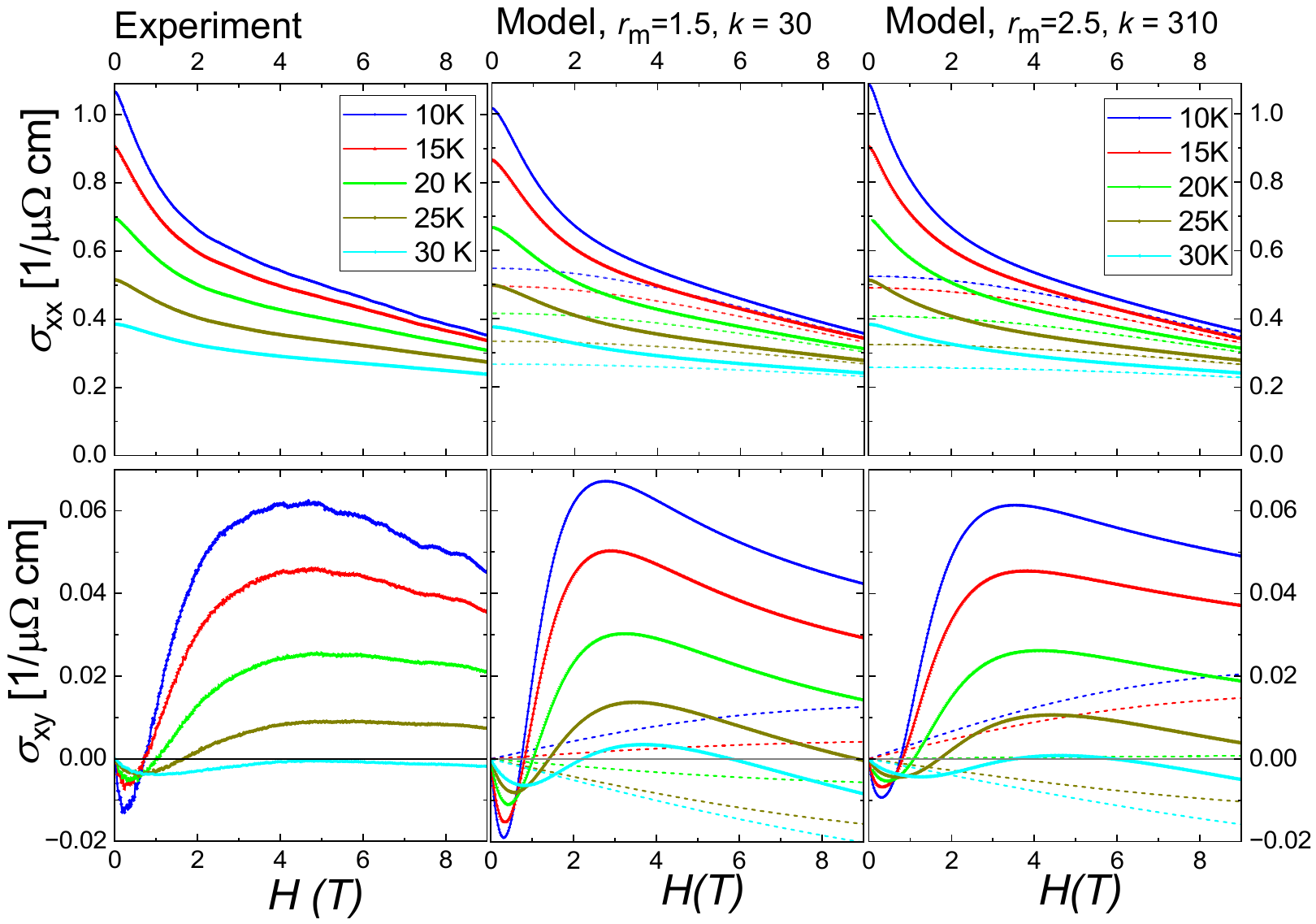}
\caption{\emph{Left column}: the experimental field dependences of the diagonal
and Hall conductivities for CsV$_{3}$Sb$_{5}$ at five different temperatures.
\emph{Middle and right column}s: modeling of these data using Eqs.~\eqref{eq:CondModel},
\eqref{eq:svHab}, \eqref{eq:Sbgxx}, and \eqref{eq:Sbgxy} with two
sets of band parameters in Eq.~\eqref{eq:svHab}, $r_{m}=1.5$ and
$\kappa=30$ (middle column) and $r_{m}=2.5$ and $\kappa=310$ (right
column). The second set looks less reasonable but gives better fits.
The dashed lines show fitted background contributions. }
\label{Fig:ExpMCondModel2sets}
\end{figure*}
\begin{figure*}
\includegraphics[width=6in]{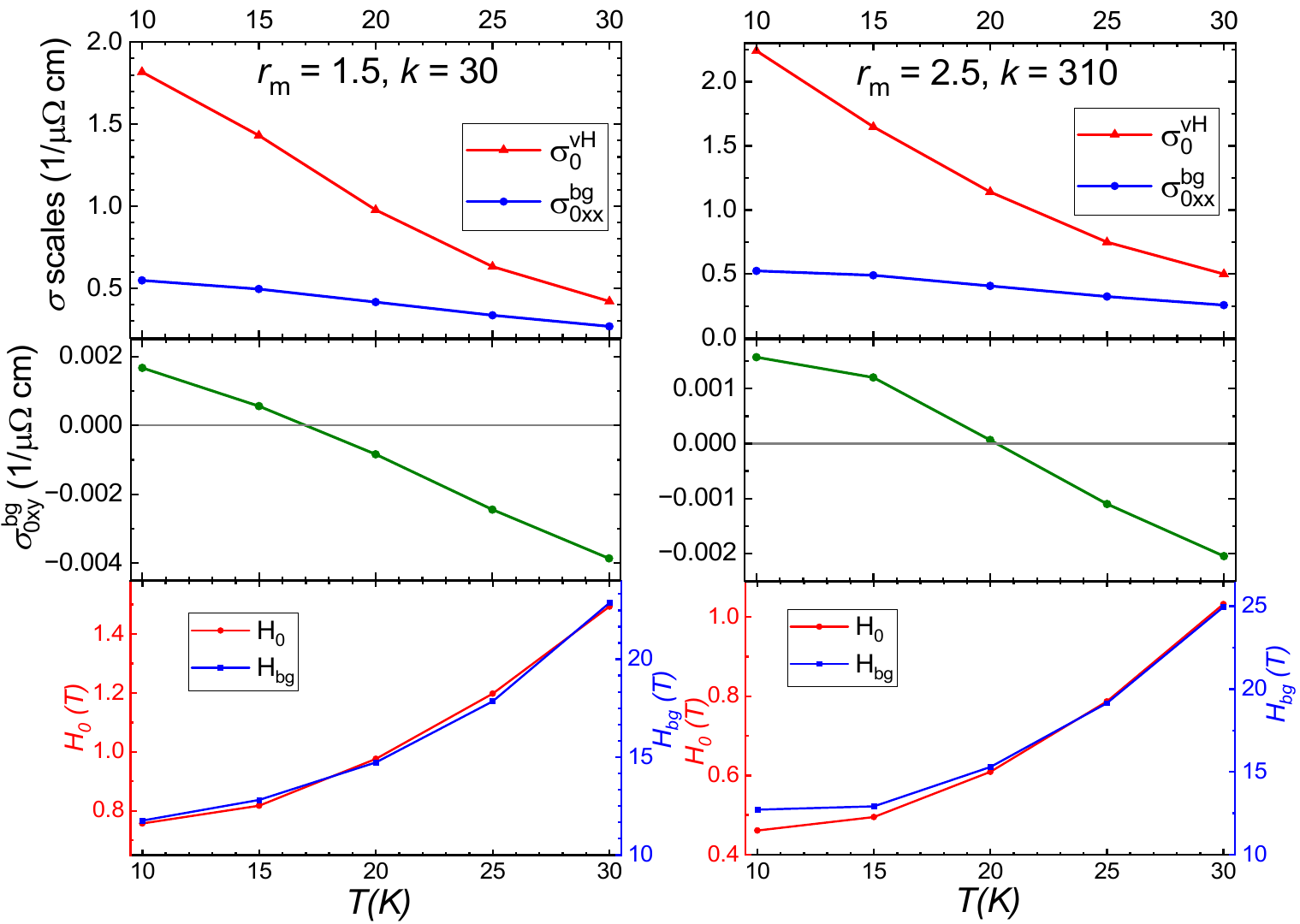}
\caption{Temperature dependences of the fitted conductivity and field scales
for two sets of the anomalous band parameters.}
\label{Fig:FitParamTDep}
\end{figure*}

As established by the DFT calculations\citep{FuPhysRevLett.127.207002,TanPhysRevLett.127.046401,OrtizPhysRevX.11.041030,KangNatPhys2022,TanNPJQM2023},
ARPES measurements
\citep{ChoPhysRevLett.127.236401,NakayamaPhysRevB.104.L161112,KangNatPhys2022,HuNatComm2022,KatoCommMat2022,LuoPhysRevLett.128.036402,LuoPhysRevB.105.L241111,LuoNatComm2022,HuPhysRevB.106.L241106},
and magnetic oscillations data \citep{OrtizPhysRevX.11.041030,FuPhysRevLett.127.207002,ShresthaPhysRevB.105.024508,ZhangPhysRevB.106.195103,HuangPhysRevB.106.064510,BroylesPhysRevLett.129.157001,ChapaiPhysRevLett.130.126401},
kagome superconductors \emph{A}V$_{3}$Sb$_{5}$ are multiple-band materials
with Fermi surfaces composed of multiple sheets. We can hardly expect
to quantitatively describe the magnetotransport of a material with such
complex band structure by taking into account contribution from only
one pocket, even if it has distinct anomalous behavior. In addition, we evaluated
the contribution from one $k_{z}$ slice, and the total conductivity is
obtained by integration over $k_{z}$ accounting for the $k_{z}$ dependence
of the band parameters. A full microscopic calculation of the magnetoconductivity
does not appear feasible. Therefore, we aim at a qualitative
modeling of the experimental data assuming that (i) all bands except
the ones originating from the large hexagon pocket give a smooth background
contribution and (ii) averaging over $k_{z}$ can be approximately
replaced by taking parameters from a typical $k_{z}$ slice, i.e., we
approximate Eq.~\eqref{CondGen} for the anomalous part dominated
by van Hove singularities as
\begin{equation}
\sigma_{\alpha\beta}^{\mathrm{vH}}\approx2e^{2}\frac{p_{z0}}{(2\pi)^{3}}\bar{S}_{\alpha\beta},\label{CondTypicalSlice}
\end{equation}
where $p_{z0}$ is the size of the large hexagon in the c-axis direction.
and $\bar{S}_{\alpha\beta}\!=\!S_{0}F_{\alpha\beta}^{\mathrm{R}}(H/H_{0},\bar{\kappa},r_{m})$
is the typical slice contribution with the parameter $\bar{\kappa}$
being the typical value of the ratio $K/p_{u0}$, see Eq.~\eqref{eq:SRab}.
The maximum value of $p_{z0}$ is obviously given by the size of the
Brillouin zone $K_{z}=2\pi/c$. Therefore, we try to model the experimental
magnetoconductivity data with the following ansatz
\begin{equation}
\sigma_{\alpha\beta}(H)=\sigma_{\alpha\beta}^{\mathrm{vH}}(H)+\sigma_{\alpha\beta}^{\mathrm{bg}}(H),\label{eq:CondModel}
\end{equation}
where $\sigma_{\alpha\beta}^{\mathrm{vH}}(H)$ is the anomalous magnetoconductivity
due to reconstructed FSs computed in the previous sections, which
we approximate as 
\begin{equation}
\sigma_{\alpha\beta}^{\mathrm{vH}}(H)=\sigma_{0}^{\mathrm{vH}}\kappa^{-2}F_{\alpha\beta}^{\mathrm{R}}\left(H/H_{0},\kappa,r_{m}\right)\label{eq:svHab}
\end{equation}
and $\sigma_{\alpha\beta}^{\mathrm{bg}}(H)$ is the background contribution
from all other bands for which we assume simple Drude shapes\begin{subequations}
\begin{align}
\sigma_{xx}^{\mathrm{bg}}(H) & =\frac{\sigma_{0xx}^{\mathrm{bg}}}{1+\left(H/H_{\mathrm{bg}}\right)^{2}},\label{eq:Sbgxx}\\
\sigma_{xy}^{\mathrm{bg}}(H) & =\frac{\sigma_{0xy}^{\mathrm{bg}}H/H_{\mathrm{bg}}}{1+\left(H/H_{\mathrm{bg}}\right)^{2}}.\label{eq:Sbgxy}
\end{align}
\end{subequations}As follows from Eqs.~\eqref{CondTypicalSlice},
the scale for the anomalous part can be estimated as
\begin{align*}
\sigma_{0}^{\mathrm{vH}} & \simeq2e^{2}\frac{p_{z0}}{(2\pi)^{3}}\frac{3\tau K^{2}}{\sqrt{m_{u}m_{v}}},
\end{align*}
while the definition of the field scale $H_{0}$ is in the Table \ref{Tbl:param}.
We observe that the product of conductivity and field scales does
not depend on scattering rate and effective masses
\begin{align}
\sigma_{0}^{\mathrm{vH}}H_{0} & \simeq2\frac{e^{2}}{\hbar}\frac{\Phi_{0}}{\pi}\frac{3K^{2}p_{z0}}{(2\pi)^{3}}.\label{eq:sigmaH}
\end{align}
Using $\frac{e^{2}}{\hbar}=\!2.433\cdot10^{-4}\,\Omega^{-1}$, $K_{z}\!=\!2\pi/c$
with $c\!=\!0.94$ nm \citep{FuPhysRevLett.127.207002}, $K\!=\!2\pi/(\sqrt{3}a)$ with $a\!=\!0.55$
nm, we estimate
\begin{equation}
\sigma_{0}^{\mathrm{vH}}H_{0}\approx11.3\frac{p_{z0}}{K_{z}}\frac{T}{\mu\Omega\cdot\mathrm{cm}}.\label{eq:sigmaHest}
\end{equation}

For modeling, we use only data below 30K which display distinctly anomalous behavior of the Hall resistivity in 
Fig.\ \ref{Fig:rhoabHExp}(b). The experimental field dependences of the diagonal and Hall conductivities
for CsV$_{3}$Sb$_{5}$ we are trying to model are shown in the left
column of Fig.~\ref{Fig:ExpMCondModel2sets}. They are obtained by inverting the resistivity matrix using experimental data presented in Fig.\ \ref{Fig:rhoabHExp}(a,b),
$\sigma_{xx}\!=\!\rho_{xx}/(\rho_{xx}^{2}\!+\!\rho_{yx}^{2})$ and $\sigma_{xy}\!=\!\rho_{yx}/(\rho_{xx}^{2}\!+\!\rho_{yx}^{2})$.
We fixed the temperature-independent
band parameters $r_{m}$ and $k$ and fitted simultaneously two experimental
curves $\sigma_{xx}(H)$ and $\sigma_{xy}(H)$ at a given temperature
using five fitting parameters for the conductivity and field scales,
$\sigma_{0}^{\mathrm{vH}}$, $\sigma_{0xx}^{\mathrm{bg}}$, $\sigma_{0xy}^{\mathrm{bg}}$,
$H_{0}$, and $H_{\mathrm{bg}}$, which vary with temperature. Presumably,
this temperature dependence mostly originates from the scattering rate.
We also modified the parameters $r_{m}$ and $k$ to find the set
giving the best modeling of the data for all temperatures. We found
that for ``naively reasonable'' parameter sets within the ranges
$r_{m}\!=\!1.2\!-\!1.8$, $k\!=\!15\!-\!50$, one can obtain qualitative
description of the data with some deviations, see, for example, series
of the simulated $\sigma_{\alpha\beta}(H)$ curves in the middle column
of Fig.~\ref{Fig:ExpMCondModel2sets} for $r_{m}\!=\!1.5$ and $k\!=\!30$.
Much better fits, however, may be achieved assuming larger mass ratio
$r_{m}=2.5$ and very large $k=310$ corresponding to very close proximity
to the van Hove point. The modeled curves with these parameters are
shown in the right column of Fig.~\ref{Fig:ExpMCondModel2sets}.
\begin{figure}
\includegraphics[width=3in]{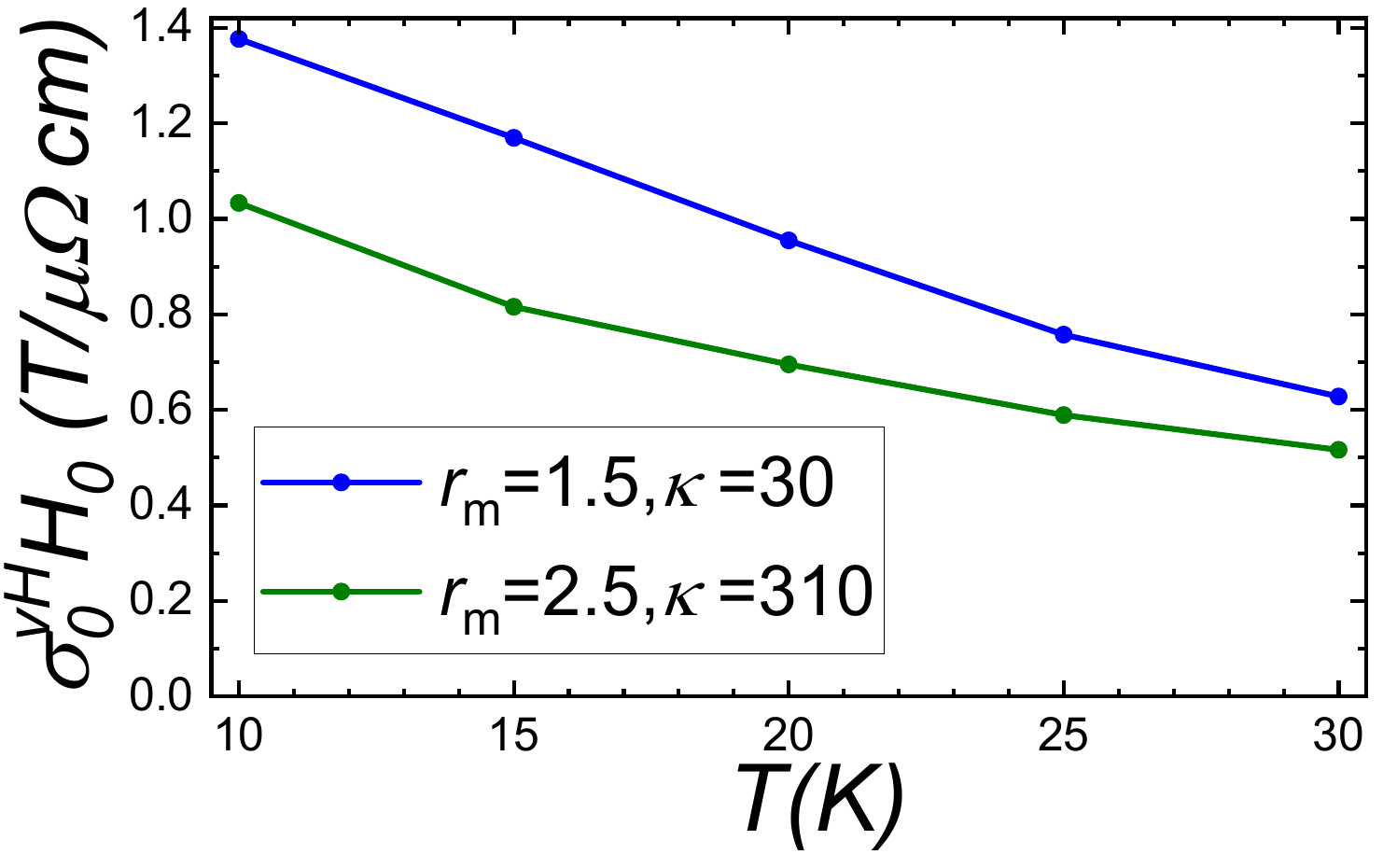}
\caption{Temperature dependences of the product $\sigma_{0}^{\mathrm{vH}}H_{0}$
for two sets of the anomalous band parameters.}
\label{Fig:sigmaHvsT}
\end{figure}

Figure \ref{Fig:FitParamTDep} shows temperature dependences of the
fitted conductivity and field scales for the two sets of the anomalous
band parameters used in Fig.~\ref{Fig:ExpMCondModel2sets}. We can
see that all parameters display regular temperature dependences. The
conductivity and field scales for two parameter sets are very close
with somewhat smaller field scales for the second set. The anomalous
and background diagonal conductivities have similar sizes. Note
that the zero-field anomalous conductivity is given by $\sigma_{\alpha\beta}^{\mathrm{vH}}(0)=0.26\sigma_{0}^{\mathrm{vH}}$
and $0.25\sigma_{0}^{\mathrm{vH}}$ for the first and second set,
respectively. The scale for the anomalous conductivity is characterized
by a stronger temperature dependence. The magnetic field scale for the anomalous
part decreases with temperature from 1.45 to 0.8 teslas for the first
set and from 1 to 0.5 teslas for the second set. The field scale for
the regular contribution is 15 -- 25 times larger. Nevertheless, both
scales have very similar temperature dependences described by the
law $a+bT^{3}$. The value of $H_{0}=1$ T together with the estimated
values of the masses $m_{u}$ and $m_{v}$ gives an estimate for the
scattering time $\tau\approx4.2\cdot10^{-12}$s. We also observe that
the background Hall conductivity changes sign with increasing temperature.

Figure \ref{Fig:sigmaHvsT} shows temperature dependences of the product
$\sigma_{0}^{\mathrm{vH}}H_{0}$ for two sets of the anomalous band
parameters. Even though Eq.~\eqref{eq:sigmaH} suggests that this product
depends only on band parameters, it does have a weak temperature
dependence. A comparison of the absolute value of this product with
the estimate in Eq.~\eqref{eq:sigmaHest} suggests that the large hexagon
occupies only small fraction of the Brillouin zone $p_{z0}/K_{z}\sim0.1$.

\begin{figure}
	\includegraphics[width=3.4in]{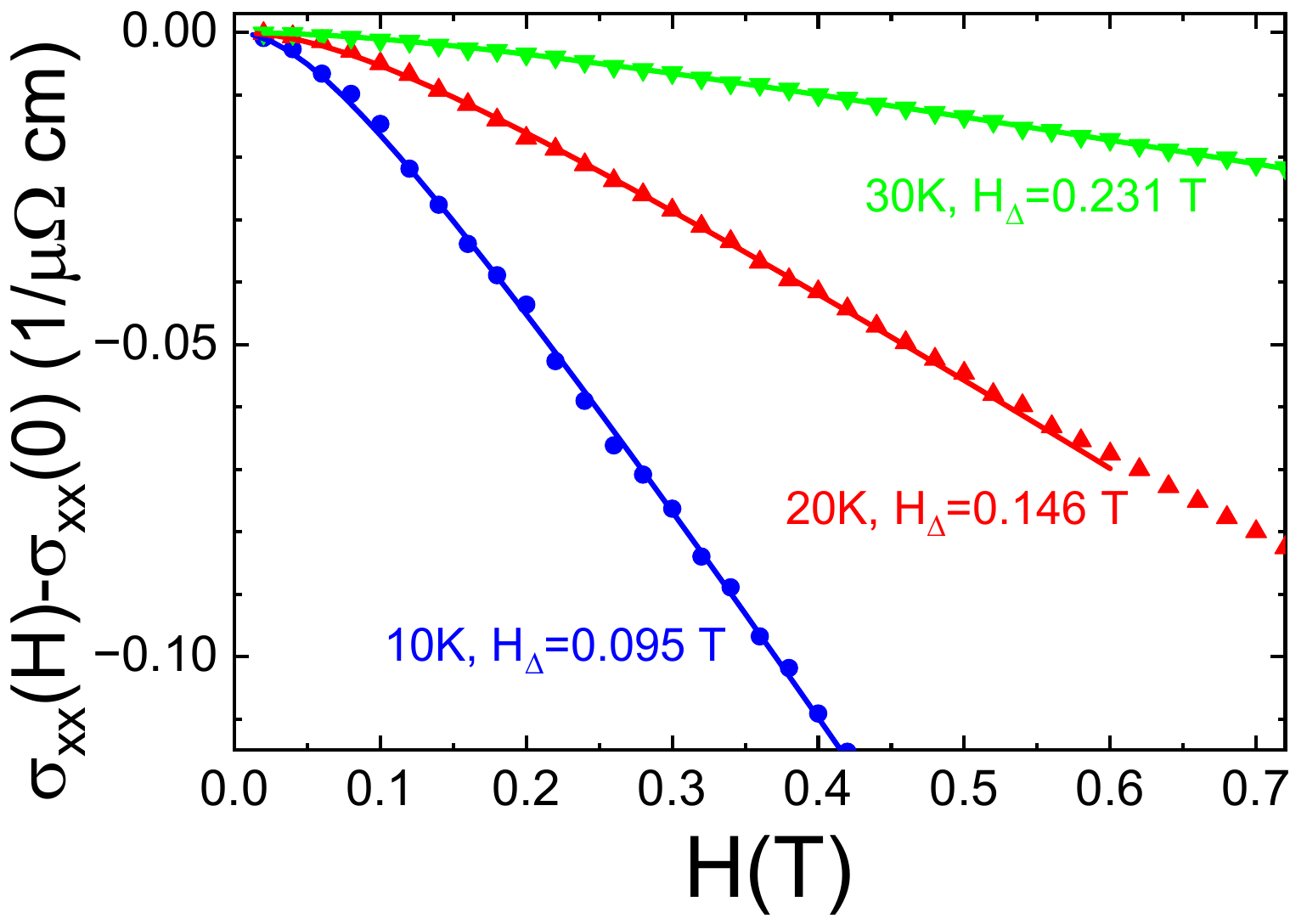}
	\caption{The field dependences of the diagonal conductivity at low magnetic fields for three temperatures displaying crossover between quadratic and linear behaviors. The solid lines show fits using Eqs.\ \eqref{DSxxHCross} and \eqref{eq:GhAppr}}
	\label{Fig:SigmaxxTLowHFits}
\end{figure}

We consider now the behavior at low magnetic fields $\lesssim 0.5$T. The linear magnetic-field dependence of the diagonal conductivity crosses over to the quadratic one for $H\rightarrow0$. This is an expected behavior because the corners at the branch crossing are not infinitely sharp but rounded due opening of the CDW gap, as described in Section \ref{sec:CornerRoundCDW}. We model the low-field behavior of the experimental magnetoconductivity using theoretical results from Ref.\ \cite{Koshelev2013} presented in section \ref{sec:CornerRoundCDW}. 
Figure \ref{Fig:SigmaxxTLowHFits} shows the representative field dependences of diagonal conductivity at low magnetic fields for three temperatures, 10, 20, and 30K. We can see that these dependences indeed display a clear crossover between quadratic and linear behaviors around 0.1--0.2 T. The solid lines show fits using Eqs.\ \eqref{DSxxHCross} and \eqref{eq:GhAppr}. We see that the crossover is very accurately described by the theory based on the CDW FS reconstruction. The extracted CDW field scale $H_{\Delta}$ defined in Eq.\ \eqref{eq:MagFieldScale} increases from 0.095 T at 10K to 0.23 T at 30K. This increase is mostly caused by the temperature dependence of the scattering time. The ratio $H_{\Delta}/H_0$ is approximately 0.125-0.15 for the parameter set in the left column of Fig.\ \ref{Fig:ExpMCondModel2sets}. According to Eq.\ \eqref{eq:HDCDWH0Ratio}, this ratio is determined by the ratio of the CDW gap $\Delta_{\mathrm{CDW}}$ and the shift of the Fermi level with respect to the van Hove energy $\varepsilon_{\mathrm{vH}}$. For $r_m\!=\!1.5$, we estimate the numerical constant in Eq.\ \eqref{eq:HDCDWH0Ratio} as $C_m\!\approx 0.64\!$. This means that our consideration suggests that $\Delta_{\mathrm{CDW}}$ is 4--5 times smaller than $\varepsilon_{\mathrm{vH}}$.

The ARPES data reported in Refs.~\citep{KangNatPhys2022,HuNatComm2022}
suggest that the van Hove energy relevant for the large hexagon is
located at 100 -- 200 meV below the Fermi level. Also, the feature at -200
meV in the tunneling spectrum of CsV$_{3}$Sb$_{5}$ has been attributed
to the van Hove singularity in Ref.\ \citep{ZhaoNature2021}. On the other hand,
the CDW gap estimated from ARPES\citep{NakayamaPhysRevB.104.L161112,WangPhysRevB.104.075148,LuoNatComm2022}
and STM\citep{JiangNatMat2021,ZhaoNature2021} is $\sim$20 meV. This means that the relation between $\Delta_{\mathrm{CDW}}$ and  $\varepsilon_{\mathrm{vH}}$ following from our qualitative analysis is reasonably consistent with the experimental data.

\section{Summary and discussion \label{sec:summary}}

In summary, we have introduced a minimal model of the large hexagon pocket in the Fermi surface of kagome metals that provides a natural interpretation for the experimentally observed anomalous magnetotransport in the CDW state of the compounds \emph{A}V$_{3}$Sb$_{5}$. The shapes of magnetoconductivity components in our model are determined by two dimensionless parameters: the mass ratio of the saddle-point spectrum and the ratio of the reciprocal-lattice vector to the minimum separation of the large-hexagon vertices from the van Hove point. 

We evaluated magnetoconductivity components for both large hexagon pocket in the pristine BZ and Fermi-surface sheets emerging from its reconstruction by the CDW order in the folded BZ, small hexagon and two large triangles. 
Two features account for the anomalous magnetotransport in \emph{A}V$_{3}$Sb$_{5}$: proximity of the vertices of a large hexagonal pocket to the van Hove points and the presence of very sharp corners in the reconstructed FS sheets. 
The magnetoconductivity components of the large hexagon are characterized by the two van Hove magnetic field scales: the lower scale $H_0$ is determined by the two effective masses of the saddle-point electronic spectrum and the upper scale $H_b$ contains an additional large logarithmic factor diverging as the Fermi energy approaches the van Hove energy. While these two scales are still present for the partial conductivities of the triangular pockets in the reconstructed FS, the third magnetic-field scale $H_\Delta$ emerges due to the sharp corners. It is proportional to the CDW gap and is the lowest scale in the problem. For CsV$_{3}$Sb$_{5}$,  our analysis yields the following hierarchy of the field scales at low temperatures: $H_\Delta\!\approx 0.1$ T, $H_0\!\approx 1$ T, and $H_b\!\approx \! 7$ T.

The sign of the linear Hall conductivity for the large hexagon and triangles is \emph{negative}, in spite of the hole nature of these Fermi sheets. The reason for this anomalous feature is the proximity of the regions with large positive curvature to the van Hove points in combination with overall concave shape of these pockets. The same feature causes strong nonmonotonic field dependence and sign change of the Hall conductivity. Therefore, our model naturally accounts for the anomalous magnetotransport behavior of the kagome family \emph{A}V$_{3}$Sb$_{5}$. Furthermore, it allows for a semi-quantitative description of the experimental magnetoconductivity in CsV$_{3}$Sb$_{5}$ with reasonable fitting parameters. 
Following the Occam's razor principle, the proposed mechanism provides a more likely explanation for a peculiar behavior of the Hall resistivity than the previously employed interpretation based on spontaneous Hall effect due to chiral CDW. We would like to point, however, that our interpretation by itself does not rule out possibility of a chiral CDW state in these materials considering that the anomalous Hall effect is not the only experimental indication for such a state. One can naturally expect that if a chiral CDW is realized, a random array of domains with opposite chiralities would form in a bulk sample. In such multidomain state a small spontaneous Hall effect will be hidden due to the averaging between the domains. This cancellation may persist in a finite magnetic field if it does not generate a single-domain state, which is not obvious \emph{a priori}. 

In conclusion, our work reveals the crucial role of van Hove singularities and CDW Fermi-surface reconstruction on the transport phenomena in the kagome superconductors \emph{A}V$_{3}$Sb$_{5}$.
Since van Hove points are ubiquitous in kagome metals, they very likely influence magnetotransport in other systems including recently discovered kagome families 
\emph{A}V$_6$Sn$_6$ (\emph{A}=Y, Gd, Ho, Sc, Sm)\citep{PokharelPhysRevB.104.235139,PengPhysRevLett.127.266401,ArachchigePhysRevLett.129.216402,HuangPhysRevMaterials.7.054403,HuNatComm2024} and
\emph{A}Ti$_3$Bi$_5$(\emph{A}=Cs, Rb)\citep{yangArXiv2022,LiuPhysRevLett.131.026701,JiangNatComm2023}. 
For example, it is highly probable that the mechanism discussed here is responsible for a nonmonotonic, sign-changing magnetic-field dependence of Hall resistivity recently reported for ScV$_6$Sn$_6$\cite{mozaffariArXiV2023}. 
Furthermore, our approach can be extended to a large family of metallic systems hosting singular features in their electronic spectrum. 

\begin{acknowledgments}
	This work was supported by the US Department of Energy, Office of Science, Basic Energy Sciences, Materials Sciences and Engineering Division. 
\end{acknowledgments}

\appendix
\section{Calculation of hyperbolic hexagon area and effective mass\label{App:AreaMass}}

\subsection{Hyperbolic hexagon \label{App:HHarea}}

With the Fermi-surface hyperbolic equation
\[
p_{F,u}(p_{v})=\sqrt{p_{v0}^{2}+p_{v}^{2}}/\sqrt{r_{m}},
\]
the hexagon area can be written as 
\begin{align*}
	A_{\mathrm{HH}} & =12\int_{0}^{p_{vc}}\left(p_{F,u}(p_{v})-\sqrt{3}p_{v}\right)dp_{v}.
\end{align*}
Using hyperbolic parametrization, we can evaluate this integral as
\begin{align*}
	A_{\mathrm{HH}} & =\frac{12p_{v0}^{2}}{\sqrt{r_{m}}}\int_{0}^{t_{c}}\cosh^{2}tdt-6\sqrt{3}p_{vc}^{2}\\
	& =\frac{6p_{v0}^{2}}{\sqrt{r_{m}}}\left(t_{c}+\frac{\sinh2t_{c}}{2}\right)-6\sqrt{3}p_{v0}^{2}\sinh^{2}t_{c}\\
	& =6p_{u0}^{2}\sqrt{r_{m}}t_{c}.
\end{align*}
Using the relation $\tanh t_{c}=\frac{1}{\sqrt{3r_{m}}}$, we finally
obtain 
\begin{equation}
	A_{\mathrm{HH}}=3p_{u0}^{2}\sqrt{r_{m}}\ln\frac{\sqrt{3r_{m}}\!+\!1}{\sqrt{3r_{m}}\!-\!1}.\label{eq:HexAreaApp}
\end{equation}
For comparison, the area of an ideal hexagon is $A_{\mathrm{hex}}=2\sqrt{3}p_{u0}^{2}\approx3.464p_{u0}^{2}$
(clearly, $A_{\mathrm{HH}}<A_{\mathrm{hex}}$ corresponding to $\sqrt{3r_{m}}\ln\frac{\sqrt{3r_{m}}\!+\!1}{\sqrt{3r_{m}}\!-\!1}<2$).

The effective mass probed by magnetic oscillations is defined as
\[
m_{\mathrm{HH}}=\frac{1}{2\pi}\frac{dA_{\mathrm{HH}}}{d\varepsilon_{F}}
\]
and from Eq.~\eqref{eq:HexAreaApp} we immediately obtain
\begin{equation}
	m_{\mathrm{HH}}=\frac{3}{\pi}\sqrt{m_{u}m_{v}}\ln\frac{\sqrt{3r_{m}}\!+\!1}{\sqrt{3r_{m}}\!-\!1}=\frac{3}{\pi}m_{u}\sqrt{r_{m}}\ln\frac{\sqrt{3r_{m}}\!+\!1}{\sqrt{3r_{m}}\!-\!1}.
	\label{eq:HexCyclMassApp}
\end{equation}

\subsection{Triangular pocket\label{sec:TArea}}

\begin{figure}
	\includegraphics[width=3.0in]{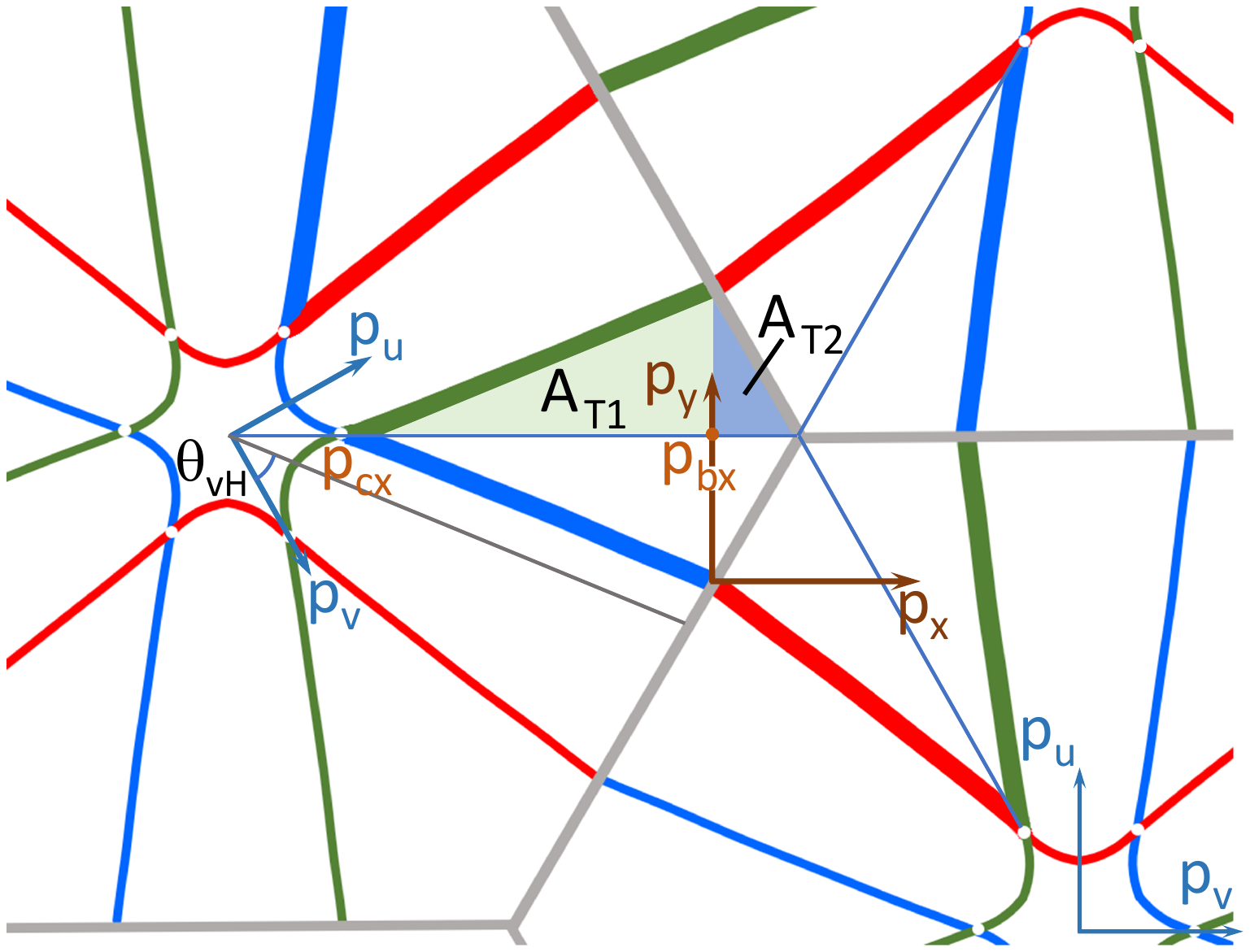}
	\caption{Triangular pocket composed of the highlighted small triangles with
		areas $A_{T1}$ and $A_{T2}$ computed in the text.}
	\label{Fig:TriArea}
\end{figure}
The triangular pocket is composed of six triangular pieces. Each piece,
in turn can be split into two smaller triangles highlighted in Fig.~\ref{Fig:TriArea}.
We compute the areas of these two pieces notated as $A_{T1}$ and
$A_{T2}$. The area of the triangle located between the horizontal point $p_{cx}$ and $p_{bx}$ shown in Fig.\ \ref{Fig:TriArea},
can be evaluated as
\begin{align*}
	A_{T1} & =\int_{p_{cx}}^{p_{bx}}p_{y}(p_{x})dp_{x}
\end{align*}
Using hyperbolic parametrization 
\begin{align*}
	p_{x} & =p_{u0}\left(\frac{1}{2}\sqrt{r_{m}}\sinh t\!+\frac{\sqrt{3}}{2}\cosh t\right),\\
	p_{y} & =p_{u0}\left(\frac{\sqrt{3}}{2}\sqrt{r_{m}}\sinh t\!-\frac{1}{2}\cosh t\right),
\end{align*}
we compute
\begin{align}
	&A_{T1} \! =\!\frac{p_{u0}^{2}}{4}\!\int\limits _{t_{c}}^{t_{b}}\!\!\left(\sqrt{3r_{m}}\sinh t\!-\!\cosh t\right)\!\left(\!\sqrt{r_{m}}\cosh t\!+\!\sqrt{3}\sinh t\right)\!dt\nonumber \\
	& =\frac{p_{u0}^{2}}{8}\!\Big[\!\left(\sqrt{3r_{m}}\sinh t_{b}\!-\!\cosh t_{b}\right)\!
	\left(\sqrt{r_{m}}\sinh t_{b}\!+\!\sqrt{3}\cosh t_{b}\right)\nonumber \\
	& -4\sqrt{r_{m}}\left(t_{b}-t_{c}\right)\Big]\label{eq:AT1}
\end{align}
The area of the second small triangle can be straightforwardly evaluated
as
\begin{align}
	A_{T2} & =\frac{1}{2}p_{by}\left(\frac{K}{\sqrt{3}}-p_{xb}\right)\label{eq:AT2}\\
	= & \frac{p_{u0}^{2}}{8\sqrt{3}}\left(\sqrt{3r_{m}}\sinh t_{b}\!-\cosh t_{b}\right)^{2}.\nonumber 
\end{align}
Therefore, the full area of the triangular pocket is given by
\begin{align}
	A_{\mathrm{T}} & =6\left(A_{T1}+A_{T2}\right)\nonumber \\
	= & p_{u0}^{2}\left[\frac{\sqrt{3}}{2}\left(3r_{m}\sinh^{2}t_{b}\!-\!\cosh^{2}t_{b}\right)\!-\!3\sqrt{r_{m}}\left(t_{b}-t_{c}\right)\right]\label{eq:TriAreaApp}
\end{align}
Using definition of the parameter $t_{b}$ in Eq.~\eqref{eq:tb},
we can also transform this result to a more transparent form 
\begin{align}
	A_{\mathrm{T}} & =\frac{\sqrt{3}}{2}\frac{\sqrt{3r_{m}}\!-\!1}{\sqrt{3r_{m}}\!+\!1}K^{2}\nonumber \\
	- & 3\sqrt{r_{m}}p_{u0}^{2}\left(t_{b}\!-t_{c}\!+\!\frac{\kappa}{\sqrt{\kappa^{2}\!+\!3r_{m}\!-\!1}\!+\!\kappa}\right).
	\label{eq:TriAreaApp1}
\end{align}
Using this result and dependences of $p_{u0}$ and $t_{b}$ on the Fermi
energy, we can also derive the following result for the effective
mass of the triangular pocket
\begin{equation}
	m_{T}=\frac{1}{2\pi}\frac{dA_{\mathrm{T}}}{d\varepsilon_{F}}\approx\!-\!\frac{3}{\pi}\left(t_{b}-t_{c}\right)\sqrt{m_{u}m_{v}}.\label{eq:TriMassApp}
\end{equation}
Due to proximity to the van Hove point, the effective mass has additional large factor $\sim t_b$.

\section{Conductivity slice $S_{x\beta}$ for arbitrary pocket with m-fold symmetry \label{sec:App-mfold}}

Consider a 2D slice of a Fermi surface having $m$-fold symmetry. We split
it into $m$ equivalent segments separated by the momenta $p_{k}$  with $k=1,\dots,m$,
located at the Fermi contour. Each segment is assumed to have mirror
symmetry. We select the local coordinate system for each segment along
and perpendicular to the mirror plane. The local axes for the segment
with index $k$ are rotated at an angle $\theta_{k}=\!2\left(k\!-\!1\right)\pi/m\!+\!\alpha$
with respect to the laboratory frame. We can split the Fermi velocity
into the longitudinal and transverse components, $v_{l}$ and $v_{t}$,
which are symmetric and antisymmetric, respectively, with respect to the mirror plane. 

To proceed, we rewrite the slice $S_{x\beta}$ in Eq.~\eqref{pzSliceFullPer} as
\begin{equation}
S_{x\beta}\!=\!\varsigma_{x\beta}\frac{c}{|e|H}\left[1\!-\!\exp\left(\!-\!\varointctrclockwise\frac{dp}{v}\frac{c}{|e|H\tau}\right)\right]^{-1}\!\!\varointctrclockwise\frac{dp}{v}v_{\beta}\mathcal{I}_{x}(p),\label{eq:Sxbmfold}
\end{equation}
where
\[
\mathcal{I}_{x}(p)=\varointctrclockwise_{p}\frac{dp^{\prime}}{v^{\prime}}v_{x}^{\prime}\exp\left(\!-\mathfrak{J}_{p}^{p^{\prime}}\right)
\]
represents the $p^{\prime}$ integral over the whole orbit for fixed initial momentum $p$ and 
\begin{equation}
\mathfrak{J}_{p}^{p^{\prime}}\equiv\int_{p}^{p^{\prime}}\frac{dp^{\prime\prime}}{v^{\prime\prime}}\frac{c}{|e|H\tau}\label{eq:Int-notation}
\end{equation}
is the orbit integral between point $p$ and $p^{\prime}$, which has an obvious property $\mathfrak{J}_{p}^{p^{\prime}}=\mathfrak{J}_{p}^{p_{0}}+\mathfrak{J}_{p_{0}}^{p^{\prime}}$
for an arbitrary intermediate point $p_{0}$ at the orbit. For $p$ located
inside the segment $[p_{1},p_{2}]$, the full integral over $p^{\prime}$
can be split into the segments as 
\begin{align*}
\mathcal{I}_{x}(p)= & \int\limits_{p}^{p_{2}}\frac{dp^{\prime}}{v^{\prime}}v_{x}^{\prime}\exp\left(\!-\mathfrak{J}_{p}^{p^{\prime}}\right)\\
+ & \sum_{k=2}^{m}\exp\left(\!-\mathfrak{J}_{p}^{p_{2}}\!-\!\sum_{j=2}^{k-1}\!\mathfrak{J}_{p_{j}}^{p_{j+1}}\!\right)\!\int\limits _{p_{k}}^{p_{k+1}}\frac{dp^{\prime}}{v^{\prime}}v_{x}^{\prime}\exp\left(\!-\mathfrak{J}_{p_{k}}^{p^{\prime}}\right)\\
+ & \exp\left(\!-\mathfrak{J}_{p}^{p_{2}}\!-\sum_{j=2}^{m}\!\int\limits _{p_{j}}^{p_{j+1}}\mathfrak{J}_{p_{j}}^{p_{j+1}}\!\right)\int\limits _{p_{1}}^{p}\frac{dp^{\prime}}{v^{\prime}}v_{x}^{\prime}\exp\left(\!-\mathfrak{J}_{p_{1}}^{p^{\prime}}\right).
\end{align*}
For the momentum $p^{\prime}$ located inside the $k$'s segment $\left[p_{k},p_{k+1}\right]$,
we can expand the velocity over the local longitudinal and transverse
components, $v_{x}^{\prime}=c_{k}v_{t}^{\prime}-s_{k}v_{l}^{\prime}$
with $c_{k}=\cos\left[2\left(k\!-\!1\right)\pi/m\!+\!\alpha\right]$
and $s_{k}=\sin\left[2\left(k\!-\!1\right)\pi/m\!+\!\alpha\right]$.
Since the segments are equivalent, the integrals\begin{subequations}
\begin{align}
\mathcal{Q} & =\mathfrak{J}_{p_{k}}^{p_{k+1}},\label{eq:mfoldQdef}\\
\mathcal{R}_{s} & \!=\!\int\limits _{p_{k}}^{p_{k+1}}\frac{dp^{\prime}}{v^{\prime}}v_{s}^{\prime}\exp\left(\!-\mathfrak{J}_{p_{k}}^{p^{\prime}}\right)\label{eq:mfoldRsDef}
\end{align}
\end{subequations}
with $s=l,t$ do not depend on the segment index $k$. This allows us to represent the $p^{\prime}$ integral as
\begin{align*}
\mathcal{I}_{x}(p)\! & =\int\limits_{p}^{p_{2}}\frac{dp^{\prime}}{v^{\prime}}\left(c_{1}v_{t}^{\prime}-s_{1}v_{l}^{\prime}\right)\exp\left(\!-\mathfrak{J}_{p}^{p^{\prime}}\right)\\
+ & \exp\left(\!-\mathfrak{J}_{p}^{p_{2}}\!\right)\sum_{k=2}^{m}\exp\left[\!-(k\!-\!2)\mathcal{Q}\right]\left(c_{k}\mathcal{R}_{t}-s_{k}\mathcal{R}_{l}\right)\\
+ & \exp\left(\!-\mathfrak{J}_{p}^{p_{2}}\!-\!(m\!-\!1)\mathcal{Q}\right)\!
\int\limits _{p_{1}}^{p}\frac{dp^{\prime}}{v^{\prime}}\left(c_{1}v_{t}^{\prime}\!-\!s_{1}v_{l}^{\prime}\right)\exp\left(\!-\!\mathfrak{J}_{p_{1}}^{p^{\prime}}\right).
\end{align*}
Further, the last term can be transformed as
\begin{align*}
 & \exp\left(\!-\mathfrak{J}_{p}^{p_{2}}\!-(m\!-\!1)\mathcal{Q}\!\right)\int\limits _{p_{1}}^{p}\frac{dp^{\prime}}{v^{\prime}}\left(c_{1}v_{t}^{\prime}-s_{1}v_{l}^{\prime}\right)\exp\left(\!-\mathfrak{J}_{p_{1}}^{p^{\prime}}\right)\\
= & \exp\left(\!-\mathfrak{J}_{p}^{p_{2}}\!-(m\!-\!1)\mathcal{Q}\!\right)\left(c_{1}\mathcal{R}_{t}-s_{1}\mathcal{R}_{l}\right)\\
- & \exp\left(\!-m\mathcal{Q}\right)\int\limits _{p}^{p_{2}}\frac{dp^{\prime}}{v^{\prime}}\left(c_{1}v_{t}^{\prime}-s_{1}v_{l}^{\prime}\right)\exp\left(\!-\mathfrak{J}_{p}^{p^{\prime}}\right),
\end{align*}
allowing us to combine it with the first two terms and obtain the
convenient presentation for $\mathcal{I}_{x}(p)$,
\begin{align}
&\mathcal{I}_{x}(p) =\left[1\!-\!\exp\left(\!-\!m\mathcal{Q}\right)\right]\int\limits _{p}^{p_{r+1}}\frac{dp^{\prime}}{v^{\prime}}\left(c_{r}v_{t}^{\prime}\!-\!s_{r}v_{l}^{\prime}\right)\exp\left(\!-\mathfrak{J}_{p_{r}}^{p^{\prime}}\right)\nonumber \\
&+\!  \exp\left(\!-\mathfrak{J}_{p}^{p_{r+1}}\right)\!\sum_{k=r+1}^{r+m}\!\exp\left[\!-(k\!-\!r\!-\!1)\mathcal{Q}\right]\left(c_{k}\mathcal{R}_{t}\!-\!s_{k}\mathcal{R}_{l}\right),\label{eq:Ipmfold-pres}
\end{align}
where we generalized the result for $p$ located inside an arbitrary
segment $[p_{r},p_{r+1}]$ and used the periodicity $c_{k+m}\!=\!c_{k}$,
$s_{k+m}\!=\!s_{k}$. We will use this result for calculation of slices
$S_{xx}$ and $S_{xy}$ from Eq.~\eqref{eq:Sxbmfold}. 

\subsection{Diagonal conductivity}

We proceed with calculation of the diagonal part of the slice conductivity,
$S_{xx}$ from Eq.~\eqref{eq:Sxbmfold}. The part of the integral in
$S_{xx}$ for the momentum $p$ inside the segment $\left[p_{r},p_{r+1}\right]$
using the result for $\mathcal{I}_{x}(p)$ in Eq.~\eqref{eq:Ipmfold-pres}
can now be written as\begin{widetext}
\begin{align*}
 & \frac{c}{|e|H}\smallint\limits _{p_{r}}^{p_{r+1}}\frac{dp}{v}v_{x}\mathcal{I}_{x}(p)\\
= & \frac{c}{|e|H}\left[1\!-\!\exp\left(\!-m\mathcal{Q}\right)\right]\smallint\limits _{p_{r}}^{p_{r+1}}\frac{dp}{v}\left(c_{r}v_{t}\!-\!s_{r}v_{l}\right)\!\int\limits _{p}^{p_{r+1}}\frac{dp^{\prime}}{v^{\prime}}\left(c_{r}v_{t}^{\prime}\!-\!s_{r}v_{l}^{\prime}\right)\exp\left(\!-\mathfrak{J}_{p}^{p^{\prime}}\right)\\
+ & \frac{c}{|e|H}\smallint\limits _{p_{r}}^{p_{r+1}}\frac{dp}{v}\left(c_{r}v_{t}\!-\!s_{r}v_{l}\right)\exp\left(\!-\mathfrak{J}_{p}^{p_{r+1}}\!\right)\sum_{k=r+1}^{m+r}\exp\left[\!-(k\!-\!r\!-\!1)\mathcal{Q}\right]\left(c_{k}\mathcal{R}_{t}\!-\!s_{k}\mathcal{R}_{l}\right).
\end{align*}
Using the symmetry properties of $v_{t}$ and $v_{l}$, the integrals in
the second line can be related to the segment integrals in Eq.~\eqref{eq:mfoldRsDef}
as \begin{subequations}
\begin{align}
 & \int\limits _{p_{r}}^{p_{r+1}}\frac{dp}{v}v_{t}\exp\left(\!-\mathfrak{J}_{p}^{p_{r+1}}\right)=-\mathcal{R}_{t},\label{eq:RvLHsuppl-1}\\
 & \int\limits _{p_{r}}^{p_{r+1}}\frac{dp}{v}v_{l}\exp\left(\!-\mathfrak{J}_{p}^{p_{r+1}}\right)=\mathcal{R}_{l}.\label{eq:RuLHsuppl-1}
\end{align}
\end{subequations} 
We also introduce the following notations
\begin{align}
\mathcal{G}_{sq}= & \int_{p_{r}}^{p_{r+1}}\!\frac{dp}{v}v_{s}\int_{p}^{p_{r+1}}\frac{dp^{\prime}}{v^{\prime}}v_{q}^{\prime}\exp\left(\!-\mathfrak{J}_{p}^{p^{\prime}}\right),\label{eq:GsrDefmfold}
\end{align}
with $s,r=l,t$ for the same-segment integrals in the second line.
They also are identical for all segments. This gives the following result for the $p$ integral over the segment $[p_r,p_{r+1}]$
\begin{align}
 & \frac{c}{|e|H}\smallint\limits _{p_{r}}^{p_{r+1}}\frac{dp}{v}v_{x}\mathcal{I}_{x}(p)\nonumber \\
= & \frac{c}{|e|H}\left[1\!-\!\exp\left(\!-m\mathcal{Q}\right)\right]\left(c_{r}^{2}\mathcal{G}_{tt}+s_{r}^{2}\mathcal{G}_{ll}-c_{r}s_{r}\left(\mathcal{G}_{tl}+\mathcal{G}_{lt}\right)\right)\nonumber \\
- & \frac{c}{|e|H}\left(c_{r}\mathcal{R}_{t}+s_{r}\mathcal{R}_{l}\right)\sum_{k=r+1}^{r+m}\exp\left[\!-(k\!-\!r\!-\!1)\mathcal{Q}\right]\left(c_{k}\mathcal{R}_{t}-s_{k}\mathcal{R}_{l}\right).\label{eq:SmfoldxxIntPres}
\end{align}
Adding all $p$ segments and using the identities $\sum_{r=1}^{m}c_{r}^{2}\!=\!\sum_{r=1}^{m}s_{r}^{2}\!=\!m/2$
and $\sum_{r=1}^{m}c_{r}s_{r}\!=\!0$, we can write the full result
as 
\begin{align}
S_{xx}\! & =\!\frac{c}{|e|H}\left\{ \frac{m}{2}\left(\mathcal{G}_{tt}+\mathcal{G}_{ll}\right)-\left[1\!-\!\exp\left(\!-\!m\mathcal{Q}\right)\right]^{-1}\right.\nonumber \\
\times & \left.\sum_{r=1}^{m}\left(c_{r}\mathcal{R}_{t}\!+\!s_{r}\mathcal{R}_{l}\right)\sum_{k=r+1}^{r+m}\exp\left[\!-(k\!-\!r\!-\!1)\mathcal{Q}\right]\left(c_{k}\mathcal{R}_{t}\!-\!s_{k}\mathcal{R}_{l}\right)\right\} ,\label{eq:SmfoldxxFullSum}
\end{align}
As $c_{k}\!=\!\cos\left[2\left(k\!-\!1\right)\pi/m\!+\!\alpha\right]$
and $s_{k}\!=\!\sin\left[2\left(k\!-\!1\right)\pi/m\!+\!\alpha\right]$,
calculation of the trigonometric sums is facilitated by the complex
presentations 
\begin{equation}
c_{k}\mathcal{R}_{t}\!-\!s_{k}\mathcal{R}_{l}\!=\mathrm{Re}\left[\exp\left(\imath\frac{2\pi}{m}\left(k\!-\!1\right)\!+\!i\alpha\right)\left(\mathcal{R}_{t}\!+\!\imath\mathcal{R}_{l}\right)\right]\label{eq:ComplPres}
\end{equation}
 and $c_{r}\mathcal{R}_{t}\!+\!s_{r}\mathcal{R}_{l}\!=\!\mathrm{Re}\left[\exp\left(\imath\frac{2\pi}{m}\left(r\!-\!1\right)\!+\!\imath\alpha\right)\left(\mathcal{R}_{t}\!-\!\imath\mathcal{R}_{l}\right)\right]\!=\frac{1}{2}\sum_{\delta=\pm1}\exp\left(\imath\frac{2\pi}{m}\delta\left(r\!-\!1\right)\!+\!\imath\delta\alpha\right)\left(\mathcal{R}_{t}\!-\!\imath\delta\mathcal{R}_{l}\right)$.
Using these presentations, we compute the sum over the index $k$
as, 
\begin{equation}
\sum_{k=r+1}^{r+m}\exp\left[\!-(k\!-\!r\!-\!1)\mathcal{Q}+\imath\frac{2\pi}{m}\left(k-1\right)+i\alpha\right]=\exp\left(i\alpha+\!\imath r\frac{2\pi}{m}\right)\frac{1-\exp\left(\!-m\mathcal{Q}\right)}{1-\exp\left(\!-\mathcal{Q}\!+\!\imath\frac{2\pi}{m}\right)}\label{eq:mfoldkSum}
\end{equation}
and obtain
\begin{align}
S_{xx}\! & =\!\frac{c}{|e|H}\left\{ \frac{m}{2}\left(\mathcal{G}_{tt}+\mathcal{G}_{ll}\right)\right.\nonumber \\
- & \frac{1}{2}\left.\mathrm{Re}\left[\sum_{\delta=\pm1}\left(\mathcal{R}_{t}\!-\!\imath\delta\mathcal{R}_{l}\right)\sum_{r=1}^{m}\exp\left[\imath\frac{2\pi}{m}\delta\left(r\!-\!1\right)\!+\!\imath r\frac{2\pi}{m}+\!\imath(\delta\!+\!1)\alpha\right]\frac{\mathcal{R}_{t}\!+\!\imath\mathcal{R}_{l}}{1-\exp\left(\!-\mathcal{Q}\!+\!\imath\frac{2\pi}{m}\right)}\right]\right\} .\label{eq:SmfoldxxPres}
\end{align}
As the sum over index $r$ vanishes for $\delta=1$ and the terms
are $r$ independent for $\delta=-1$ , we arrive at the final result
\begin{align}
S_{xx}^{(m)} & =\frac{m}{2}\frac{c}{|e|H}\left\{ \mathcal{G}_{tt}+\mathcal{G}_{ll}-\mathrm{Re}\left[\frac{\left(\mathcal{R}_{t}\!+\!\imath\mathcal{R}_{l}\right)^{2}}{\exp\left(-\imath\frac{2\pi}{m}\right)\!-\!\exp\left(\!-\mathcal{Q}\right)}\right]\right\} \label{eq:SmfoldxxApp}\\
 & =\frac{m}{2}\frac{c}{|e|H}\left\{ \mathcal{G}_{tt}+\mathcal{G}_{ll}-\frac{\left(\mathcal{R}_{t}^{2}-\mathcal{R}_{l}^{2}\right)\left[\cos\left(\frac{2\pi}{m}\right)-\exp\left(\!-\mathcal{Q}\right)\right]\!-\!2\mathcal{R}_{t}\mathcal{R}_{l}\sin\left(\frac{2\pi}{m}\right)}{1-2\cos\left(\frac{2\pi}{m}\right)\exp\left(\!-\mathcal{Q}\right)+\exp\left(\!-2\mathcal{Q}\right)}\right\} ,\nonumber 
\end{align}
which expresses the diagonal conductivity slice in terms of segment
integrals defined in Eqs.~\eqref{eq:mfoldQdef},~\eqref{eq:mfoldRsDef},
and \eqref{eq:GsrDefmfold}.

\subsection{Hall conductivity}

The calculation of the Hall component $S_{xy}$ is again based on
the presentation in Eq.~\eqref{eq:Sxbmfold} in combination with
$\mathcal{I}_{x}(p)$ in Eq.~\eqref{eq:Ipmfold-pres}. Using the
expansion $v_{y}=s_{r}v_{t}\!+c_{r}v_{l}$, the $p$ integral for
segment $\left[p_{r},p_{r+1}\right]$ in $S_{xy}$ can be represented
as
\begin{align*}
 & \frac{c}{|e|H}\!\smallint\limits _{p_{r}}^{p_{r+1}}\frac{dp}{v}v_{y}\mathcal{I}_{x}(p)\\
= & \frac{c}{|e|H}\left[1\!-\!\exp\left(\!-m\mathcal{Q}\right)\right]\smallint\limits _{p_{r}}^{p_{r+1}}\frac{dp}{v}\left(s_{r}v_{t}\!+c_{r}v_{l}\right)\int\limits _{p}^{p_{r+1}}\frac{dp^{\prime}}{v^{\prime}}\left(c_{r}v_{t}^{\prime}\!-\!s_{r}v_{l}^{\prime}\right)\exp\left(\!-\mathfrak{J}_{p}^{p^{\prime}}\right)\\
+ & \frac{c}{|e|H}\smallint\limits _{p_{r}}^{p_{r+1}}\frac{dp}{v}\left(s_{r}v_{t}\!+c_{r}v_{l}\right)\exp\left(\!-\!\mathfrak{J}_{p}^{p_{r+1}}\right)\sum_{k=r+1}^{m+r}\exp\left[\!-(k\!-\!r\!-\!1)\mathcal{Q}\right]\left(c_{k}\mathcal{R}_{t}-s_{k}\mathcal{R}_{l}\right)\\
= & \frac{c}{|e|H}\left[1\!-\!\exp\left(\!-m\mathcal{Q}\right)\right]\left(c_{r}^{2}\mathcal{G}_{lt}-s_{r}^{2}\mathcal{G}_{tl}+c_{r}s_{r}\left(\mathcal{G}_{tt}-\mathcal{G}_{ll}\right)\right)\\
+ & \frac{c}{|e|H}\left(-s_{r}\mathcal{R}_{t}\!+c_{r}\mathcal{R}_{l}\right)\sum_{k=r+1}^{m+r}\exp\left[\!-(k\!-\!r\!-\!1)\mathcal{Q}\right]\left(c_{k}\mathcal{R}_{t}-s_{k}\mathcal{R}_{l}\right)
\end{align*}
Therefore, for the full Hall conductivity slice, we obtain
\begin{align}
S_{xy}\!= & \varsigma_{xy}\!\frac{c}{|e|H}\left\{ \frac{m}{2}\left(\mathcal{G}_{lt}-\mathcal{G}_{tl}\right)+\left[1\!-\!\exp\left(\!-m\mathcal{Q}\right)\right]^{-1}\right.\nonumber\\
\times & \!\left.\!\sum_{r=1}^{m}\left(-s_{r}\mathcal{R}_{t}\!+c_{r}\mathcal{R}_{l}\right)\sum_{k=r+1}^{r+m}
\exp\left[-(k\!-\!r\!-\!1)\mathcal{Q}\right]\left(c_{k}\mathcal{R}_{t}\!-\! s_{k}\mathcal{R}_{l}\right)\right\} 
\label{eq:SxymfoldInt}
\end{align}
Using the complex presentation in Eq.~\eqref{eq:ComplPres} and $-s_{r}\mathcal{R}_{t}\!+\!c_{r}\mathcal{R}_{l}\!=\!\frac{1}{2}\sum_{\delta=\pm1}\exp\left(\imath\delta\frac{2\pi}{m}(r\!-\!1)\!+\!i\delta\alpha\right)\left(\imath\delta\mathcal{R}_{t}\!+\!\mathcal{R}_{l}\right)$,
as well as Eq.~\eqref{eq:mfoldkSum} for the sum in the second line, we derive 
\begin{align*}
S_{xy}\!= & \varsigma_{xy}\!\frac{c}{|e|H}\Bigg\{ \frac{m}{2}\left(\mathcal{G}_{lt}-\mathcal{G}_{tl}\right)\\
+\frac{1}{2} & \left.\mathrm{Re}\left[\sum_{\delta=\pm1}\sum_{r=1}^{m}\exp\left(\imath\delta\frac{2\pi}{m}(r\!-\!1)\!+\!\imath r\frac{2\pi}{m}+i(\delta\!+\!1)\alpha\right)\frac{\left(\imath\delta\mathcal{R}_{t}\!+\!\mathcal{R}_{l}\right)\left(\mathcal{R}_{t}\!+\!\imath\mathcal{R}_{l}\right)}{1-\exp\left[\!-\mathcal{Q}\!+\imath\frac{2\pi}{m}\right]}\right]\right\} .
\end{align*}
Again, only the term with $\delta=-1$ is finite yielding the final result 
\begin{align}
S_{xy}^{(m)} & =\varsigma_{xy}\!\frac{m}{2}\frac{c}{|e|H}\left[\mathcal{G}_{lt}\!-\!\mathcal{G}_{tl}\!+\!\mathrm{Im}\left[\frac{\left(\mathcal{R}_{t}\!+\!\imath\mathcal{R}_{l}\right)^{2}}{\exp\left(-\imath\frac{2\pi}{m}\right)\!-\!\exp\left(\!-\mathcal{Q}\right)}\right]\right]\label{eq:mfoldSxyApp}\\
= & \varsigma_{xy}\frac{m}{2}\frac{c}{|e|H}\left\{ \mathcal{G}_{lt}\!-\!\mathcal{G}_{tl}+\frac{\sin\left(\frac{2\pi}{m}\right)\left(\mathcal{R}_{t}^{2}\!-\!\mathcal{R}_{l}^{2}\right)\!+\!2\left[\cos\left(\frac{2\pi}{m}\right)\!-\exp\left(\!-\mathcal{Q}\right)\right]\mathcal{R}_{l}\mathcal{R}_{t}}{1-2\exp\left(-\mathcal{Q}\right)\cos\left(\frac{2\pi}{m}\right)+\exp\left(-2\mathcal{Q}\right)}\right\} ,\nonumber 
\end{align}
which have a form similar to the diagonal slice in Eq.~\eqref{eq:mfoldSxx}.
\end{widetext}

\section{Integrations along hyperbolic Fermi surface near van Hove point for
large hexagon \label{subsec:Integration-hyperbolicLH}}

In this appendix, we analytically calculate the segment integrals
$\mathcal{R}_{k}^{\mathrm{LH}}$ and $\mathcal{G}_{km}^{\mathrm{LH}}$
in Eqs.~\eqref{eq:mfoldRsDef-1} and \eqref{eq:mfoldGsrDef} determining
the single-slice contribution to conductivity in Eq.~\eqref{pzSliceFullPer}.
The Fermi pocket of the large concave hexagon is approximated by segments
of hyperbola, Eq.~\eqref{eq:PieceSpectrumLargeHex}, see Fig.~\ref{Fig:LargeHex}.
The momenta tracing the Fermi surface can be presented in the local
coordinates as 
\begin{subequations}
\begin{align}
p_{x} & =K_{j,x}+p_{v}c_{j}\!-p_{u}s_{j},\label{eq:LHRotCoord-px}\\
p_{y} & =K_{j,y}+p_{v}s_{j}\!+p_{u}c_{j}.\label{eq:LHRotCoord-py}
\end{align}
\end{subequations}
for $\frac{\pi}{3}(j\!-\!1)<\theta<\frac{\pi}{3}j$,
where $p_{x}\!=\!p\cos\theta$, $p_{y}\!=\!p\sin\theta$, $\theta_{j}\!=\left(\pi/3\right)(j\!-\!2)$,
with $j=(1,2,\ldots,6)$, $c_{j}\!\equiv\cos\theta_{j}$, and $s_{j}\!\equiv\sin\theta_{j}$.
For the integrations along the Fermi surface in Eqs.~\eqref{eq:mfoldRsDef-1}
and \eqref{eq:mfoldGsrDef}, it is convenient to use the hyperbolic
parametrization defined in Eq.~\eqref{eq:HypParam}. Correspondingly,
the velocity components in real space are expanded as\begin{subequations}
\begin{align}
v_{x} & =v_{v}c_{j}\!-v_{u}s_{j},\label{eq:LHRotCord-vx}\\
v_{y} & =v_{v}s_{j}\!+v_{u}c_{j},\label{eq:LHRotCord-vy}
\end{align}
\end{subequations}
where $v_{v}=-\frac{p_{v}}{m_{v}}$ and $v_{u}=\frac{p_{u}}{m_{u}}$
are velocity components in hyperbolic basis, which can be parametrized
as
\begin{equation}
v_{v}=v_{v0}\sinh t,\,v_{u}=-v_{u0}\cosh t,\label{eq:VelHypParam}
\end{equation}
with
\begin{align}
v_{u0} & =\frac{p_{u0}}{m_{u}},\,v_{v0}=\frac{p_{v0}}{m_{v}},\,\label{eq:LHvso}
\end{align}
so that $v_{v0}/v_{u0}\!=\!p_{u0}/p_{v0}\!=\!1/\sqrt{r_{m}}$. Therefore,
for the total in-plane velocity we obtain 
\begin{align*}
v & =\sqrt{v_{u0}^{2}\cosh^{2}t+v_{v0}^{2}\sinh^{2}t}.
\end{align*}
For the integrations in the exponent of Eq.~\eqref{pzSliceFullPer},
we obtain the relation
\begin{align}
\frac{dp}{v} & =\frac{\sqrt{\left(dp_{v}/dt\right)^{2}+\left(dp_{u}/dt\right)^{2}}}{\sqrt{v_{u0}^{2}\cosh^{2}t+v_{v0}^{2}\sinh^{2}t}}dt\nonumber \\
 & =\frac{p_{u0}}{v_{v0}}dt=\sqrt{m_{u}m_{v}}dt\label{eq:IntElemHypPar}
\end{align}
yielding
\begin{align}
\!\int\limits_{p}^{p^{\prime}}\!\frac{dp^{\prime\prime}}{v^{\prime\prime}}\frac{c}{|e|H\tau} & =\frac{t^{\prime}\!-\!t}{\omega_{h}}\nonumber \\
= & -\frac{\mathrm{arcsinh}\left(p_{v}^{\prime}/p_{v0}\right)\!-\!\mathrm{arcsinh}\left(p_{v}/p_{v0}\right)}{\omega_{h}}\label{eq:IntInExp-1}
\end{align}
In particular, for the whole-orbit integral in Eq.~\eqref{pzSliceFullPer}, we have
\[
\varointctrclockwise\frac{dp}{v}\frac{c}{|e|H\tau}\!=\!\frac{12t_{b}}{\omega_{h}}\!=\!6\mathcal{Q}_{\mathrm{LH}},
\]
where the limiting hyperbolic parameter $t_{b}$ is defined in Eq.~\eqref{eq:tb}.

Both $\mathcal{R}_{k}^{\mathrm{LH}}$ and $\mathcal{G}_{km}^{\mathrm{LH}}$
in Eqs.~\eqref{eq:mfoldRsDef-1} and \eqref{eq:mfoldGsrDef} are
determined by the integrals
\begin{equation}
\mathcal{J}_{k}(p)=\int_{p}^{p_{2}}\frac{dp^{\prime}}{v^{\prime}}v_{k}^{\prime}\exp\left(\!-\!\int_{p}^{p^{\prime}}\frac{dp^{\prime\prime}}{v^{\prime\prime}}\frac{c}{|e|H\tau}\right)\label{eq:JkInt}
\end{equation}
with $k=v,u$. Using the hyperbolic parametrization for the point
at the Fermi surface, $p(t)$, and for the Fermi velocities, Eqs.~\eqref{eq:HypParam}
and \eqref{eq:VelHypParam}, these integrals can be evaluated analytically
\begin{subequations}
\begin{align}
\mathcal{J}_{v}(t) & =p_{u0}G_{v}(t_{b},t),\label{eq:JvRes}\\
\mathcal{J}_{u}(t) & =-p_{v0}G_{u}(t_{b},t),\label{eq:JuRes}
\end{align}
\end{subequations}where the functions $G_{v}(t_{2},t_{1})$ and $G_{u}(t_{2},t_{1})$
are defined and evaluated as
\begin{widetext}
\begin{subequations}
\begin{align}
G_{v}(t_{2},t_{1}) & \equiv\int_{t_{1}}^{t_{2}}dt^{\prime}\sinh t^{\prime}\exp\left(\!-\frac{t^{\prime}\!-t_{1}}{\omega_{h}}\right)\nonumber \\
= & \frac{\left(\cosh t_{2}\!+\frac{1}{\omega_{h}}\sinh t_{2}\right)\exp\left(\!-\frac{t_{2}-t_{1}}{\omega_{h}}\right)\!-\!\left(\cosh t_{1}\!+\frac{1}{\omega_{h}}\sinh t_{1}\right)}{1\!-\!\omega_{h}^{-2}},\label{eq:Gvt1t2}\\
G_{u}(t_{2},t_{1}) & \equiv\int_{t_{1}}^{t_{2}}dt^{\prime}\cosh t^{\prime}\exp\left(\!-\frac{t^{\prime}\!-t_{1}}{\omega_{h}}\right)\nonumber \\
= & \frac{\left(\sinh t_{2}\!+\frac{1}{\omega_{h}}\cosh t_{2}\right)\exp\left(\!-\frac{t_{2}-t_{1}}{\omega_{h}}\right)\!-\!\left(\sinh t_{1}\!+\frac{1}{\omega_{h}}\cosh t_{1}\right)}{1\!-\!\omega_{h}^{-2}},\label{eq:Gut1t2}
\end{align}
\end{subequations}
and we also used the relations $\sqrt{m_{u}m_{v}}v_{v0}=p_{u0}$,
$\sqrt{m_{u}m_{v}}v_{u0}=p_{v0}$. 

The integrals over the whole segment $[p_{1},p_{2}]$ defined in Eq.~\eqref{eq:mfoldRsDef-1}
are given by $\mathcal{R}_{s}^{\mathrm{LH}}=\mathcal{J}_{s}(-t_{b})$
with $s=u,v$. Introducing shortened notations $G_{bs}\!\equiv\!G_{s}(t_{b},-t_{b})$,
we can present $\mathcal{R}_{v}^{\mathrm{LH}}=p_{u0}G_{bv}$ and $\mathcal{R}_{u}^{\mathrm{LH}}=-p_{v0}G_{bu}$.
The explicit presentations for the functions $G_{bv}$ and $G_{bu}$
directly following from Eqs.~\eqref{eq:Gvt1t2} and \eqref{eq:Gut1t2}
are given in Eqs.~ \eqref{eq:Gbv} and \eqref{eq:Gbu}. One can also
derive alternative presentations\begin{subequations}
\begin{align}
G_{bv} & =-2\exp\left(-\frac{t_{b}}{\omega_{h}}\right)\frac{\cosh t_{b}\sinh\left(t_{b}/\omega_{h}\right)\!-\!\frac{1}{\omega_{h}}\sinh t_{b}\cosh\left(t_{b}/\omega_{h}\right)}{1\!-\!\omega_{h}^{-2}}\label{eq:GbvLHAlt}\\
G_{bu} & =2\exp\left(-\frac{t_{b}}{\omega_{h}}\right)\frac{\sinh t_{b}\cosh\left(t_{b}/\omega_{h}\right)\!-\!\frac{1}{\omega_{h}}\cosh t_{b}\sinh\left(t_{b}/\omega_{h}\right)\!}{1\!-\!\omega_{h}^{-2}}\label{eq:GbuLHAlt}
\end{align}
\end{subequations}These functions have the following asymptotics\begin{subequations}
\begin{align}
G_{bv} & \simeq\begin{cases}
-\omega_{h}\left(\sinh t_{b}-\omega_{h}\cosh t_{b}\right), & \mathrm{for}\,\omega_{h}\ll1\\
-\frac{2}{\omega_{h}}\left(t_{b}\cosh t_{b}\!-\!\sinh t_{b}\right)\!, & \mathrm{for}\,\omega_{h}\gg2t_{b}
\end{cases},\label{eq:GbvAs}\\
G_{bu} & \simeq\begin{cases}
\omega_{h}\left(\cosh t_{b}-\omega_{h}\sinh t_{b}\right), & \mathrm{for}\,\omega_{h}\ll1\\
2\sinh t_{b}\left(1\!-\frac{t_{b}}{\omega_{h}}\right)\!, & \mathrm{for}\,\omega_{h}\gg2t_{b}
\end{cases},\label{eq:GbuAs}
\end{align}
\end{subequations}In calculations, we will also need the supplemental
integrals over $p$ defined as\begin{subequations}
\begin{align}
\bar{\mathcal{R}}_{v}^{\mathrm{LH}} & =\int\limits _{p_{1}}^{p_{2}}\frac{dp}{v}v_{v}\exp\left(\!-\!\int\limits _{p}^{p_{2}}\frac{dp^{\prime\prime}}{v^{\prime\prime}}\frac{c}{|e|H\tau}\!\right)=-\mathcal{R}_{v}^{\mathrm{LH}},\label{eq:RvLHsuppl}\\
\mathcal{\bar{R}}_{u}^{\mathrm{LH}} & =\int\limits _{p_{1}}^{p_{2}}\frac{dp}{v}v_{u}\exp\left(\!-\!\int\limits _{p}^{p_{2}}\frac{dp^{\prime\prime}}{v^{\prime\prime}}\frac{c}{|e|H\tau}\!\right)=\mathcal{R}_{u}^{\mathrm{LH}}.\label{eq:RuLHsuppl}
\end{align}
\end{subequations}

We now evaluate the same-segment integrals $\mathcal{G}_{km}^{\mathrm{LH}}$
in Eq.~\eqref{eq:mfoldGsrDef}. The term $\mathcal{G}_{vv}^{\mathrm{LH}}$
is connected with the integral $\mathcal{J}_{v}(p)$ defined in Eq.~\eqref{eq:JkInt}
as
\begin{align*}
\mathcal{G}_{vv}^{\mathrm{LH}}= & \int_{p_{1}}^{p_{2}}\frac{dp}{v}v_{v}\mathcal{J}_{v}(p).
\end{align*}
Using again the hyperbolic parametrization for for $p$ and $v_{v}$
together with the result for $\mathcal{J}_{v}(p)$ in Eqs.~\eqref{eq:JvRes}
and \eqref{eq:Gvt1t2}, we compute
\begin{align}
 & \mathcal{G}_{vv}^{\mathrm{LH}}=p_{u0}^{2}K_{bvv}\nonumber \\
 & K_{bvv}=\int_{-t_{b}}^{t_{b}}dt\sinh tG_{v}(t_{b},t)\nonumber \\
= & \frac{1}{1\!-\!\omega_{h}^{-2}}\left[\left(\cosh t_{b}\!+\!\frac{1}{\omega_{h}}\sinh t_{b}\right)\frac{\cosh t_{b}\!-\!\omega_{h}^{-2}\sinh t_{b}\!-\!\exp\left(\!-\frac{2}{\omega_{h}}t_{b}\right)\left(\cosh t_{b}\!+\!\frac{1}{\omega_{h}}\sinh t_{b}\right)}{1\!-\!\omega_{h}^{-2}}\right.\nonumber \\
 & \left.-\!\frac{1}{2\omega_{h}}\left(\sinh2t_{b}-2t_{b}\right)\right]\nonumber \\
= & \frac{1}{1\!-\!\omega_{h}^{-2}}\left[\sinh^{2}t_{b}+\frac{1-\exp\left(\!-\frac{2}{\omega_{h}}t_{b}\right)\left(\cosh t_{b}\!+\!\frac{1}{\omega_{h}}\sinh t_{b}\right)^{2}}{1\!-\!\omega_{h}^{-2}}-\!\frac{1}{2\omega_{h}}\left(\sinh2t_{b}-2t_{b}\right)\right].\label{eq:AppGvvLH}
\end{align}
Even though this result contains $1\!-\!\omega_{h}^{-2}$ denominators,
one can check that it does not diverge and approaches a finite limit
for $\omega_{h}\rightarrow1$,
\[
K_{bvv}\!\rightarrow\frac{1}{2}\left\{ \frac{\sinh\left(2t_{b}\right)}{2}\left(1\!+\frac{\exp\left(-2t_{b}\right)}{2}\right)\!-\!\exp\left(-2t_{b}\right)t_{b}-\frac{t_{b}\left(1\!+\!2t_{b}\right)}{2}\right\} .
\]
In a similar way, one can evaluate the term 
\[
\mathcal{G}_{uu}^{\mathrm{LH}}=\int_{p_{1}}^{p_{2}}\frac{dp}{v}v_{u}\mathcal{J}_{u}(p)
\]
using the result for $\mathcal{J}_{u}(p)$ in Eqs.~\eqref{eq:JuRes}
and \eqref{eq:Gut1t2}. Explicit calculation gives
\begin{align}
 & \mathcal{G}_{uu}^{\mathrm{LH}}=p_{v0}^{2}K_{buu}\nonumber \\
 & K_{buu}=\int_{-t_{b}}^{t_{b}}dt\cosh tG_{u}(t_{b},t)\nonumber \\
= & \frac{1}{1\!-\!\omega_{h}^{-2}}\left[\left(\sinh t_{b}\!+\frac{1}{\omega_{h}}\cosh t_{b}\right)\frac{\sinh t_{b}\!-\frac{1}{\omega_{h}^{2}}\cosh t_{b}+\exp\left(\!-\!\frac{2}{\omega_{h}}t_{b}\right)\left(\sinh t_{b}\!+\frac{1}{\omega_{h}}\cosh t_{b}\right)}{1\!-\!\omega_{h}^{-2}}\right.\nonumber \\
\!- & \left.\!\frac{1}{\omega_{h}}\left(\frac{\sinh2t_{b}}{2}\!+\!t_{b}\right)\right]\nonumber \\
= & \frac{1}{1\!-\!\omega_{h}^{-2}}\left[\cosh^{2}t_{b}-\frac{1\!-\!\exp\left(\!-\!\frac{2}{\omega_{h}}t_{b}\right)\left(\sinh t_{b}\!+\frac{1}{\omega_{h}}\cosh t_{b}\right)^{2}}{1\!-\!\omega_{h}^{-2}}\!-\!\frac{1}{\omega_{h}}\left(\frac{\sinh2t_{b}}{2}\!+\!t_{b}\right)\right].\label{eq:AppGuuLH}
\end{align}
The functions $\mathcal{K}_{bss}$ have the following asymptotics.
\begin{align}
\mathcal{K}_{bvv}\simeq & \begin{cases}
\omega_{h}\left[\left(\frac{\sinh2t_{b}}{2}-\!t_{b}\right)-\omega_{h}\sinh^{2}t_{b}\right], & \mathrm{for}\,\omega_{h}\ll1\\
\frac{1}{\omega_{h}}\left[t_{b}\left(\cosh2t_{b}+2\right)\!-\!\frac{3\sinh2t_{b}}{2}\right]\!, & \mathrm{for}\,\omega_{h}\gg2t_{b}
\end{cases},\label{eq:KbvvAs}\\
\mathcal{K}_{buu}\simeq & \begin{cases}
\omega_{h}\left[\left(\frac{\sinh2t_{b}}{2}\!+\!t_{b}\right)-\omega_{h}\cosh^{2}t_{b}\right], & \mathrm{for}\,\omega_{h}\ll1\\
2\sinh^{2}t_{b}\!+\frac{1}{\omega_{h}}\left(\frac{\sinh2t_{b}}{2}\!-t_{b}\left(2\sinh^{2}t_{b}\!+\!1\right)\right), & \mathrm{for}\,\omega_{h}\gg2t_{b}
\end{cases},\label{eq:KbuuAs}
\end{align}

The Hall component also contains the off-diagonal function,
\begin{align*}
\mathcal{G}_{uv}^{\mathrm{LH}}= & \int_{p_{1}}^{p_{2}}\frac{dp}{v}v_{u}\mathcal{J}_{v}(p),
\end{align*}
which we evaluate as 
\begin{align}
\mathcal{G}_{uv}^{\mathrm{LH}} & =-p_{u0}p_{v0}K_{buv}\nonumber \\
K_{buv}= & \int\limits _{-t_{b}}^{t_{b}}dt\cosh tG_{v}(t_{b,},t)\nonumber \\
= & \frac{1}{1\!-\!\omega_{h}^{-2}}\left[\left(\cosh t_{b}\!+\!\frac{1}{\omega_{h}}\sinh t_{b}\right)\frac{\left(\sinh t_{b}-\!\frac{1}{\omega_{h}}\cosh t_{b}\right)+\left(\sinh t_{b}+\!\frac{1}{\omega_{h}}\cosh t_{b}\right)\exp\left(-\frac{2}{\omega_{h}}t_{b}\right)}{1\!-\!\omega_{h}^{-2}}-\frac{\sinh2t_{b}}{2}-t_{b}\right]\nonumber \\
= & \frac{1}{1\!-\!\omega_{h}^{-2}}\left[\frac{-\!\frac{1}{\omega_{h}}+\left[\left(1\!+\!\omega_{h}^{-2}\right)\sinh\left(2t_{b}\right)/2\!+\!\frac{1}{\omega_{h}}\cosh\left(2t_{b}\right)\right]\exp\left(-\frac{2}{\omega_{h}}t_{b}\right)}{1\!-\!\omega_{h}^{-2}}-t_{b}\right].\label{eq:GuvAppLH}
\end{align}
This function has the following asymptotics 
\begin{align}
\mathcal{K}_{buv}\simeq & \begin{cases}
\omega_{h}^{2}\left(t_{b}-\omega_{h}\right), & \mathrm{for}\,\omega_{h}\ll1\\
\frac{1}{2}\sinh\left(2t_{b}\right)-t_{b}\!-\frac{1}{\omega_{h}}\left[1\!+\!t_{b}\sinh\left(2t_{b}\right)\!-\cosh\left(2t_{b}\right)\right], & \mathrm{for}\,\omega_{h}\gg2t_{b}
\end{cases},\label{eq:KbuvAs}
\end{align}
\end{widetext}
One can demonstrate antisymmetry with respect to the index
switching, $\mathcal{G}_{vu}^{\mathrm{LH}}=-\mathcal{G}_{uv}^{\mathrm{LH}}$.

In addition to the asymptotic limit $\omega_{h}\gg2t_{b}$, the functions
$G_{bs}$ and $\mathcal{K}_{bst}$ are also characterized by more
general asymptotic limits valid in the whole range $\omega_{h}\gg1$
\begin{subequations}
\begin{align}
G_{bv} & \!\simeq\!-\cosh t_{b}\left(1\!-\eta_{\mathrm{LH}}\right),\,G_{bu}\!\simeq\!\sinh t_{b}\left(1\!+\!\eta_{\mathrm{LH}}\right),\label{eq:GbvsInt}\\
\mathcal{K}_{bvv} & \!\simeq\cosh^{2}t_{b}\left(1\!-\eta_{\mathrm{LH}}\right),\,\mathcal{K}_{buu}\!\simeq\!\sinh^{2}t_{b}\left(1\!+\!\eta_{\mathrm{LH}}\right),\label{eq:KbssInt}\\
\mathcal{K}_{buv} & \!\simeq-t_{b}\!+\!\eta_{\mathrm{LH}}\sinh t_{b}\cosh t_{b} \label{eq:KbuvInt}
\end{align}
\end{subequations}
with $\eta_{\mathrm{LH}}\!\equiv\!\exp\left(-2t_{b}/\omega_{h}\right)$.
These asymptotics also describe behavior in the intermediate field
range $1\ll\omega_{h}\ll2t_{b}$. 

\section{Magnetoconductivity contribution for triangular pockets\label{app:TrianPockets}}

In this appendix we evaluate the conductivity slices for the triangular
pocket illustrated in Figs.~\ref{Fig:HypHexagon}(c) and \ref{Fig:TriangPocketAxes} 
based on the general presentations in Eqs.~\eqref{eq:SxxTri} and \eqref{eq:SxyTri}.
For this, we need to evaluate the segment integrals in Eqs.~\eqref{eq:mfoldQdef-1},
\eqref{eq:mfoldRsDef-1}, and \eqref{eq:mfoldGsrDef}. In our model,
the segment is the triangle side, which is composed of ``outgoing''
and ``incoming''  hyperbola branches. 

We will again use the local hyperbolic coordinates $p_{v}$ and $p_{u}$
for every branch. However, contrary to the hexagon pockets, the transverse
and longitudinal directions do not coincide with these coordinates.
This means that we need to expand the transverse and longitudinal
velocity components $v_{t}$ and $v_{l}$ over the hyperbolic components
$v_{v}$ and $v_{u}$ for every hyperbola branch. We write these expansions in a unified way
as 
\begin{equation}
v_{s}=\alpha_{a,s}v_{u}+\beta_{a,s}v_{v}\label{eq:vlt-vuv}
\end{equation}
where $s\!=\!t,l$, the index $a\!=\!o,i$ corresponds to the outgoing and incoming
branches, and the coefficients are given by
\begin{align*}
\alpha_{o,l} & \!=\!\alpha_{i,l}\!=\beta_{o,t}\!=\beta_{i,t}\!=\frac{\sqrt{3}}{2},\\
\alpha_{o,t} & \!=\!-\alpha_{i,t}\!=\,-\beta_{o,l}\!=\!\beta_{i,l}\!=\frac{1}{2}.
\end{align*}
The presentations in Eq.~\eqref{eq:vlt-vuv} can be also rewritten
in a compact complex way as $v_{t}+\imath v_{l}=\exp\left(-\imath\delta\frac{\pi}{6}\right)\left(v_{v}+\imath v_{u}\right)$
with $\delta=1$($-1$) for the outgoing (incoming) branch. We will
use the hyperbolic parametrization defined in Eq.~\eqref{eq:HypParam},
in which the hyperbolic parameter $t$ varies in the range $t_{c}\!<\!t\!<\!t_{b}$ for the outgoing branch and $-t_{b}\!<\!t\!<\!-t_{c}$ for the incoming branch. 
The hyperbolic velocity components $v_{v}$ and $v_{u}$ are parametrized identically with the large hexagon in Eq.\ \eqref{eq:VelHypParam}. 
Note that $v_{l}$ is continuous and $v_{t}$ has a jump at the matching
point between outgoing and incoming branches. 

Relations in Eqs.~\eqref{eq:IntElemHypPar} and \eqref{eq:IntInExp-1}
allow us straightforwardly evaluate the parameter $\mathcal{Q}_{\mathrm{T}}\!=\!\int\limits _{p_{k}}^{p_{k+1}}\frac{dp^{\prime}}{v^{\prime}}\frac{c}{|e|H\tau}$
yielding the result in Eq.~\eqref{eq:nuT}. We evaluate the integrals
$\mathcal{R}_{s}^{\mathrm{T}}$ in Eq.~\eqref{eq:SxxTri} for the segment between $p_k$ and $p_{k+1}$ as
\begin{align}
\mathcal{R}_{s}^{\mathrm{T}} & =\!\int\limits _{p_{k}}^{p_{b}}
\frac{dp^{\prime}}{v^{\prime}}\left(\alpha_{o,s}v_{u}^{\prime}+\beta_{o,s}v_{v}^{\prime}\right)\exp\left(\!-\!\mathfrak{J}_{p_{k}}^{p^{\prime}}\right)\nonumber \\
 & +\exp\left(\!-\!\mathfrak{J}_{p_{k}}^{p_{b}}\right)\int\limits _{p_{b}}^{p_{k+1}}
 \frac{dp^{\prime}}{v^{\prime}}\left(\alpha_{i,s}v_{u}^{\prime}+\beta_{i,s}v_{v}^{\prime}\right)\exp\left(\!-\!\mathfrak{J}_{p_{b}}^{p^{\prime}}\right)\label{eq:RTsInt}
\end{align}
where $p_{b}$ is the momentum at the matching point between outgoing
and incoming branches and we again employ the notation $\mathfrak{J}_{p}^{p^{\prime}}$
for the orbit integral between the points $p$ and $p^{\prime}$ as defined
in Eq.~\eqref{eq:Int-notation}. Introducing the contributions from the outgoing
and incoming branches
\begin{subequations}
\begin{align}
\mathcal{R}_{o,s}^{\mathrm{T}} & =\!\int\limits _{p_{k}}^{p_{b}}
\frac{dp^{\prime}}{v^{\prime}}v_{s}^{\prime}\exp\left(\!-\!\mathfrak{J}_{p_{k}}^{p^{\prime}}\right),\label{eq:RTis}\\
\mathcal{R}_{i,s}^{\mathrm{T}} & =\int\limits _{p_{b}}^{p_{k+1}}\frac{dp^{\prime}}{v^{\prime}}v_{s}^{\prime}\exp\left(\!-\!\mathfrak{J}_{p_{b}}^{p^{\prime}}\right),\label{eq:RTos}
\end{align}
\end{subequations}
we rewrite the presentation in Eq.~\ref{eq:RTsInt} as
\begin{equation}
\mathcal{R}_{s}^{\mathrm{T}}  =\mathcal{R}_{o,s}^{\mathrm{T}}+\sqrt{\eta_{T}}\mathcal{R}_{i,s}^{\mathrm{T}}.\label{eq:RTsPres}
\end{equation}
with $\eta_{T}\!\equiv\!\exp\left(\!-\mathcal{Q}_{\mathrm{T}}\right)$.
Substituting the hyperbolic parametrization, we obtain\begin{subequations}
\begin{align}
\mathcal{R}_{o,s}^{\mathrm{T}}\! & =\!\!\int\limits _{t_{c}}^{t_{b}}\!\!dt^{\prime}\!
\left(\!-\alpha_{o,s}p_{v0}\cosh t^{\prime}\!+\!\beta_{o,s}p_{u0}\sinh t^{\prime}\right)
\exp\!\left(\!-\frac{t^{\prime}\!-\!t_{c}}{\omega_{h}}\right)\nonumber \\
 & =\!-\alpha_{o,s}p_{v0}G_{bcu}\!+\!\beta_{o,s}p_{u0}G_{bcv},\label{eq:RTisResult}\\
\mathcal{R}_{i,s}^{\mathrm{T}}\! & =\!\!\int\limits _{-t_{b}}^{-t_{c}}\!\!dt^{\prime}\!
\left(\!-\alpha_{i,s}p_{v0}\cosh t^{\prime}\!+\!\beta_{i,s}p_{u0}\sinh t^{\prime}\right)
\exp\!\left(\!-\frac{t^{\prime}\!+\!t_{b}}{\omega_{h}}\right)\nonumber \\
 & =\!-\alpha_{i,s}p_{v0}G_{cbu}\!+\!\beta_{i,s}p_{u0}G_{cbv},\label{eq:RTosResult}
\end{align}
\end{subequations}
where we introduced notations
\[
G_{bck}=G_{k}(t_{b},t_{c}),\,G_{cbk}=G_{k}(-t_{c},-t_{b})
\]
with $k=u,v$ and the functions $G_{k}(t_{2},t_{1})$ with $k=v,u$ defined as
\begin{align*}
G_{v}(t_{2},t_{1}) & =\int_{t_{1}}^{t_{2}}\!dt^{\prime}\sinh t^{\prime}\exp\left(\!-\frac{t^{\prime}\!-t_{1}}{\omega_{h}}\right),\\
G_{u}(t_{2},t_{1}) & =\int_{t_{1}}^{t_{2}}\!dt^{\prime}\cosh t^{\prime}\exp\left(\!-\frac{t^{\prime}\!-t_{1}}{\omega_{h}}\right)
\end{align*}
 are evaluated in Eqs.~\eqref{eq:Gvt2t1Result} and \eqref{eq:Gut1t2Result}.
The functions $G_{bck}$ and $G_{cbk}$ have the following asymptotic
limits
\begin{align}
G_{bcv} & \simeq\omega_{h}\sinh t_{c},\,G_{bcu}\simeq\omega_{h}\cosh t_{c},\label{eq:GbckSmallH}\\
G_{cbv} & \simeq-\omega_{h}\sinh t_{b},\,G_{cbu}\simeq\omega_{h}\cosh t_{b},\label{eq:GcbkSmallH}
\end{align}
for $\omega_{h}\ll1$ and 
\begin{align}
G_{bcv} & \simeq\cosh t_{b}\sqrt{\eta_{T}}\!-\!\cosh t_{c},\,G_{bcu}\simeq\sinh t_{b}\sqrt{\eta_{T}}\!-\!\sinh t_{c}\label{eq:GbckHighH}\\
G_{cbv} & \simeq\cosh t_{c}\sqrt{\eta_{T}}\!-\!\cosh t_{b},\,G_{cbu}\simeq-\sinh t_{c}\sqrt{\eta_{T}}\!+\!\sinh t_{b}\label{eq:GcbkHighH}
\end{align}
for $\omega_{h}\gg1$. 

Using complex presentation for the trigonometric coefficients 
\begin{align*}
\alpha_{o,t}\!+\!\imath\alpha_{o,l} & \!=\imath\exp\left(-\imath\frac{\pi}{6}\right),\,\alpha_{i,t}\!+\!\imath\alpha_{i,l}\!=\imath\exp\left(\imath\frac{\pi}{6}\right),\\
\beta_{o,t}\!+\!\imath\beta_{o,l} & \!=\exp\left(-\imath\frac{\pi}{6}\right),\,\beta_{i,t}\!+\!\imath\beta_{i,l}\!=\exp\left(\imath\frac{\pi}{6}\right),
\end{align*}
and introducing two complex functions $\mathcal{G}_{bc}$ and $\mathcal{G}_{cb}$
in Eq.~\eqref{eq:ComplexGbcGcb}, we obtain complex combination which
determines the conductivity slices in Eqs.~\eqref{eq:SxxTri} and
\eqref{eq:SxyTri}.
\begin{equation}
\mathcal{R}_{t}^{\mathrm{T}}\!+\!\imath\mathcal{R}_{l}^{\mathrm{T}}\!=\!p_{u0}\exp\left(-\imath\frac{\pi}{6}\right)
\!\left[\mathcal{G}_{bc}\!+\!\exp\left(\!-\!\frac{\mathcal{Q}_{\mathrm{T}}}{2}\!+\!\imath\frac{\pi}{3}\right)\mathcal{G}_{cb}\right]\!.
\label{eq:TranComplexR}
\end{equation}
For the combination in Eqs.~\eqref{eq:SxxTri} and \eqref{eq:SxyTri}, we find
\begin{align}
 & \frac{\left(\mathcal{R}_{t}^{\mathrm{T}}\!+\!\imath\mathcal{R}_{l}^{\mathrm{T}}\right)^{2}}{\exp\left(-\imath\frac{2\pi}{3}\right)\!-\exp\left(\!-\mathcal{Q}_{\mathrm{T}}\right)}=p_{u0}^{2}\nonumber \\
 & \times\!\frac{\exp\left(\!-\imath\frac{\pi}{3}\right)\mathcal{G}_{bc}^{2}\!+\!2\exp\left(\!-\frac{\mathcal{Q}_{\mathrm{T}}}{2}\right)\mathcal{G}_{bc}\mathcal{G}_{cb}\!+\!\exp\left(\!-\mathcal{Q}_{\mathrm{T}}\!+\!\imath\frac{\pi}{3}\right)\mathcal{G}_{cb}^{2}}{\exp\left(-\imath\frac{2\pi}{3}\right)-\exp\left(\!-\mathcal{Q}_{\mathrm{T}}\right)}.\label{eq:TrianRComb}
\end{align}

The same-segment integrals $\mathcal{G}_{sr}^{\mathrm{T}}$ defined
in Eq.~\eqref{eq:mfoldGsrDef}, we also split into contributions
from incoming and outgoing segments,
\begin{align*}
\mathcal{G}_{sr}^{\mathrm{T}}= & \int_{p_{k}}^{p_{b}}\!\frac{dp}{v}v_{s}\int_{p}^{p_{b}}\frac{dp^{\prime}}{v^{\prime}}v_{r}^{\prime}\exp\left(\!-\!\mathfrak{J}_{p}^{p^{\prime}}\right)\\
+ & \int_{p_{k}}^{p_{b}}\!\frac{dp}{v}v_{s}\exp\left(\!-\mathfrak{J}_{p}^{p_{b}}\right)\int_{p_{b}}^{p_{k+1}}\frac{dp^{\prime}}{v^{\prime}}v_{r}^{\prime}\exp\left(\!-\!\mathfrak{J}_{p_{b}}^{p^{\prime}}\right)\\
+ & \int_{p_{b}}^{p_{k+1}}\!\frac{dp}{v}v_{s}\int_{p}^{p_{k+1}}\frac{dp^{\prime}}{v^{\prime}}v_{r}^{\prime}\exp\left(\!-\!\mathfrak{J}_{p}^{p^{\prime}}\right).
\end{align*}
The middle term here can be transformed as 
\[
\int\limits_{p_{k}}^{p_{b}}\!\frac{dp}{v}v_{s}\exp\left(\!-\mathfrak{J}_{p}^{p_{b}}\right)\int\limits_{p_{b}}^{p_{k+1}}\frac{dp^{\prime}}{v^{\prime}}v_{r}^{\prime}\exp\left(\!-\!\mathfrak{J}_{p_{b}}^{p^{\prime}}\right)
=\delta_{s}\mathcal{R}_{i,s}^{\mathrm{T}}\mathcal{R}_{i,r}^{\mathrm{T}}
\]
with $\delta_{l}=1$, $\delta_{t}=-1$, where the integrals $\mathcal{R}_{i,s}^{\mathrm{T}}$
are defined in Eq.~\eqref{eq:RTos} and evaluated in Eq.~\eqref{eq:RTosResult}.
Here we used an alternative presentation for $\mathcal{R}_{i,s}^{\mathrm{T}}$,
$\mathcal{R}_{i,s}^{\mathrm{T}}\!=\!\delta_{s}\int_{p_{k}}^{p_{b}}\!\frac{dp}{v}v_{s}\exp\left(\!-\mathfrak{J}_{p}^{p_{b}}\right)$ following from the symmetry properties of the velocity components. 
Substituting hyperbolic parametrization, we obtain
\begin{widetext}
\begin{align*}
\mathcal{G}_{sr}^{\mathrm{T}}= & \int\limits _{t_{c}}^{t_{b}}dt
\left(\alpha_{o,s}p_{v0}\cosh t\!-\!\beta_{o,s}p_{u0}\sinh t\right)
\left(\alpha_{o,r}p_{v0}G_{u}(t_{b},t)\!-\!\beta_{o,r}p_{u0}G_{v}(t_{b},t)\right)
+ \delta_{s}\mathcal{R}_{i,s}^{\mathrm{T}}\mathcal{R}_{i,r}^{\mathrm{T}} \\
+ & \int\limits _{-t_{b}}^{-t_{c}}dt
\left(\alpha_{i,s}p_{v0}\cosh t\!-\!\beta_{i,s}p_{u0}\sinh t\right)
\left(\alpha_{i,r}p_{v0}G_{u}(-t_{c},t)\!-\!\beta_{i,r}p_{u0}G_{v}(-t_{c},t)\right),
\end{align*}
where we used identities
\begin{align*}
\int_{t_{c}}^{t_{b}}dt\sinh t\exp\left(\!-\frac{t_{b}\!-t}{\omega_{h}}\right) & =-G_{v}(-t_{c},-t_{b}),\\
\int_{t_{c}}^{t_{b}}dt\cosh t\exp\left(\!-\frac{t_{b}\!-t}{\omega_{h}}\right) & =G_{u}(-t_{c},-t_{b}).
\end{align*}
Introducing the integrals $\mathcal{K}_{\alpha\beta}(t_{2},t_{1})$ as
\begin{align*}
\mathcal{K}_{vv}(t_{2},t_{1}) & \!=\!\int_{t_{1}}^{t_{2}}\!dt\sinh t\, G_{v}(t_{2},t),\: 
\mathcal{K}_{uu}(t_{2},t_{1})  \!=\!\int_{t_{1}}^{t_{2}}\!dt\cosh t\,G_{u}(t_{2},t),\\
\mathcal{K}_{vu}(t_{2},t_{1}) & \!=\!\int_{t_{1}}^{t_{2}}\!dt\sinh t\,G_{u}(t_{2},t),\: 
\mathcal{K}_{uv}(t_{2},t_{1})  \!=\!\int_{t_{1}}^{t_{2}}\!dt\cosh t\,G_{v}(t_{2},t),
\end{align*}
we arrive at the finite presentation for the same-segment integrals,
\begin{align}
\mathcal{G}_{sr}^{\mathrm{T}}= & \left(\alpha_{o,s}\alpha_{o,r}\!+\!\alpha_{i,s}\alpha_{i,r}\right)p_{v0}^{2}\mathcal{K}_{uu}(t_{b},t_{c})\!+\!\left(\beta_{o,s}\beta_{o,r}\!+\!\beta_{i,s}\beta_{i,r}\right)p_{u0}^{2}\mathcal{K}_{vv}(t_{b},t_{c})\nonumber \\
- & p_{v0}p_{u0}\left[\left(\alpha_{o,s}\beta_{o,r}-\beta_{i,s}\alpha_{i,r}\right)\mathcal{K}_{uv}(t_{b},t_{c})\!
+\!\left(\beta_{o,s}\alpha_{o,r}-\alpha_{i,s}\beta_{i,r}\right)\mathcal{K}_{vu}(t_{b},t_{c})\right]
+ 
\delta_{s}\mathcal{R}_{i,s}^{\mathrm{T}}\mathcal{R}_{i,r}^{\mathrm{T}} 
\label{eq:GsrTri}
\end{align}
The functions $\mathcal{K}_{\alpha\beta}(t_{2},t_{1})$ can be analytically
evaluated as
\begin{align*}
\mathcal{K}_{vv}(t_{2},t_{1})= & \frac{1}{1\!-\!\omega_{h}^{-2}}\left\{ \left(\cosh t_{2}\!+\!\frac{1}{\omega_{h}}\sinh t_{2}\right)\frac{\cosh t_{2}\!-\!\frac{1}{\omega_{h}}\sinh t_{2}-\exp\left(\!-\frac{1}{\omega_{h}}\left(t_{2}\!-t_{1}\right)\right)\left(\cosh t_{1}\!-\!\frac{1}{\omega_{h}}\sinh t_{1}\right)}{1\!-\!\omega_{h}^{-2}}\right.\\
 & -\!\frac{\cosh\left(2t_{2}\right)-\cosh\left(2t_{1}\right)}{4}-\!\frac{1}{2\omega_{h}}\left[\frac{\sinh\left(2t_{2}\right)\!-\sinh\left(2t_{1}\right)}{2}-t_{2}\!+t_{1}\right]\Bigg\},
\end{align*}
\begin{align*}
\mathcal{K}_{uu}(t_{2},t_{1})= & \frac{1/2}{1\!-\!\frac{1}{\omega_{h}^{2}}}\left\{ 2\left(\sinh t_{2}\!+\!\frac{1}{\omega_{h}}\cosh t_{2}\right)\frac{\sinh t_{2}\!-\frac{1}{\omega_{h}}\cosh t_{2}-\exp\left[\!-\!\frac{1}{\omega_{h}}\left(t_{2}\!-\!t_{1}\right)\right]\left(\sinh t_{1}\!-\frac{1}{\omega_{h}}\cosh t_{1}\right)}{1\!-\!\omega_{h}^{-2}}\!\right.\\
 & \!-\!\frac{\cosh\left(2t_{2}\right)\!-\!\cosh\left(2t_{1}\right)}{2}\!-\!\frac{1}{\omega_{h}}\left(\frac{\sinh\left(2t_{2}\right)-\sinh\left(2t_{1}\right)}{2}\!+t_{2}\!-t_{1}\right)\Bigg\},
\end{align*}
\begin{align*}
\mathcal{K}_{vu}(t_{2},t_{1}) & =\frac{1}{1\!-\!\omega_{h}^{-2}}\left\{ \frac{\frac{1}{\omega_{h}}\!-\exp\left(\!-\frac{t_{2}-t_{1}}{\omega_{h}}\right)\left(\sinh t_{2}+\frac{1}{\omega_{h}}\cosh t_{2}\right)\left(\cosh t_{1}\!-\frac{1}{\omega_{h}}\sinh t_{1}\right)}{1\!-\frac{1}{\omega_{h}^{2}}}\right.\\
+ & \!\left.\frac{\sinh\left(2t_{2}\right)\!+\!\sinh\left(2t_{1}\right)}{4}\!+\frac{t_{2}-t_{1}}{2}\!-\frac{1}{\omega_{h}}\frac{\cosh\left(2t_{2}\right)-\cosh\left(2t_{1}\right)}{4}\right\} ,
\end{align*}
\begin{align*}
\mathcal{K}_{uv}(t_{2},t_{1})= & \frac{1}{1\!-\frac{1}{\omega_{h}^{2}}}\Bigg\{-\frac{\!\frac{1}{\omega_{h}}\!+\exp\left(\!-\frac{t_{2}-t_{1}}{\omega_{h}}\right)\left(\cosh t_{2}\!+\frac{1}{\omega_{h}}\sinh t_{2}\right)\left(\sinh t_{1}\!-\frac{1}{\omega_{h}}\cosh t_{1}\right)}{1\!-\frac{1}{\omega_{h}^{2}}}\!\\
 & +\frac{\sinh\left(2t_{2}\right)\!+\!\sinh\left(2t_{1}\right)}{4}\!-\frac{t_{2}\!-t_{1}}{2}\!-\frac{1}{\omega_{h}}\frac{\cosh\left(2t_{2}\right)-\cosh\left(2t_{1}\right)}{4}\Bigg\}.
\end{align*}
These functions have the symmetry properties
\begin{align*}
\mathcal{K}_{ss}(-t_{1},-t_{2}) & =\mathcal{K}_{ss}(t_{2},t_{1}),\\
\mathcal{K}_{vu}(-t_{1},-t_{2}) & =-\mathcal{K}_{uv}(t_{2},t_{1}),
\end{align*}
which we have already used to simplify Eq.~\eqref{eq:GsrTri}. 
The functions
$\mathcal{K}_{\alpha\beta}(t_{b},t_{c})$ have the following behavior
in different limits:
\begin{align}
\mathcal{K}_{vv}(t_{b},t_{c})\simeq & \frac{\omega_{h}}{2}\left(\frac{\sinh\left(2t_{b}\right)\!-\!\sinh\left(2t_{c}\right)}{2}-t_{b}\!+t_{c}\!\right)\nonumber \\
- & \frac{\omega_{h}^{2}}{2}\left(\frac{\cosh\left(2t_{b}\right)\!+\!\cosh\left(2t_{c}\right)}{2}-1\right),\label{eq:KvvbcSmallH}\\
\mathcal{K}_{uu}(t_{b},t_{c})\simeq & \frac{\omega_{h}}{2}\left(\frac{\sinh\left(2t_{b}\right)\!-\!\sinh\left(2t_{c}\right)}{2}\!+t_{b}\!-t_{c}\right)\nonumber \\
- & \frac{\omega_{h}^{2}}{2}\left(\frac{\cosh\left(2t_{b}\right)\!+\!\cosh\left(2t_{c}\right)}{2}+1\right),\label{eq:KuubcSmallH}\\
\mathcal{K}_{vu}(t_{b},t_{c})= & \frac{\omega_{h}}{4}\left(\cosh\left(2t_{b}\right)\!-\!\cosh\left(2t_{c}\right)\right)\nonumber \\
- & \frac{\omega_{h}^{2}}{4}\left(\sinh\left(2t_{b}\right)\!+\!\sinh\left(2t_{c}\right)\!+2\left(t_{b}\!-\!t_{c}\right)\right)\label{eq:KvubcSmallH}
\end{align}
for $\omega_{h}\ll1$ and 
\begin{align}
\mathcal{K}_{vv}(t_{b},t_{c}) & \simeq\frac{\cosh^{2}t_{b}\!+\!\cosh^{2}t_{c}}{2}\!-\!\cosh t_{b}\cosh t_{c}\sqrt{\eta_{T}},\label{eq:KvvbcHighH}\\
\mathcal{K}_{uu}(t_{b},t_{c}) & \simeq\frac{\sinh^{2}t_{b}+\sinh^{2}t_{c}}{2}\!-\sinh t_{b}\sinh t_{c}\sqrt{\eta_{T}}\label{eq:KuubcHighH}\\
\mathcal{K}_{vu}(t_{b},t_{c}) & \simeq\frac{\sinh\left(2t_{b}\right)\!+\!\sinh\left(2t_{c}\right)}{4}\!+\frac{t_{b}-t_{c}}{2}\!-\sqrt{\eta_{T}}\sinh t_{b}\cosh t_{c}\label{eq:KvubcHighH}
\end{align}
for $\omega_{h}\gg1$. 

We proceed with evaluation of the diagonal conductivity slice $S_{xx}^{\mathrm{T}}$
based on the complex presentation in Eq.~\ref{eq:SxxTri}. For combination
in the first line of Eq.~\ref{eq:SxxTri}, we obtain from Eq.~\eqref{eq:GsrTri}
\begin{align}
\mathcal{G}_{tt}^{\mathrm{T}}+\mathcal{G}_{ll}^{\mathrm{T}}= & 2p_{v0}^{2}\mathcal{K}_{uu}(t_{b},t_{c})\!+\!2p_{u0}^{2}\mathcal{K}_{vv}(t_{b},t_{c})\nonumber \\
+ & \frac{1}{2}p_{v0}^{2}G_{cbu}^{2}\!-\!\sqrt{3}p_{v0}p_{u0}G_{cbu}G_{cbv}\!-\!\frac{1}{2}p_{u0}^{2}G_{cbv}^{2}.\label{eq:TriGttGllSum}
\end{align}
Using the complex function $\mathcal{G}_{cb}$ introduced in Eq.~\eqref{eq:ComplexGbcGcb},
we can rewrite the second line in this equation as
\[
\frac{1}{2}p_{v0}^{2}G_{cbu}^{2}\!-\!\sqrt{3}p_{v0}p_{u0}G_{cbu}G_{cbv}\!-\!\frac{1}{2}p_{u0}^{2}G_{cbv}^{2}=-p_{u0}^{2}\mathrm{Re}\left[\exp\left(\imath\frac{\pi}{3}\right)\mathcal{G}_{cb}^{2}\right].
\]
Substituting the above results and Eq. \eqref{eq:TrianRComb} into
Eq.~\ref{eq:SxxTri}, we obtain
\begin{align*}
S_{xx}^{\mathrm{T}} & =\frac{3}{2}\frac{c}{|e|H}p_{u0}^{2}\left\{ 2\mathcal{K}_{vv}(t_{b},t_{c})+\!2r_{m}\mathcal{K}_{uu}(t_{b},t_{c})\right.\\
- & \left.\mathrm{Re}\left[\exp\left(\imath\frac{\pi}{3}\right)\mathcal{G}_{cb}^{2}+\frac{\exp\left(-\imath\frac{\pi}{3}\right)\mathcal{G}_{bc}^{2}\!+\!2\mathcal{G}_{bc}\exp\left(\!-\!\frac{\mathcal{Q}_{\mathrm{T}}}{2}\right)\mathcal{G}_{cb}+\exp\left(\!-\!\mathcal{Q}_{\mathrm{T}}+\!\imath\frac{\pi}{3}\right)\mathcal{G}_{cb}^{2}}{\exp\left(-\imath\frac{2\pi}{3}\right)\!-\!\exp\left(\!-\mathcal{\mathcal{Q}_{\mathrm{T}}}\right)}\right]\right\} ,
\end{align*}
which is identical to the result in Eq.~\ref{eq:SxxTriResult}
of the main text.

For the Hall conductivity slice in Eq.~\eqref{eq:SxyTri}, we
need the function $\mathcal{G}_{lt}^{\mathrm{T}}$, which we evaluate
from the general presentation in Eq.~\eqref{eq:GsrTri} as
\begin{align}
\mathcal{G}_{lt}^{\mathrm{T}}= & -\!p_{v0}p_{u0}\left[\mathcal{K}_{uv}(t_{b},t_{c})\!-\!\mathcal{K}_{vu}(t_{b},t_{c})\right]\!\nonumber \\
- & \frac{\sqrt{3}}{4}\left(p_{v0}^{2}G_{cbu}^{2}\!-\!p_{u0}^{2}G_{cbv}^{2}\right)\!-\!\frac{1}{2}p_{v0}p_{u0}G_{cbv}G_{cbu}\label{eq:GltTri}
\end{align}
Using the complex function $\mathcal{G}_{cb}$ in Eq.~\eqref{eq:ComplexGbcGcb},
the second line can be presented as
\[
-\frac{\sqrt{3}}{4}\left(p_{v0}^{2}G_{cbu}^{2}\!-\!p_{u0}^{2}G_{cbv}^{2}\right)\!-\!\frac{1}{2}p_{v0}p_{u0}G_{cbv}G_{cbu}=\frac{p_{u0}^{2}}{2}\mathrm{Im}\left[\exp\left(\imath\frac{\pi}{3}\right)\mathcal{G}_{cb}^{2}\right].
\]
As $\mathcal{G}_{tl}^{\mathrm{T}}=-\mathcal{G}_{lt}^{\mathrm{T}}$,
substituting the above results and Eq.~\eqref{eq:TrianRComb} into
Eq.~\eqref{eq:SxyTri} yields
\begin{align*}
S_{xy}^{\mathrm{T}} & =-\frac{3}{2}\frac{c}{|e|H}p_{u0}^{2}\left\{ \!2\sqrt{r_{m}}\left[\mathcal{K}_{vu}(t_{b},t_{c})\!-\!\mathcal{K}_{uv}(t_{b},t_{c})\right]\right.\\
+ & \left.\mathrm{Im}\left[\exp\left(\imath\frac{\pi}{3}\right)\mathcal{G}_{cb}^{2}+\frac{\exp\left(-\imath\frac{\pi}{3}\right)\mathcal{G}_{bc}^{2}\!+\!2\mathcal{G}_{bc}\exp\left(\!-\!\frac{\mathcal{Q}_{\mathrm{T}}}{2}\right)\mathcal{G}_{cb}+\exp\left(\!-\!\mathcal{Q}_{\mathrm{T}}+\!\imath\frac{\pi}{3}\right)\mathcal{G}_{cb}^{2}}{\exp\left(-\imath\frac{2\pi}{3}\right)\!-\!\exp\left(\!-\mathcal{\mathcal{Q}_{\mathrm{T}}}\right)}\right]\right\} .
\end{align*}
This result is equivalent to Eq.~\eqref{eq:SxyTriResult} of the
main text. Note that the difference $\mathcal{K}_{vu}(t_{b},t_{c})\!-\!\mathcal{K}_{uv}(t_{b},t_{c})$
in the first line can be represented in alternative form
\begin{align*}
 & \mathcal{K}_{vu}(t_{b},t_{c})-\mathcal{K}_{uv}(t_{b},t_{c})\\
= & -\frac{\exp\left[\left(1\!-\frac{1}{\omega_{h}}\right)\left(t_{b}\!-\!t_{c}\right)\right]\!-\!1}{2\left(1\!-\frac{1}{\omega_{h}}\right)^{2}}+\frac{\exp\left[\!-\left(1\!+\frac{1}{\omega_{h}}\right)\left(t_{b}\!-\!t_{c}\right)\right]\!-\!1\!}{2\left(1\!+\frac{1}{\omega_{h}}\right)^{2}}\!+\frac{t_{b}\!-\!t_{c}}{1\!-\frac{1}{\omega_{h}^{2}}},
\end{align*}
which is somewhat more convenient for numerical evaluation.\end{widetext}

\bibliography{MagnCondKagome}

\end{document}